\documentclass[9pt,twocolumn,twoside]{pnas-new}

\templatetype{pnasresearcharticle} 

\usepackage{mathrsfs} \usepackage{amssymb} \usepackage{graphicx,color} 
\usepackage{bbm} \usepackage{multirow} \usepackage{bm} \usepackage{mathrsfs} 
\usepackage{commath} \usepackage{stmaryrd} 
\usepackage{tikz}\usetikzlibrary{shapes,arrows,positioning,calc}
\usepackage{comment}\usepackage{bbold}
\usepackage{esint}\usepackage{upgreek} 
\usepackage{braket} 

\let\a=\alpha \let\b=\beta \let\g=\gamma \let\d=\delta
 \let\z=\zeta  
 \let\m=\mu   \let\p=\pi
\let\s=\sigma   \let\c=\chi

\let\Si=\Sigma   
  \let\th=\theta \let\io=\infty
\let\om=\omega
\def\ie{{\textit{i.e.} }}\def\eg{{\textit{e.g.} }}

\def\MM{{\cal M}}

\def\LL{{\cal L}}  \def \OO{{\cal O}}
\def\DD{{\cal D}} 
\def\UU{{\cal U}}

\def\Im{{\rm Im}\,}

\def\to{\rightarrow} \def\la{\left\langle} \def\ra{\right\rangle}

\def\ZZZ{\mathbbm{Z}}

\def\dd{\mathrm{d}}
\def\id{\mathbbm{1}}
\def\Tr{\mathrm{Tr}}

\newcommand{\beq}{\begin{equation}} \newcommand{\eeq}{\end{equation}}
 \newcommand{\wt}{\widetilde}

\title{Impact of jamming criticality on low-temperature anomalies in structural glasses}

\author[a]{Silvio Franz}
\author[b,1]{Thibaud Maimbourg} 
\author[c,d,e]{Giorgio Parisi}
\author[b,f]{Antonello Scardicchio}

\affil[a]{LPTMS, CNRS, Universit\'e Paris-Sud, Universit\'e Paris-Saclay, 91405 Orsay, France}
\affil[b]{The Abdus Salam International Centre for Theoretical Physics, Strada Costiera 11, 34151 Trieste, Italy}
\affil[c]{Dipartimento di Fisica, Sapienza Universit\`a di Roma, Piazzale A. Moro 2, I-00185 Rome, Italy}
\affil[d]{Nanotec-CNR, UOS Rome, Sapienza Universit\`a di Roma, Piazzale A. Moro 2, I-00185 Rome, Italy}
\affil[e]{INFN, Sezione di Roma 1, Piazzale A. Moro 2, 00185, Rome, Italy}
\affil[f]{INFN, Sezione di Trieste, Via Valerio 2, 34126 Trieste, Italy}

\leadauthor{Franz} 

\significancestatement{
The thermodynamics of glasses (disordered solids) exhibits experimentally a universal behaviour upon cooling below 10 K that is very different from regular solids. 
Theoretical interpretations of this discrepancy are notably hard to vindicate from first principles. 
Here through the analytic solution of a simple microscopic model of a glass we propose an alternative scenario to quantum tunneling between similar energy states. We suggest that the anomalous scaling gets broader in temperature range as the system is uncompressed towards the jamming transition where, classically, soft spherical particles would be in contact in their ground state. This scaling is thus most easily detected close to jamming and fades out away from it.
}
\correspondingauthor{\textsuperscript{1}To whom correspondence should be addressed. E-mail: thibaud.maimbourg@lptms.u-psud.fr}
\keywords{low-temperature glasses $|$ jamming $|$ marginal stability $|$ Debye model} 

\begin{abstract}
We present a novel mechanism for the anomalous behaviour of the specific heat 
in low-temperature amorphous solids. The analytic solution of a mean-field model belonging to the same universality class as high-dimensional glasses, the spherical perceptron, suggests that there exists a crossover temperature above which the specific heat scales linearly with temperature while below it a cubic scaling is displayed. This relies on two crucial features of the phase diagram: \textit{(i)} The marginal stability of the free-energy landscape, which induces a gapless phase responsible for the emergence of a power-law scaling \textit{(ii)} The vicinity of the classical jamming critical point, as 
the crossover temperature gets lowered when approaching it. This scenario arises from a direct study of the thermodynamics of the system in the quantum regime, where we show that, contrary to crystals, the Debye approximation does not hold. 
\end{abstract}

\begin{document}

\maketitle
\thispagestyle{firststyle}
\ifthenelse{\boolean{shortarticle}}{\ifthenelse{\boolean{singlecolumn}}{\abscontentformatted}{\abscontent}}{}


\dropcap{A}morphous solids exhibit many puzzling differences with respect to crystals, their ordered counterparts. 
One of the prominent enigma is the behaviour at cryogenic temperatures ($T< 10$ K) of thermodynamic quantities such as the specific heat $C_V$, 
measured in Zeller and Pohl's (ZP) seminal experiment~\cite{ZP71}. It revealed that for most glassformers $C_V$ scales universally 
linearly in $T$, being in great excess with respect to the usual cubic dependence in crystals well explained by Debye's theory of phononic excitations around a periodic lattice~\cite{kittel}.
Almost half a century later, the situation is still much controversial. Experimentally, 
the most recent studies claim that hyperaged amber obey ZP's scaling while ultrastable vapor-deposited glasses conform to the Debye law~\cite{LQMKH14,PCRTRVR14,PCJRR14,TCBSE16}. 
Theory-wise, a possible interpretation was devised right after ZP's results, 
based on the idea that in amorphous matter atoms or groups of atoms 
may arrange themselves equally well in different metastable configurations and at very low $T$ would tunnel between these energy levels, 
giving rise to disordered effective two-level systems (TLS), 
responsible of the linear scaling~\cite{AHV72,Ph87,YL88}. 
Although this explanation was later extended through the Soft-Potential Model, allowing to quantitatively reproduce experimental 
data at the expense of the introduction of more fitting parameters~\cite{KKI83,Kl88,BGGPRS92,Pa93}, 
the nature of the TLS together with their interaction are still elusive~\cite{YL88,KvMWZ07,LV13,PCRTRVR14,Lu18}. 
Another viewpoint on the basic TLS degrees of freedom has been suggested by the mosaic picture of the
Random First-Order Transition theory~\cite{LW01,LW07,Lu18}, which suffers from the same difficulty to root it in first principles.

In this work we explore a different mechanism provided by the mean-field (MF) theory of structural glasses (SG).
The idea relies on two ingredients. First, it has been suggested by the study of SG in infinite spatial dimension 
that low-$T$ amorphous solids lie in a free-energy landscape riddled with minima and barriers whose distribution is hierarchical~\cite{KPUZ13,CKPUZ14,nature,BU16,SBZ17}. 
A crucial property of such a landscape is that many local minima are only marginally stable~\cite{BW09,ML11,DKP12,XLN17,CKPUZ17}
(referred hereafter as landscape marginally stable, LMS), 
inducing soft vibrational modes~\cite{WSNW05,XVLN10,DGLFDLW14}. LMS phases are greatly universal in MF glasses~\cite{MW15,FLD18} and 
display clear thermodynamic signatures in the quantum regime~\cite{S05,AM12}.
Second, finite-ranged soft-core or hard-core particle systems present a jamming transition at $T=0$ (for soft potentials) and high enough density, where particles cannot satisfy anymore the non-overlapping requisite~\cite{LNvSW11,BC19}. At this transition the system becomes isostatic~\cite{OHSLN03,LNvSW11}, meaning that the number of mechanical constraints is exactly matched by the number of degrees of freedom. This property implies that the system is mechanically marginally stable (MMS), in the sense that the removal 
of one constraint (a contact between particles) causes a flow of particles along a soft mode, without energy cost~\cite{Wy12,LDW13,MW15,XLN17}. As a result the vicinity of a jamming critical point enhances further the number of soft modes. This classical physical picture near jamming is found qualitatively and quantitatively both in infinite-dimensional systems and in the Effective Medium Theory, 
irrespectively of dimension. Despite their MF flavour, most of its aspects are observed in numerical simulations~\cite{OHSLN03,DGLFDLW14,DGLBW14,nature,CCPZ12,CCPZ15,CKPUZ17}. 

The interest of the MF limit is that its outcomes are microscopically grounded. Nevertheless 
making progress within the theory is filled with challenges. The presence of seemingly universal features then
suggests to simplify the approach by building an ensemble of continuous constraint satisfaction problems
belonging to the same universality class as large-dimensional SG close to jamming. It turns out 
the spherical perceptron model of neural networks~\cite{GD88} fulfills the role of the simplest model in this class~\cite{FP16,FPUZ15,AFP16,FPSUZ17}.
In this paper, we surmise that the predictions from this model bring useful insights into 
the MF theory of SG in the corresponding phases. 
We first compute the specific heat in a Debye harmonic approximation at low temperature. 
We then resort to a more controlled derivation directly from the exact free energy of the model. The main result is that the LMS phase produces a gapless scaling of the self-energy which entails a power-law scaling of the specific heat. Besides, the presence of a jamming transition inside this phase in the classical model controls a crossover temperature $T_{\rm cut}$ which vanishes at jamming to leading order in $\hbar$. One finds $C_V\propto T^3$ for $T\ll T_{\rm cut}$ and $C_V\propto T$ for $T\gg T_{\rm cut}$; therefore, if $T_{\rm cut}$ is small enough, \ie close to jamming, one can observe this linear dependence.

~~\\
\noindent{\sffamily\bfseries The model: random obstacles on a sphere --~~}%
The spherical perceptron is defined as follows. 
One considers a particle $\bm X=(X^1,\dots,X^N)$ constrained to move on a $N$-dimensional sphere of radius\footnote{The unit of length $\DD$, set to 1, has the interpretation of a typical
inter-particle  distance in a glass.} $\sqrt N$ and interacting with an assembly of $M=\a N$ obstacles $\bm\xi_\m$ randomly placed 
on the sphere (\ie each component is independently normal distributed with zero mean and
unit variance), through a soft-sphere pair potential\footnote{$\th$ represents the Heaviside step function.} $v(h)=\varepsilon h^2\th(-h)/2$.
The spirit of the model is to replace the physics of a collection of
particles in a glassy configuration, whose disorder is self-generated by the interactions, with a tagged particle evolving in a disordered background. 
The potential energy of the system is $H_{\rm cl}=\sum_{\m=1}^M v(h_\m)$, implying an energy cost if a variable $h_\m=\frac{1}{\sqrt N}\bm X\cdot \bm\xi_\m-\s$ is negative, and zero energy if it is positive. 
The $h_\m$'s are thus microscopic geometric constraints to satisfy relatively to each obstacle. For $\s>0$ a ground state is obtained when the particle is closer than some distance to every $\bm\xi_\m$, therefore wandering in a convex volume, 
while for $\s<0$ a zero energy means staying away from any obstacle, as in a liquid configuration, allowed positions of the particle forming a non-convex region. The latter regime is the interesting one 
from a glassy perspective, as the particle's dynamics is impeded by the obstacles. 

\begin{figure}[h!]
\centering
\includegraphics[width=0.98\linewidth]{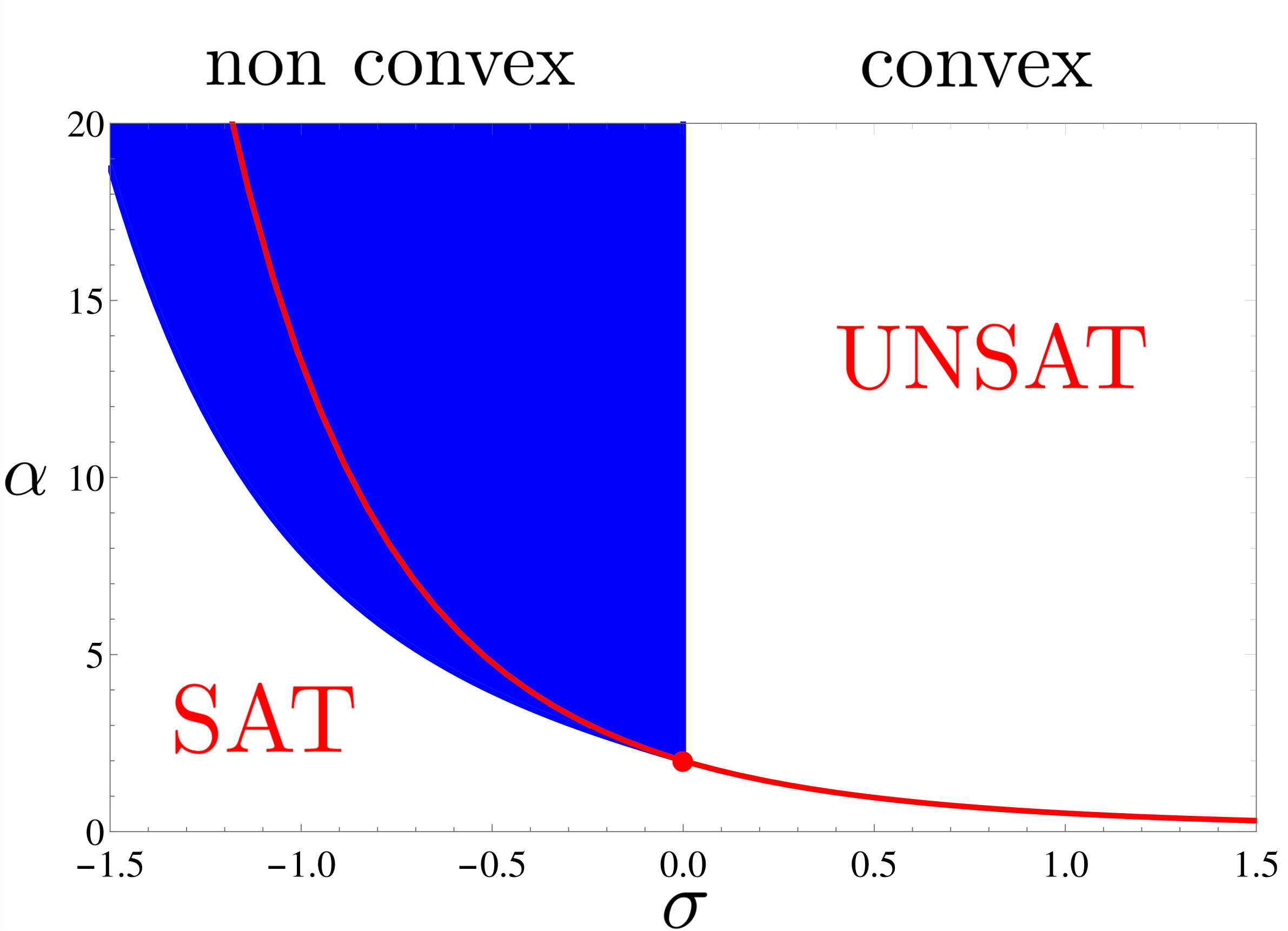}
\caption{Phase diagram of the classical $T=0$ model~\cite{FP16,FPSUZ17}.
$\s=0$ delimits the non-convex and convex regimes. 
The SAT and UNSAT phases are separated by a SAT-UNSAT ``jamming'' line $\a_J^{\rm cl}(\s)$ (red solid line). 
The blue region is the LMS phase. 
In the UNSAT phase, the low-$T$ specific heat scales as $C_V\sim T^3/(T_{\rm cut}^2+T^2)$ in the LMS regime, and is exponentially small outside (gapped).
See Ref.~\cite{FPSUZ17} for a more complete phase diagram in the SAT phase (shown at RS level here).}
\label{fig:phd}
\end{figure}

This is confirmed by the study of the classical $T=0$ phase diagram $(\s,\a)$ of the model, which can be computed exactly in the thermodynamic limit $N\to\io$~\cite{FP16,FPSUZ17} (depicted in Fig.~\ref{fig:phd}). 
The system must be in a ground state, meaning that the constraints are enforced as much as possible.
At low density of obstacles (small enough $\a$), typical configurations can satisfy all the constraints (SAT phase), whereas this is not the case anymore at high enough density (UNSAT phase). 
These two phases are separated by a sharp SAT-UNSAT transition line. In the convex regime $\s>0$, the free energy landscape consists in a single well, whereas the non-convex phase $\s<0$ 
contains a high-density region where the landscape is very rugged and marginally stable, typical of MF SG close to jamming~\cite{nature,BU16,CKPUZ17,SBZ17}. The SAT-UNSAT  line in this
regime falls deep within this LMS phase, and may be referred to as a ``jamming line'' since it is in the same universality class as the MF jamming transition of hard spheres 
where the system cannot find a stable configuration without overlaps~\cite{FP16,FPSUZ17,CKPUZ14,nature}.

The study of quantum spin glasses shows that the presence of a LMS phase affects deeply the behaviour of the system, and in particular the scaling of the specific heat~\cite{S05,CGSS01,AM12}. 
In the following we shall pinpoint an analogous phenomenon in SG, which in addition is affected by the criticality at jamming.

\noindent{\sffamily\bfseries Debye approximation in the UNSAT phase --~~}%
An early, successful application of quantum mechanics to low-temperature physics is the correct prediction of the specific heat in ordered solids by Debye's model~\cite{kittel}. 
It describes the system at low temperature $T=1/\b$ by small oscillations around a minimum of the potential-energy landscape, 
and only the density of states (DOS) for such harmonic modes enters the calculation: one only needs to compute the Hessian of the Hamiltonian around its minima and quantize the ensuing vibrational modes. 
The same line of  thought can be applied to the spherical perceptron. 
The Hessian $\partial^2H_{\rm cl}/\partial X^i\partial X^j$ is a positive semi-definite random matrix, whose spectral density in the UNSAT phase has been computed in~\cite{FPUZ15}. 
The frequencies of such ``phonons'' $\om$ are related to the eigenvalues of the Hessian $\lambda$ as $\lambda=\MM\om^2$ ($\MM$ being the particle mass). 
The DOS $D(\om)$ has its support in a positive interval $[\om_-,\om_+]$:
\begin{equation}\label{eq:DOS}
D(\om)\propto \om\frac{\sqrt{(\om^2-\om_-^2)(\om_+^2-\om^2)}}{\om^2+\zeta/\MM}
\end{equation}
where $\om_->0$ in the convex region, but for $\s\leqslant0$, $\om_-=0$. 
The parameters are $\MM\om_\pm^2=\left(\sqrt{[1]}\pm1\right)^2-\zeta$ (energies are expressed in units of $\varepsilon$) and $\zeta=[h^2]+\s[h]$ with $[h^n]=\overline{\la\frac1N\sum_{\m=1}^M (h_\m)^n\th(-h_\m) \ra}$ 
the moments of the configuration in a typical potential-energy minimum~\cite{FPUZ15}.
The overline represents the disorder average over the obstacles, whereas the brackets stand for thermal average. 
For $\s\leqslant0$, $\z\geqslant0$ defines a cutoff frequency 
$\om_*=\sqrt{\zeta/\MM}$ related to excess vibrational modes (with respect to a crystal) and the boson peak~\cite{FPUZ15,OHSLN03,WSNW05,XVLN10,DGLFDLW14}.
The quantization of these harmonic vibrations brings an average energy per mode $\hbar\om f_{\rm B}(\om)$ (removing the zero-point energy), where $f_{\rm B}(\om)=(e^{\b\hbar\om}-1)^{-1}$ is the Bose-Einstein distribution, and relates the internal energy $\UU_{\rm Debye}$ to the DOS as
\begin{equation}
 \frac{\UU_{\rm Debye}}{N}=\int_0^\io \dd \om\, D(\om)\hbar\om f_{\rm B}(\om)
\end{equation}
From the latter equation one may compute the low-$T$ specific heat $C_V=\partial\UU/\partial T$. 
The stability of the single-well free energy is related to the fact that $\om_->0$ in the corresponding region ($\s>0$). The spectrum has a gap $\hbar\om_-$, which entails $C_V\propto e^{-\hbar\om_-/T}$ when $T\to0$. 
Conversely, the LMS phase implies that the landscape has flat directions along which soft modes can flow, and the gap closes, \ie $\om_-=0$. 
There, away from the jamming line $\om_*\neq0$, which directly gives a cubic  
$C_V\propto T^3$ (see SI Appendix for derivation details).
On the jamming line, the constraints are on the verge of satisfiability $h_\m=0$, making instead $\om_*$ vanish and one gets a linear 
$C_V\propto T$. The different scaling is caused by the emergence of a large number of MMS modes at jamming related to isostaticity, 
which induce a flat DOS at small frequency instead of $D(\om)\propto\om^2$ in the rest of the LMS UNSAT phase. \\
\indent However the legitimacy of the Debye approximation to include quantum effects in this model may be questioned. 
First of all, LMS implies flat directions that cannot be treated in a harmonic approximation. Second, the jamming transition is  a purely classical phenomenon, which becomes a crossover if quantum effects are included.
One needs a fully quantum-mechanical treatment from the start, which we set up to do in the next section.

~~\\
\noindent{\sffamily\bfseries Quantization and thermodynamics --~~}%
A more accurate study can be performed by investigating directly the free energy of the system. 
We first quantize the model 
by supplying a momentum to the particle and imposing the canonical commutation relations. The Hamiltonian becomes
\begin{equation}\label{eq:originalH}
 \hat H=\frac{\hat{\bm P}^2}{2\MM}+\sum_{\m=1}^M v(h_\m(\hat{\bm X}))\ ,\quad \textrm{with} \quad  [\hat X^i,\hat P^j]=i\hbar\d_{ij}\hat\id
\end{equation}
The thermodynamics of the system is derived from the Feynman path integral representation of the partition function 
$\overline{\Tr\, \exp(-\b \hat H)}$, 
where the sum is performed over periodic trajectories of the particle in imaginary time with period $\b\hbar$, which defines the Matsubara frequencies $\om_n=2\p n/(\b\hbar)$~\cite{Kleinert}. 
The disorder average is computed through standard disordered systems tools~\cite{MPV87,BM80b}, \ie the introduction of $n$ replicas of the particle $\bm X_a(t)$, $a=1,\dots,n$, where $n\to0$ at the end 
of the computation. In the limit $N\to\io$, this method provides the free energy as a functional of the overlap order 
parameter $Q_{ab}(t,t')=\overline{\la\bm X_a(t)\cdot\bm X_b(t')\ra}/N$. 

A first angle of attack is to start from the convex UNSAT phase and analyze what happens when approaching the LMS phase boundary, 
classically located on the $\s=0$ line. 
In this region the free energy is expected to display a single well, which translates in the replica framework~\cite{MPV87,KMRSZ07,FP16,FPSUZ17,BM80b,CGSS01} into a 
replica-symmetric (RS) ansatz $Q_{aa}(t,t')=q_d(t-t')$ and $Q_{a\neq b}=q$ (a purely static inter-replica overlap). With this assumption one gets the thermodynamic energy 
\begin{equation}\label{eq:energyRS}
\begin{split}
 \frac{\UU_{\rm RS}}{N}=&\overbrace{\frac{1}{2\b}\sum_{n\in\ZZZ} \frac{\m+\wt\Si(\om_n)} {\MM\om_n^2+\m+\wt\Si(\om_n)}}^{u_{\rm kin}}\\
 &+\underbrace{\a\int\dd h\,\g_q(h+\s)\la v(r(t)+h) \ra_v}_{u_{\rm pot}}
 \end{split}
\end{equation}
The first and second terms are respectively kinetic- and potential-energy contributions. $\g_q(h)$ is a Gaussian weight of variance $q$, $\m$ is a Lagrange multiplier imposing the spherical constraint, 
and $\Si$ is a self-energy arising from the interactions with the obstacles ($\wt\Si$ stands for its Fourier components). 
The above quantities are fixed by saddle-point (variational) equations stemming from the large-$N$ limit. 
In this limit the quantum thermodynamics of the model is mapped onto an effective one-dimensional quantum degree of freedom $r=\bm X\cdot\bm\xi_\mu/\sqrt N$, 
representing the typical overlap with an obstacle. This is a fluctuating quantity even for large $N$. 
The average in the potential-energy term of~\eqref{eq:energyRS} is done over its quantum dynamics, and in the 
following we shall call such averages $\la\bullet\ra_v$ \textit{effective averages}. 
Namely, the main variational equations are 
\begin{equation}\label{eq:Schehrlike}
 \begin{split}
  \la \bullet \ra_v=&\frac{\la \bullet\, e^{-\frac1\hbar\int_0^{\b\hbar}\dd t\,v(r(t)+h)}\ra_r}{\la e^{-\frac1\hbar\int_0^{\b\hbar}\dd t\,v(r(t)+h)}\ra_r}\,,\\
\Si(t)=&-\a\b\int\dd h\, \g_q(h+\s)\\
&\times\la  v'(r(t)+h)v'(r(0)+h)-\hbar\d(t)v''(r(t)+h)\ra_v\,,\\
\frac{1}{\b\wt G(\om_n)}=&\MM\om_n^2+\m+\wt\Si(\om_n)\quad\forall n\neq0\quad \textrm{(Dyson equation)}
\end{split}
\end{equation}
where $r(t)$ is a centered Gaussian average with covariance $\la r(t)r(t')\ra_r=G(t-t')$, related to the overlap order parameter by $G(t)=q_d(t)-q$, and determined by the self-energy through the Dyson equation. 
The self-energy is related self-consistently to the variance by the average over the effective quantum process.
Finally, the Lagrange multiplier is fixed by the spherical constraint $G(0)=1-q$, and another variational equation sets the overlap between different replicas $q$. 

~~\\
\noindent{\sffamily\bfseries Schehr-Giamarchi-Le Doussal expansion --~~}%
The computation of the thermodynamic energy from Eqs.~(\ref{eq:energyRS}),(\ref{eq:Schehrlike}) is a difficult problem without further approximations. 
Fortunately, in a series of papers, Schehr, Giamarchi and Le Doussal have devised an expansion and identified a general mechanism within it providing
the low-$T$ scaling of the specific heat of quantum MF models~\cite{SGLD04,SGLD05,S05}. It amounts to perform an expansion $\hbar\to0$ with $\b\hbar$ \textit{kept fixed}, which retains quantum fluctuations 
while being analytically closer to a semiclassical computation and therefore simpler than a direct $T\to0$ expansion. 
At each order in the expansion, the analytical properties of the self-energy $\wt\Si(\om)$ around $\om=0$ prescribe the scaling of $C_V$ at the same order. An analytic self-energy is a sign of a gapped phase, whereas a singularity generically occurs in a LMS phase, ensuring a power-law behaviour of $C_V$. This singularity has been recognized as a direct consequence of 
the LMS condition. An extra simplification comes from the fact that this mechanism is identically found  at each order in perturbation, yielding the same scaling in $T$ to all orders~\cite{S05}. 

Following this scheme, we expand all quantities  $\OO=\sum_{n\geqslant0}\hbar^n \OO_{(n)}(\b\hbar)$, as well as the variational equations whose structure permits to fix the observables systematically from the previous orders. 
At the zeroth order, one retrieves the classical $T=0$ quantities and phase diagram of~\cite{FP16,FPSUZ17}, which is the starting point of the expansion. To begin our analysis, we place ourselves in the RS UNSAT phase in the following from which we approach the LMS line. 
Effective averages are systematically computed order by order using an asymptotic expansion 
around the saddle point of the dynamical action appearing in the exponent of the definition~\eqref{eq:Schehrlike}, this exponent being proportional to the diverging factor $1/\hbar$. 

First we investigate  all physical quantities at zeroth order, \ie their classical $T=0$ value.
In the classical limit observables become imaginary-time independent thus $G(t)=G(0)=1-q^{\rm cl}$ from the spherical constraint. 
In the UNSAT phase, the overlap's classical value is $q^{\rm cl}=1-\chi T+O(T^2)$~\cite{FP16,FPSUZ17}, 
meaning that at low $T$ (taken \textit{after} $\hbar\to0$ here), $\wt G(\om_n)\sim\chi T\d_{n0}\to0$. 
This implies $\wt G_{(0)}(\om_n)=0$ and $q_{(0)}=1$, hence one needs to look at the next order. 
The lowest-order expansion of the variational equation for the static overlap $q$ 
fixes the value of $\wt G_{(1)}(0)$. One finds, as expected, that $\b\hbar \wt G_{(1)}(0)=\chi$ where
$\chi$ is the above classical rate of deviation of $q^{\rm cl}$ from 1 when $T>0$, determined by $(1+1/\chi)^2=\a/\a_J^{\rm cl}(\s)$ with $\a_J^{\rm cl}(\s)$  the classical RS jamming transition line (see Fig.~\ref{fig:phd}), where $\chi\to\io$.
For the other frequencies, $\wt G_{(1)}(\om_n\neq0)$ is directly given by the Dyson equation~[\ref{eq:Schehrlike}]. The expansion of the variational equations combined with the expansion of the effective averages 
yields, for $n\neq0$, $\m_{(0)}+\wt\Si_{(0)}(\om_n)=1/\chi+\wt I_{(0)}(\om_n)$, 
that is a \textit{mass} term plus a renormalized self-energy $\wt I$, verifying $\wt I_{(0)}(0)=0$. 
Hence the lowest-order Dyson equation
\begin{equation}\label{eq:G1alln}
 \forall n\in\ZZZ\,,\quad\frac{1}{\b\hbar\wt{G}_{(1)}(\om_n)}=\MM\om_n^2+\frac{1}{\chi}+\wt I_{(0)}(\om_n)
\end{equation}
Since the expansion of the effective averages used to calculate $\m_{(0)}+\wt\Si_{(0)}$ depends self-consistently on $\wt G_{(1)}(\om_n)$, one gets a closed equation for $\wt I_{(0)}$:
\begin{equation}\label{eq:computeSi0}
\wt I_{(0)}(\om_n)^2+\wt I_{(0)}(\om_n)\left[K(\s,\a)+\MM\om_n^2\right]
=C(\s,\a)\MM\om_n^2\\
\end{equation}
where $C$ and $K$ are simple constants (SI Appendix, Sec. 5.F).

The location of the LMS phase can be derived by \eg studying the stability of the RS ansatz. The resulting equation -- so-called \textit{marginal condition} -- is expanded in the same way, 
and at lowest order one finds that the line $(\s=0,\a\geqslant\a_J^{\rm cl}(0)=2)$ is solution, which is the classical LMS boundary line in the RS UNSAT phase (see Fig.~\ref{fig:phd}), as one expects.

Now that we have all the lowest-order variational quantities, let us go back to the energy, expanded from~\eqref{eq:energyRS} as $\UU_{\rm RS}/N=u_{(0)}+\hbar\, u_{(1)}(\b\hbar)+O(\hbar^2)$ with 
$u_{(1)}=u_{(1)}^{\rm kin}+u_{(1)}^{\rm pot}$ defined below:
\begin{equation}\label{eq:u0u1}
\begin{split}
u_{(1)}^{\rm kin}(\b\hbar)=&\frac{1}{2\b\hbar}\sum_{n\in\ZZZ}\frac{\frac{1}{\chi}+\wt I_{(0)}(\om_n)}{\MM\om_n^2+\frac{1}{\chi}+\wt I_{(0)}(\om_n)}\\
  u_{(1)}^{\rm pot}(\b\hbar)=& \frac{A_1(\s,\a)}{\b\hbar}\sum_{n\in\ZZZ}\frac{1}{1+\frac{1}{\chi}+\MM\om_n^2+\wt I_{(0)}(\om_n)}\\
  &-\frac{A_2(\s,\a)}{\b\hbar}\sum_{n\in\ZZZ}\frac{1}{\MM\om_n^2+\frac{1}{\chi}+\wt I_{(0)}(\om_n)}
 \end{split}
\end{equation}
where constants $A_1$, $A_2$ are provided in the SI Appendix (Sec. 5.E).
$u_{(0)}=1/(2\chi^2)$ is actually independent of $\b\hbar$: it is the ground-state energy of $H_{\rm cl}$, the classical model (strictly positive by definition in the UNSAT phase). The first $T$-dependent contribution 
comes from $u_{(1)}$. Using the Poisson summation formula, the sums over frequencies are converted into integrals whose 
$T\to0$ scaling depends on the properties of the function $\wt I_{(0)}(\om)$ in the complex plane near $\om=0$~\cite{Bruus}. These are known from~\eqref{eq:computeSi0} 
with $\om_n\to\om$, which is a quadratic equation: the solution is expressed with a square root of a quartic polynomial in $\om$, and as a result this renormalized 
self-energy gets many branch cuts in the complex plane, including two symmetric ones on the imaginary axis. 
There are three cases: \\ 
(\textit{i}) Out of the LMS line (\ie $\s>0$), $K>0$ hence for small $\om$, $\wt I_{(0)}(\om)\sim C\MM\om^2/K$ is analytic. Correspondingly, all 
the Matsubara sums scale as $\exp(-\b\hbar\om_-)$, and so does $C_V$. This is the fingerprint of a gapped Hamiltonian of the full system (where the energy gap is $\hbar\om_-$).\\
(\textit{ii}) On the LMS line, $K(\s=0,\a)=0$ and the small frequency behaviour changes: 
$\wt I_{(0)}(\om)\sim\sqrt{C\MM}|\om|$ is singular, because here the imaginary axis branch cuts merge at the origin. 
This singular absolute value scaling in the LMS phase is expected from the generic scenario of Refs.~\cite{SGLD04,SGLD05,S05}, as well as from the study of the quantum Sherrington-Kirkpatrick model in a transverse field in Ref.~\cite{AM12} via an independent strategy.
\eqref{eq:u0u1} becomes
\begin{equation}\label{eq:gapless}
 \begin{split}
  u_{(1)}(\b\hbar)\sim&\,\textrm{constant}+
  \MM\int_0^\io\frac{\dd \om}{\p}\om^2 \rho_1(\om)f_{\rm B}(\om)\\
  &+\a\int_0^\io\frac{\dd \om}{2\p} \left(\rho_2(\om)-\frac{\rho_1(\om)}{(1+\chi)^2}\right)f_{\rm B}(\om)\\
 \rho_1(\om)=&\Im\left[\b\hbar\wt{G}_{(1)}(-i\om+0^+)\right]\sim\c^2\sqrt{C\MM}\,\om\\
   \rho_2(\om)=&\Im\left[ \frac{1}{1+\frac{1}{\b\hbar\wt{G}_{(1)}(-i\om+0^+)}}\right]
   \sim\frac{\sqrt{C\MM}\,\om}{(1+1/\chi)^2}
 \end{split}
\end{equation}
where we gave the small-$\om$ asymptotics useful for the present discussion. 
In the first line, the term $\propto\MM$ comes from the kinetic energy and the one $\propto\a$ from the potential energy. 
The limit $T\to0$ is provided by the small-$\om$ behaviour of the integrands, and once again we recover features described in Refs~\cite{SGLD04,SGLD05,S05}:
the kinetic contribution to $C_V$ scales like $T^3$, while the potential term is a difference of two terms linear in $T$ with the same prefactor. 
This cancellation gives likewise a  $T^3$ scaling of the potential term. The conclusion for the specific heat at this order is  $C_V/N=A_{\rm RS}(\s,\a)\MM^{\frac32}(T/\hbar)^3$ on the LMS line except
at the jamming point $(\s=0,\a=2)$.\\
(\textit{iii}) At this jamming critical point (red point in Fig.~\ref{fig:phd}), the situation is the same as in the rest of the LMS line except that the \textit{mass} $1/\chi$ vanishes (criticality). 
As a result, both terms scale linearly and $C_V/N=\frac{2\p}{3}\sqrt\MM\, T/\hbar$.

Debye's approximation and the lowest order in the $\hbar\to0$, fixed $\b\hbar$  expansion 
provide the same results in the RS UNSAT phase, \ie the same gap, power law exponents and prefactors. 
Indeed in a RS regime the system should be well approximated at low $T$ by harmonic oscillations around an energy minimum.

%
~~\\
\noindent{\sffamily\bfseries Extension to the whole LMS UNSAT phase --~~}%
The results can be extended to the rest of the LMS UNSAT phase ($\s<0$). There, however, the simple RS ansatz is unstable: the replica symmetry gets broken (RSB), a usual characteristic of MF glassy landscapes. 
Different replicas of the system occupy many different metastable states, and as a consequence the overlap between any two replicas varies continuously. The description of this rugged phase then consists in parametrizing the inter-replica overlap matrix $Q_{a\neq b}$ by a  function $q(x)$ with $x\in[0,1]$~\cite{MPV87}. 
The energy and variational equations are still given by Eqs. [\ref{eq:energyRS}],[\ref{eq:Schehrlike}], with the substitutions 
$q\to q_M$ and $\g_q(h+\s)\to P(q_M,h)$, where $q_M$ is the overlap of replicas in the same metastable state, called Edwards–Anderson order parameter, and $P(q,h)$ is a probability distribution governed by the Parisi partial differential equation~\cite{MPV87} depending on the effective quantum problem. 
Within the lowest order in the expansion, the same mechanism as on the marginal line $(\s=0,\a\geqslant2)$ is at work and the same power-law scalings are retrieved (with different prefactors) for the kinetic-energy term; in the potential-energy term the cancellation of the linear in $T$ dependence away from jamming cannot be checked analytically but numerically, yet we expect it from the generic scenario~\cite{S05} and the results on the $\s=0$ boundary.
~~\\
\noindent{\sffamily\bfseries Extension to all orders in perturbation --~~}%
One can extrapolate the results to finite $\hbar$ following Schehr's argument~\cite{S05}. 
Indeed, after having recovered the exact same basic attributes at lowest order, whose validity relies on the LMS condition applied to a similar thermodynamic energy, we will assume here that this general mechanism holds at all orders. This allows us to perform a direct $T\to0$ expansion.
We focus in this limit on the kinetic term $u_{\rm kin}$ in the energy of the LMS UNSAT phase, given in the first line of~\eqref{eq:energyRS}, since the potential-energy term scales in the same way due to the marginality condition~\cite{S05}.
We write similarly $\m+\wt\Si(\om_n)=\Xi+\wt I(\om_n)$ for $n\neq0$ with ${\wt I(0)=0}$. From~\cite{S05} we know that $\wt I(\om\to0)\sim B|\om|+B'\om^2$ at all orders in $\hbar$; 
we have explicitly computed $\Xi_{(0)}=1/\chi$, $B_{(0)}$ and $B'_{(0)}$. Therefore, as in~\eqref{eq:gapless}, at leading order of the power-law dependence in $T$, the 
Matsubara sum can be written as an integral over an energy DOS 
dominated by $\om\to0$  
\begin{equation}\label{eq:intapproxukin}
 u_{\rm kin}\underset{T\to0}{\sim} \int_0^\io\frac{\dd \om}{\p}\, 
 \frac{B\MM\om^2}{\Xi^2+(b\,\om)^2}\hbar\om f_{\rm B}(\om)
\end{equation}
where $b=\sqrt{B^2-2\Xi(\MM+B')}$. The temperature dependence being now contained only in the Bose-Einstein factor, the low-$T$ limit is easily obtained by rescaling ${\b\hbar\om\to\om'}$. 
The scaling of~\eqref{eq:intapproxukin} depends on the comparison between both terms in the  denominator, which provides a cutoff temperature $T_{\rm cut}=\hbar\, \Xi/b$. 
$u_{\rm kin}$ contributes to the specific heat as $C_V^{\rm kin}/N\sim \frac{B\MM T}{b^2\hbar}\LL(T/T_{\rm cut})$ 
with a scaling function
$\LL(x)\propto x^2$ at small argument which increases to reach a constant value at large argument.
For $T\ll T_{\rm cut}$, one has $C_V^{\rm kin}\propto T^3$ whereas for $T\gg T_{\rm cut}$, $C_V^{\rm kin}\propto T$. 
One can estimate this crossover temperature as follows for $\s<0$ close to jamming: 
the leading order of the \textit{mass} is $\Xi\simeq\Xi_{(0)}=1/\chi\propto \om_*\propto\sqrt{\d\epsilon}$ where $\d\epsilon$ is the distance to the jamming line in the $(\s,\a)$ plane, a purely classical quantity analyzed in Refs.~\cite{FP16,FPSUZ17,FPUZ15}. Thus $T_{\rm cut}\propto \om_*$, while its order of magnitude is fixed by dimensional analysis, 
reintroducing the dimensional units (including the length $\DD$). A similar analysis for $b\simeq b_{(0)}$ yields, up to a purely numerical factor,
\begin{equation}\label{eq:Tc_estim}
  T_{\rm cut}\propto \sqrt{\frac{\varepsilon}{\MM}}\frac{\hbar}{k_B\DD}\sqrt{\d\epsilon}
\end{equation}
If we surmise that such a cutoff temperature exists in SG, we get an order of magnitude by inputting values for typical molecular glassformers~\cite{ZP71,BRH90} in~\eqref{eq:Tc_estim}: $T_{\rm cut}\propto1\  {\rm K}\cdot\sqrt{\d\epsilon}$. The next-order quantities such as $\hbar\,\Xi_{(1)}$ give finite but small 
corrections to this scaling, estimated similarly as 
${T_{\rm cut}(\d\epsilon=0)\propto {\hbar^2}/({k_B\MM\DD^2})\approx 0.01}$~K (see SI Appendix, Sec. 8). Interestingly, upon approaching jamming the value of $T_{\rm cut}$ gets considerably lowered, suggesting a mechanism to observe the linear scaling. However the finite $\hbar$ corrections ensure that $T_{\rm cut}\neq0$, so that the linear scaling is eluded at extremely low $T$ (unlike in Debye's approximation). 
This reflects the physical fact that the jamming transition itself must be avoided in the quantum regime $\hbar>0$, becoming a crossover, analogously to what happens classically from thermal fluctuations ${T>0}$. 
Indeed, the particle's position $\bm X$ cannot be a conserved quantity of the ground state as it does not commute with the Hamiltonian~\eqref{eq:originalH}, and thus cannot take a definite value (\ie a jammed state would violate Heisenberg's inequality).

~~\\
\noindent{\sffamily\bfseries Conclusions --~~}%
We have studied analytically the low-$T$ thermodynamics of a simple model with soft interaction potential relevant to the MF SG theory close to jamming, taking into account the quantum effects. The main result is that a power-law scaling of $C_V$ is entailed by the presence of a LMS phase, and that there is a crossover temperature $T_{\rm cut}$ above which $C_V\propto T$ whereas below $C_V\propto T^3$. This result has been found within the UNSAT phase, meaning that the system sits in a dense (jammed) phase. $T_{\rm cut}$ scales as the square root of the distance to unjamming (with small quantum corrections that make it always non zero), exposing a possible parameter for the observation of ZP's scaling, 
and unveiling yet another connection between jamming and glassy physics~\cite{LN98,LNvSW11}.
Moreover we found that Debye's approximation is semiclassical in the sense that it gives the right result only at the lowest order in $\hbar$ in the RS phase where a single energy minimum dominates the low-$T$ dynamics. 
The actual DOS receives quantum corrections with respect to the classical vibrational one which modify the scaling of $C_V$ especially close to the jamming transition.

We now may wonder about the relevance to finite-dimensional SG. 
On the one hand, jamming criticality seems super-universal (independent of dimensionality) for ${d\geqslant2}$~\cite{CCPZ15,GLN12}. On the other hand, the existence of a LMS phase (the Gardner phase~\cite{GKS85,Ga85}) is still an open question~\cite{UB15,BU16,CY17,Lu18,HWGM18,SBZ17,SRdPZ18,GLL18,CHRSY18}, yet the evidence of its presence is stronger close to the jamming transition~\cite{BCJPSZ16,SD16,JY17,CCFTvdN18,SZ18,SBZ18,LB18}. 
Nonetheless specific low-dimensional features, suppressed in MF, may well play a role. First, realistic glasses do possess standard phononic excitations at lower frequencies than the ones described by~\eqref{eq:DOS}, owing to translation invariance~\cite{DGLFDLW14,FPUZ15,CCPZ15}, which contribute to $C_V$. Second, localized excitations are known to arise at low $T$~\cite{GC03,GPS03,SBZ17,SRdPZ18,CCPZ15,CCPZ12,Lu18,LDB16,MSI17,WNGBSF18,APPR18} and may act as TLS at cryogenic temperatures~\cite{BJMMPPG15,GL16}. 

Finite-dimensional effects might be analyzed in a still tractable way considering models on finite-connectivity graphs~\cite{KvMWZ07}. It would also be useful to understand further the phase diagram of the present quantum model, although a true many-body model of spheres remains the main goal. In particular, approaching jamming from the SAT phase deserves a careful study~\cite{AFP16,FPSUZ17}.
The dependence of $T_{\rm cut}$ on the actual cooling protocol and preparation is another essential aspect~\cite{LQMKH14,PCRTRVR14,PCJRR14,TCBSE16,BU16,SBZ17,LB17,SBZ18,SRdPZ18}. Finally, generalizing the perceptron to model ellipsoids with a small asymmetry has recently been achieved, and based on the low-frequency characteristics of its vibrational DOS we expect similar conclusions would hold~\cite{BIUWZ18}.

\acknow{
We gratefully thank Gr\'egory Schehr for useful discussions about his and the present work, as well as Sungmin Hwang and Antonio Sclocchi for locating an error in a preliminary calculation of the classical specific heat. This project was supported by grants from the Simons Foundation (No.454941, S.F.; No.454949, G.P.) and the European Research Council (No.694925, LoTGlasSy).}

\showacknow{} 

\bibliography{HS}

\begin{thebibliography}{10}

\bibitem{ZP71}
Zeller R, Pohl R (1971) Thermal conductivity and specific heat of
  noncrystalline solids.
\newblock {\em Physical Review B} 4(6):2029.

\bibitem{kittel}
Kittel C, McEuen P (1996) {\em Introduction to solid state physics}.
\newblock (Wiley New York) Vol.{}~8.

\bibitem{LQMKH14}
Liu X, Queen DR, Metcalf TH, Karel JE, Hellman F (2014) Hydrogen-free amorphous
  silicon with no tunneling states.
\newblock {\em Phys. Rev. Lett.} 113(2):025503.

\bibitem{PCRTRVR14}
P{\'e}rez-Casta{\~n}eda T, Rodr{\'\i}guez-Tinoco C, Rodr{\'\i}guez-Viejo J,
  Ramos MA (2014) Suppression of tunneling two-level systems in ultrastable
  glasses of indomethacin.
\newblock {\em Proceedings of the National Academy of Sciences}
  111(31):11275--11280.

\bibitem{PCJRR14}
P\'erez-Casta\~neda T, Jim\'enez-Riob\'oo RJ, Ramos MA (2014) Two-level systems
  and boson peak remain stable in 110-million-year-old amber glass.
\newblock {\em Phys. Rev. Lett.} 112(16):165901.

\bibitem{TCBSE16}
Tylinski M, Chua YZ, Beasley MS, Schick C, Ediger MD (2016) Vapor-deposited
  alcohol glasses reveal a wide range of kinetic stability.
\newblock {\em J. Chem. Phys.} 145(17):174506.

\bibitem{AHV72}
Anderson PW, Halperin B, Varma CM (1972) Anomalous low-temperature thermal
  properties of glasses and spin glasses.
\newblock {\em Philosophical Magazine} 25(1):1--9.

\bibitem{Ph87}
Phillips W (1987) Two-level states in glasses.
\newblock {\em Reports on Progress in Physics} 50(12):1657.

\bibitem{YL88}
Yu CC, Leggett A (1988) Low temperature properties of amorphous materials:
  through a glass darkly.
\newblock {\em Comments on Condensed Matter Physics} 14(4):231--251.

\bibitem{KKI83}
Karpov V, Klinger I, Ignat’Ev F (1983) Theory of the low-temperature
  anomalies in the thermal properties of amorphous structures.
\newblock {\em Zh. Eksp. Teor. Fiz} 84:760--775.

\bibitem{Kl88}
Klinger MI (1988) Glassy disordered systems: topology, atomic dynamics and
  localized electron states.
\newblock {\em Physics reports} 165(5-6):275--397.

\bibitem{BGGPRS92}
Buchenau U, et~al. (1992) Interaction of soft modes and sound waves in glasses.
\newblock {\em Physical Review B} 46(5):2798--2808.

\bibitem{Pa93}
Parshin DA (1993) Soft potential model and universal properties of glasses.
\newblock {\em Physica Scripta} T49A:180--185.

\bibitem{KvMWZ07}
Kühn R, van Mourik J, Weigt M, Zippelius A (2007) Finitely coordinated models
  for low-temperature phases of amorphous systems.
\newblock {\em J. Phys. A: Math. Theor.} 40(31):9227--9252.

\bibitem{LV13}
Leggett AJ, Vural DC (2013) {\textquotedblleft}tunneling two-level
  systems{\textquotedblright} model of the low-temperature properties of
  glasses: Are {\textquotedblleft}smoking-gun{\textquotedblright} tests
  possible?
\newblock {\em J Phys. Chem. B} 117(42):12966--12971.

\bibitem{Lu18}
Lubchenko V (2018) Low-temperature anomalies in disordered solids: a cold case
  of contested relics?
\newblock {\em Advances in Physics: X} 3(1):1510296.

\bibitem{LW01}
Lubchenko V, Wolynes PG (2001) Intrinsic quantum excitations of low temperature
  glasses.
\newblock {\em Phys. Rev. Lett.} 87(19):195901.

\bibitem{LW07}
Lubchenko V, Wolynes PG (2007) Theory of structural glasses and supercooled
  liquids.
\newblock {\em Annual Review of Physical Chemistry} 58(1):235--266.

\bibitem{KPUZ13}
Kurchan J, Parisi G, Urbani P, Zamponi F (2013) Exact theory of dense amorphous
  hard spheres in high dimension. {II}. {T}he high density regime and the
  gardner transition.
\newblock {\em J. Phys. Chem. B} 117:12979--12994.

\bibitem{CKPUZ14}
Charbonneau P, Kurchan J, Parisi G, Urbani P, Zamponi F (2014) Exact theory of
  dense amorphous hard spheres in high dimension. iii. the full replica
  symmetry breaking solution.
\newblock {\em Journal of Statistical Mechanics: Theory and Experiment}
  2014(10):P10009.

\bibitem{nature}
Charbonneau P, Kurchan J, Parisi G, Urbani P, Zamponi F (2014) Fractal free
  energies in structural glasses.
\newblock {\em Nature Communications} 5:3725.

\bibitem{BU16}
Biroli G, Urbani P (2016) Breakdown of elasticity in amorphous solids.
\newblock {\em Nature Physics} 12(12):1130--1133.

\bibitem{SBZ17}
Scalliet C, Berthier L, Zamponi F (2017) Absence of marginal stability in a
  structural glass.
\newblock {\em Physical Review Letters} 119(20):205501.

\bibitem{BW09}
Berthier L, Witten TA (2009) Glass transition of dense fluids of hard and
  compressible spheres.
\newblock {\em Phys. Rev. E} 80(2):021502.

\bibitem{ML11}
Manning ML, Liu AJ (2011) Vibrational modes identify soft spots in a sheared
  disordered packing.
\newblock {\em Physical Review Letters} 107(10):108302.

\bibitem{DKP12}
Dasgupta R, Karmakar S, Procaccia I (2012) Universality of the plastic
  instability in strained amorphous solids.
\newblock {\em Physical Review Letters} 108(7):075701.

\bibitem{XLN17}
Xu N, Liu AJ, Nagel SR (2017) Instabilities of jammed packings of frictionless
  spheres under load.
\newblock {\em Physical Review Letters} 119(21):215502.

\bibitem{CKPUZ17}
Charbonneau P, Kurchan J, Parisi G, Urbani P, Zamponi F (2017) Glass and
  jamming transitions: From exact results to finite-dimensional descriptions.
\newblock {\em Annual Review of Condensed Matter Physics} 8(1):265--288.

\bibitem{WSNW05}
Wyart M, Silbert L, Nagel S, Witten T (2005) {Effects of compression on the
  vibrational modes of marginally jammed solids}.
\newblock {\em Physical Review E} 72(5):051306.

\bibitem{XVLN10}
Xu N, Vitelli V, Liu AJ, Nagel SR (2010) Anharmonic and quasi-localized
  vibrations in jammed solids{\textemdash}modes for mechanical failure.
\newblock {\em {EPL} (Europhysics Letters)} 90(5):56001.

\bibitem{DGLFDLW14}
DeGiuli E, Laversanne-Finot A, Düring G, Lerner E, Wyart M (2014) Effects of
  coordination and pressure on sound attenuation, boson peak and elasticity in
  amorphous solids.
\newblock {\em Soft Matter} 10(30):5628.

\bibitem{MW15}
Müller M, Wyart M (2015) Marginal stability in structural, spin, and electron
  glasses.
\newblock {\em Annual Review of Condensed Matter Physics} 6(1):177--200.

\bibitem{FLD18}
Fyodorov YV, Doussal PL (2018) Hessian spectrum at the global minimum of
  high-dimensional random landscapes.
\newblock {\em J. Phys. A: Math. Theor.} 51(47):474002.

\bibitem{S05}
Schehr G (2005) Low-temperature specific heat of some quantum mean-field glassy
  phases.
\newblock {\em Physical Review B} 71(18):184204.

\bibitem{AM12}
Andreanov A, M{\"u}ller M (2012) Long-range quantum ising spin glasses at t= 0:
  Gapless collective excitations and universality.
\newblock {\em Physical review letters} 109(17):177201.

\bibitem{LNvSW11}
Liu AJ, Nagel SR, van Saarloos W, Wyart M (2011) The jamming
  scenario{\textemdash}an introduction and outlook in {\em Dynamical
  Heterogeneities in Glasses, Colloids, and Granular Media}.
\newblock (Oxford University Press), pp. 298--340.

\bibitem{BC19}
Behringer RP, Chakraborty B (2019) The physics of jamming for granular
  materials: a review.
\newblock {\em Reports on Progress in Physics} 82(1):012601.

\bibitem{OHSLN03}
O'Hern CS, Silbert LE, Liu AJ, Nagel SR (2003) Jamming at zero temperature and
  zero applied stress: The epitome of disorder.
\newblock {\em Phys. Rev. E} 68(1):011306.

\bibitem{Wy12}
Wyart M (2012) Marginal stability constrains force and pair distributions at
  random close packing.
\newblock {\em Physical review letters} 109(12):125502.

\bibitem{LDW13}
Lerner E, Düring G, Wyart M (2013) Low-energy non-linear excitations in sphere
  packings.
\newblock {\em Soft Matter} 9(34):8252.

\bibitem{DGLBW14}
DeGiuli E, Lerner E, Brito C, Wyart M (2014) Force distribution affects
  vibrational properties in hard-sphere glasses.
\newblock {\em Proc Natl Acad Sci} 111(48):17054--17059.

\bibitem{CCPZ12}
Charbonneau P, Corwin EI, Parisi G, Zamponi F (2012) Universal microstructure
  and mechanical stability of jammed packings.
\newblock {\em Physical Review Letters} 109(20):205501.

\bibitem{CCPZ15}
Charbonneau P, Corwin EI, Parisi G, Zamponi F (2015) Jamming criticality
  revealed by removing localized buckling excitations.
\newblock {\em Phys. Rev. Lett.} 114(12):125504.

\bibitem{GD88}
Gardner E, Derrida B (1988) Optimal storage properties of neural network
  models.
\newblock {\em Journal of Physics A: Mathematical and General} 21(1):271--284.

\bibitem{FP16}
Franz S, Parisi G (2016) The simplest model of jamming.
\newblock {\em Journal of Physics A: Mathematical and Theoretical}
  49(14):145001.

\bibitem{FPUZ15}
Franz S, Parisi G, Urbani P, Zamponi F (2015) Universal spectrum of normal
  modes in low-temperature glasses.
\newblock {\em Proc Natl Acad Sci} 112(47):14539--14544.

\bibitem{AFP16}
Altieri A, Franz S, Parisi G (2016) The jamming transition in high dimension:
  an analytical study of the {TAP} equations and the effective thermodynamic
  potential.
\newblock {\em Journal of Statistical Mechanics: Theory and Experiment}
  2016(9):093301.

\bibitem{FPSUZ17}
Franz S, Parisi G, Sevelev M, Urbani P, Zamponi F (2017) {Universality of the
  SAT-UNSAT (jamming) threshold in non-convex continuous constraint
  satisfaction problems}.
\newblock {\em SciPost Phys.} 2(3):019.

\bibitem{CGSS01}
Cugliandolo LF, Grempel DR, da~Silva~Santos CA (2001) Imaginary-time replica
  formalism study of a quantum spherical p-spin-glass model.
\newblock {\em Physical Review B} 64(1):014403.

\bibitem{Kleinert}
Kleinert H (2009) {\em Path Integrals in Quantum Mechanics, Statistics, Polymer
  Physics, and Financial Markets}.
\newblock (World Scientific Publishing Co. Pte. Ltd.).

\bibitem{MPV87}
M\'ezard M, Parisi G, Virasoro MA (1987) {\em Spin glass theory and beyond}.
\newblock (World Scientific).

\bibitem{BM80b}
Bray AJ, Moore MA (1980) Replica theory of quantum spin glasses.
\newblock {\em Journal of Physics C: Solid State Physics} 13(24):L655--L660.

\bibitem{KMRSZ07}
Krzakala F, Montanari A, Ricci-Tersenghi F, Semerjian G, Zdeborova L (2007)
  {Gibbs states and the set of solutions of random constraint satisfaction
  problems}.
\newblock {\em Proceedings of the National Academy of Sciences} 104(25):10318.

\bibitem{SGLD04}
Schehr G, Giamarchi T, Doussal PL (2004) Specific heat of the quantum bragg
  glass.
\newblock {\em Europhysics Letters ({EPL})} 66(4):538--544.

\bibitem{SGLD05}
Schehr G, Giamarchi T, Doussal PL (2005) Specific heat of quantum elastic
  systems pinned by disorder.
\newblock {\em The European Physical Journal B} 44(4):521--534.

\bibitem{Bruus}
Bruus H, Flensberg K (2004) {\em Many-body quantum theory in condensed matter
  physics: an introduction}.
\newblock (Oxford university press).

\bibitem{BRH90}
Barrat JL, Roux JN, Hansen JP (1990) Diffusion, viscosity and structural
  slowing down in soft sphere alloys near the kinetic glass transition.
\newblock {\em Chemical Physics} 149(1-2):197--208.

\bibitem{LN98}
Liu AJ, Nagel SR (1998) Jamming is not just cool any more.
\newblock {\em Nature} 396(6706):21--22.

\bibitem{GLN12}
Goodrich CP, Liu AJ, Nagel SR (2012) Finite-size scaling at the jamming
  transition.
\newblock {\em Physical Review Letters} 109(9):095704.

\bibitem{GKS85}
Gross DJ, Kanter I, Sompolinsky H (1985) Mean-field theory of the potts glass.
\newblock {\em Physical Review Letters} 55(3):304--307.

\bibitem{Ga85}
Gardner E (1985) Spin glasses with p-spin interactions.
\newblock {\em Nuclear Physics B} 257:747--765.

\bibitem{UB15}
Urbani P, Biroli G (2015) Gardner transition in finite dimensions.
\newblock {\em Phys. Rev. B} 91(10):100202.

\bibitem{CY17}
Charbonneau P, Yaida S (2017) Nontrivial critical fixed point for
  replica-symmetry-breaking transitions.
\newblock {\em Physical Review Letters} 118(21):215701.

\bibitem{HWGM18}
Hicks C, Wheatley M, Godfrey M, Moore M (2018) Gardner transition in physical
  dimensions.
\newblock {\em Physical Review Letters} 120(22):225501.

\bibitem{SRdPZ18}
Seoane B, Reid DR, de~Pablo JJ, Zamponi F (2018) Low-temperature anomalies of a
  vapor deposited glass.
\newblock {\em Physical Review Materials} 2(1):015602.

\bibitem{GLL18}
Geirhos K, Lunkenheimer P, Loidl A (2018) Johari-goldstein relaxation far below
  tg : Experimental evidence for the gardner transition in structural glasses?
\newblock {\em Phys. Rev. Lett.} 120(8):085705.

\bibitem{CHRSY18}
Charbonneau P, Hu Y, Raju A, Sethna JP, Yaida S (2019) Morphology of
  renormalization-group flow for the de almeida--thouless--gardner universality
  class.
\newblock {\em Phys. Rev. E} 99(2):022132.

\bibitem{BCJPSZ16}
Berthier L, et~al. (2016) Growing timescales and lengthscales characterizing
  vibrations of amorphous solids.
\newblock {\em Proceedings of the National Academy of Sciences}
  113(30):8397--8401.

\bibitem{SD16}
Seguin A, Dauchot O (2016) Experimental evidence of the gardner phase in a
  granular glass.
\newblock {\em Physical Review Letters} 117(22):228001.

\bibitem{JY17}
Jin Y, Yoshino H (2017) Exploring the complex free-energy landscape of the
  simplest glass by rheology.
\newblock {\em Nature Communications} 8:14935.

\bibitem{CCFTvdN18}
Charbonneau P, Corwin EI, Fu L, Tsekenis G, van~der Naald M (2019) Glassy,
  gardner-like phenomenology in minimally polydisperse crystalline systems.
\newblock {\em Phys. Rev. E} 99(2):020901.

\bibitem{SZ18}
Seoane B, Zamponi F (2018) Spin-glass-like aging in colloidal and granular
  glasses.
\newblock {\em Soft Matter} 14(25):5222--5234.

\bibitem{SBZ18}
Scalliet C, Berthier L, Zamponi F (2019) Marginally stable phases in mean-field
  structural glasses.
\newblock {\em Phys. Rev. E} 99(1):012107.

\bibitem{LB18}
Liao Q, Berthier L (2019) Hierarchical landscape of hard disk glasses.
\newblock {\em Phys. Rev. X} 9(1):011049.

\bibitem{GC03}
Gurarie V, Chalker JT (2003) Bosonic excitations in random media.
\newblock {\em Phys. Rev. B} 68(13):134207.

\bibitem{GPS03}
Gurevich VL, Parshin DA, Schober HR (2003) Anharmonicity, vibrational
  instability, and the boson peak in glasses.
\newblock {\em Physical Review B} 67(9):094203.

\bibitem{LDB16}
Lerner E, Düring G, Bouchbinder E (2016) Statistics and properties of
  low-frequency vibrational modes in structural glasses.
\newblock {\em Physical Review Letters} 117(3):035501.

\bibitem{MSI17}
Mizuno H, Shiba H, Ikeda A (2017) Continuum limit of the vibrational properties
  of amorphous solids.
\newblock {\em Proceedings of the National Academy of Sciences} p. 201709015.

\bibitem{WNGBSF18}
Wang L, et~al. (2019) Low-frequency vibrational modes of stable glasses.
\newblock {\em Nature Communications} 10(1).

\bibitem{APPR18}
Angelani L, Paoluzzi M, Parisi G, Ruocco G (2018) Probing the non-debye
  low-frequency excitations in glasses through random pinning.
\newblock {\em Proc Natl Acad Sci} 115(35):8700--8704.

\bibitem{BJMMPPG15}
Baity-Jesi M, Mart{\'{\i}}n-Mayor V, Parisi G, Perez-Gaviro S (2015) Soft
  modes, localization, and two-level systems in spin glasses.
\newblock {\em Physical Review Letters} 115(26):267205.

\bibitem{GL16}
Gartner L, Lerner E (2016) Nonlinear modes disentangle glassy and goldstone
  modes in structural glasses.
\newblock {\em {SciPost} Physics} 1(2):016.

\bibitem{LB17}
Lerner E, Bouchbinder E (2017) Effect of instantaneous and continuous quenches
  on the density of vibrational modes in model glasses.
\newblock {\em Physical Review E} 96(2):020104.

\bibitem{BIUWZ18}
Brito C, Ikeda H, Urbani P, Wyart M, Zamponi F (2018) Universality of jamming
  of nonspherical particles.
\newblock {\em Proceedings of the National Academy of Sciences}
  115(46):11736--11741.

\end{thebibliography}


\begin{thebibliography}{10}

\bibitem{FPSUZ17}
Franz S, Parisi G, Sevelev M, Urbani P, Zamponi F (2017) {Universality of the
  SAT-UNSAT (jamming) threshold in non-convex continuous constraint
  satisfaction problems}.
\newblock {\em SciPost Phys.} 2(3):019.

\bibitem{FP16}
Franz S, Parisi G (2016) The simplest model of jamming.
\newblock {\em Journal of Physics A: Mathematical and Theoretical}
  49(14):145001.

\bibitem{An58}
Anderson PW (1958) Absence of diffusion in certain random lattices.
\newblock {\em Physical Review} 109(5):1492--1505.

\bibitem{BAA06}
Basko D, Aleiner I, Altshuler B (2006) Metal{\textendash}insulator transition
  in a weakly interacting many-electron system with localized single-particle
  states.
\newblock {\em Annals of Physics} 321(5):1126--1205.

\bibitem{FPUZ15}
Franz S, Parisi G, Urbani P, Zamponi F (2015) Universal spectrum of normal
  modes in low-temperature glasses.
\newblock {\em Proc Natl Acad Sci} 112(47):14539--14544.

\bibitem{AFP16}
Altieri A, Franz S, Parisi G (2016) The jamming transition in high dimension:
  an analytical study of the {TAP} equations and the effective thermodynamic
  potential.
\newblock {\em Journal of Statistical Mechanics: Theory and Experiment}
  2016(9):093301.

\bibitem{MPV87}
M\'ezard M, Parisi G, Virasoro MA (1987) {\em Spin glass theory and beyond}.
\newblock (World Scientific).

\bibitem{CC05}
Castellani T, Cavagna A (2005) Spin glass theory for pedestrians.
\newblock {\em Journal of Statistical Mechanics: Theory and Experiment}
  2005:P05012.

\bibitem{BM80b}
Bray AJ, Moore MA (1980) Replica theory of quantum spin glasses.
\newblock {\em Journal of Physics C: Solid State Physics} 13(24):L655--L660.

\bibitem{FHS10}
Feynman RP, Hibbs AR, Styer D (2010) {\em Quantum mechanics and path
  integrals}.
\newblock (Dover Publications).

\bibitem{Kleinert}
Kleinert H (2009) {\em Path Integrals in Quantum Mechanics, Statistics, Polymer
  Physics, and Financial Markets}.
\newblock (World Scientific Publishing Co. Pte. Ltd.).

\bibitem{CGSS01}
Cugliandolo LF, Grempel DR, da~Silva~Santos CA (2001) Imaginary-time replica
  formalism study of a quantum spherical p-spin-glass model.
\newblock {\em Physical Review B} 64(1):014403.

\bibitem{CKPUZ14}
Charbonneau P, Kurchan J, Parisi G, Urbani P, Zamponi F (2014) Exact theory of
  dense amorphous hard spheres in high dimension. iii. the full replica
  symmetry breaking solution.
\newblock {\em Journal of Statistical Mechanics: Theory and Experiment}
  2014(10):P10009.

\bibitem{Du81}
Duplantier B (1981) Comment on parisi's equation for the sk model for spin
  glasses.
\newblock {\em Journal of Physics A: Mathematical and General} 14(1):283.

\bibitem{Appel}
Appel W, Kowalski E (2007) {\em Mathematics for physics and physicists}.
\newblock (Princeton University Press Princeton, NJ, USA; Oxford, UK).

\bibitem{MP91}
M{\'e}zard M, Parisi G (1991) Replica field theory for random manifolds.
\newblock {\em Journal de Physique I} 1(6):809--836.

\bibitem{AS06}
Altland A, Simons B (2006) {\em Condensed Matter Field Theory}.
\newblock (Cambridge University Press).

\bibitem{Mahan}
Mahan GD (2000) {\em Many-Particle Physics}.
\newblock (Springer {US}).

\bibitem{Bruus}
Bruus H, Flensberg K (2004) {\em Many-body quantum theory in condensed matter
  physics: an introduction}.
\newblock (Oxford university press).

\bibitem{KMRSZ07}
Krzakala F, Montanari A, Ricci-Tersenghi F, Semerjian G, Zdeborova L (2007)
  {Gibbs states and the set of solutions of random constraint satisfaction
  problems}.
\newblock {\em Proceedings of the National Academy of Sciences} 104(25):10318.

\bibitem{SGLD04}
Schehr G, Giamarchi T, Doussal PL (2004) Specific heat of the quantum bragg
  glass.
\newblock {\em Europhysics Letters ({EPL})} 66(4):538--544.

\bibitem{SGLD05}
Schehr G, Giamarchi T, Doussal PL (2005) Specific heat of quantum elastic
  systems pinned by disorder.
\newblock {\em The European Physical Journal B} 44(4):521--534.

\bibitem{S05}
Schehr G (2005) Low-temperature specific heat of some quantum mean-field glassy
  phases.
\newblock {\em Physical Review B} 71(18):184204.

\bibitem{De12}
Debye P (1912) Zur theorie der spezifischen wärmen.
\newblock {\em Annalen der Physik} 344(14):789--839.

\bibitem{kittel}
Kittel C, McEuen P (1996) {\em Introduction to solid state physics}.
\newblock (Wiley New York) Vol.{}~8.

\bibitem{AM}
Ashcroft NW, Mermin ND (1976) Solid state physics.

\bibitem{DP1819}
Dulong PL, Petit AT (1819) {\em Recherches sur quelques points importants de la
  th{\'e}orie de la chaleur}.
\newblock (Paris - Crochard).

\bibitem{NR98}
Nieuwenhuizen TM, Ritort F (1998) Quantum phase transition in spin glasses with
  multi-spin interactions.
\newblock {\em Physica A: Statistical Mechanics and its Applications}
  250(1-4):8--45.

\bibitem{AM12}
Andreanov A, M{\"u}ller M (2012) Long-range quantum ising spin glasses at t= 0:
  Gapless collective excitations and universality.
\newblock {\em Physical review letters} 109(17):177201.

\bibitem{peskin}
Peskin M, Schroeder D (1995) {\em An introduction to quantum field theory}.
\newblock (Addison-Wesley Publishing Company).

\bibitem{nature}
Charbonneau P, Kurchan J, Parisi G, Urbani P, Zamponi F (2014) Fractal free
  energies in structural glasses.
\newblock {\em Nature Communications} 5:3725.

\bibitem{CKPUZ17}
Charbonneau P, Kurchan J, Parisi G, Urbani P, Zamponi F (2017) Glass and
  jamming transitions: From exact results to finite-dimensional descriptions.
\newblock {\em Annual Review of Condensed Matter Physics} 8(1):265--288.

\bibitem{OHSLN03}
O'Hern CS, Silbert LE, Liu AJ, Nagel SR (2003) Jamming at zero temperature and
  zero applied stress: The epitome of disorder.
\newblock {\em Phys. Rev. E} 68(1):011306.

\bibitem{WSNW05}
Wyart M, Silbert L, Nagel S, Witten T (2005) {Effects of compression on the
  vibrational modes of marginally jammed solids}.
\newblock {\em Physical Review E} 72(5):051306.

\bibitem{XVLN10}
Xu N, Vitelli V, Liu AJ, Nagel SR (2010) Anharmonic and quasi-localized
  vibrations in jammed solids{\textemdash}modes for mechanical failure.
\newblock {\em {EPL} (Europhysics Letters)} 90(5):56001.

\bibitem{DGLFDLW14}
DeGiuli E, Laversanne-Finot A, Düring G, Lerner E, Wyart M (2014) Effects of
  coordination and pressure on sound attenuation, boson peak and elasticity in
  amorphous solids.
\newblock {\em Soft Matter} 10(30):5628.

\bibitem{ZP71}
Zeller R, Pohl R (1971) Thermal conductivity and specific heat of
  noncrystalline solids.
\newblock {\em Physical Review B} 4(6):2029.

\bibitem{BRH90}
Barrat JL, Roux JN, Hansen JP (1990) Diffusion, viscosity and structural
  slowing down in soft sphere alloys near the kinetic glass transition.
\newblock {\em Chemical Physics} 149(1-2):197--208.

\bibitem{PCRTRVR14}
P{\'e}rez-Casta{\~n}eda T, Rodr{\'\i}guez-Tinoco C, Rodr{\'\i}guez-Viejo J,
  Ramos MA (2014) Suppression of tunneling two-level systems in ultrastable
  glasses of indomethacin.
\newblock {\em Proceedings of the National Academy of Sciences}
  111(31):11275--11280.

\bibitem{PCJRR14}
P\'erez-Casta\~neda T, Jim\'enez-Riob\'oo RJ, Ramos MA (2014) Two-level systems
  and boson peak remain stable in 110-million-year-old amber glass.
\newblock {\em Phys. Rev. Lett.} 112(16):165901.

\bibitem{FG70}
Fonda L, Ghirardi GC (1970) Symmetry principles in quantum physics.

\bibitem{KS97}
Kleinert H, Shabanov SV (1997) Proper dirac quantization of a free particle on
  a d-dimensional sphere.
\newblock {\em Physics Letters A} 232(5):327--332.

\bibitem{EF14}
Efthimiou C, Frye C (2014) {\em Spherical Harmonics in p Dimensions}.
\newblock ({WORLD} {SCIENTIFIC}).

\bibitem{Cohen-Tannoudji}
Cohen-Tannoudji C, Diu B, Lalo{\"e} F (1977) Quantum mechanics.

\bibitem{Sc02}
Scardicchio A (2002) Classical and quantum dynamics of a particle constrained
  on a circle.
\newblock {\em Physics Letters A} 300(1):7--17.

\bibitem{ABUZ18}
Agoritsas E, Biroli G, Urbani P, Zamponi F (2018) Out-of-equilibrium dynamical
  mean-field equations for the perceptron model.
\newblock {\em Journal of Physics A: Mathematical and Theoretical}
  51(8):085002.

\bibitem{SK75}
Sherrington D, Kirkpatrick S (1975) Solvable model of a spin-glass.
\newblock {\em Physical Review Letters} 35(26):1792--1796.

\bibitem{SS81}
Shukla P, Singh S (1981) A quantum spherical model of spin glass.
\newblock {\em Physics Letters A} 81(8):477--479.

\bibitem{KTJ76}
Kosterlitz JM, Thouless DJ, Jones RC (1976) Spherical model of a spin-glass.
\newblock {\em Physical Review Letters} 36(20):1217--1220.

\bibitem{these}
Maimbourg T (2016) Theses ({PSL Research University}).

\bibitem{vivo}
Livan G, Novaes M, Vivo P (2018) {\em Introduction to Random Matrices: Theory
  and Practice}.
\newblock (Springer).

\bibitem{mehta}
Mehta ML (2004) {\em Random matrices}.
\newblock (Elsevier) Vol.{} 142.

\end{thebibliography}

\end{document}




\SItext




\setcounter{tocdepth}{2}
\tableofcontents
\newpage

In this Supplementary Information we derive the equations for the quantum spherical perceptron introduced in the main text. The notations will be close to the ones in Ref.~\cite{FPSUZ17}, 
in particular we define the function 
\begin{equation}
 \Th(x):=\frac{1+\erf(x)}{2}
\end{equation}
whereas $\th$ stands for Heaviside's step function. \\
Lengths are dimensionless (a lengthscale can be reinstated \eg through the radius of the sphere) and the Boltzmann constant is $k_{\rm B}=1$. \\

We consider $M=\a N$ gap variables~\cite{FP16,FPSUZ17}
\begin{equation}
 h_\m(\bm X)=\frac{\bm X\cdot \bm\xi_\m}{\sqrt N}-\s
\end{equation}
with respect to the particle living on the $N$-dimensional sphere of radius $\sqrt N$, \ie $\bm X^2=\sum_{i=1}^N(X^i)^2=N$. 
$\xi_\m^i$ are independent Gaussian centered random variables with variance 1, so that by the central limit theorem
${\bm \xi}_\m^2=\sum_{i=1}^N(\xi_\m^i)^2\sim N$, \ie for large $N$ each obstacle lies effectively on the sphere, 
and since $\bm X$ is isotropically distributed, $\bm X\cdot \bm\xi_\m/\sqrt N=O(1)$. 
The Hamiltonian reads
\begin{equation}\label{eq:originalH}
\hat H=\frac{\hat{\bm P}^2}{2\MM}+\sum_{\m=1}^M v(h_\m(\hat{\bm X}))
\end{equation}
with a soft harmonic spheres potential 
\begin{equation}
 v(h)=\varepsilon\frac{h^2}{2}\th(-h)
\end{equation}

In the following we focus on the partition function on the model, but note that in the quantum regime the standard ergodic assumptions of statistical mechanics may well break down in such a model owing to localization effects~\cite{An58,BAA06}; here we are interested in the relation found at the level of the free energy between this model and the infinite-dimensional limit of structural glassy systems~\cite{FP16,FPUZ15,AFP16,FPSUZ17}.

\section{The partition function of the model}\label{sec:Z}
We wish to compute the free energy at temperature $T=1/\b$ from its $n$-times replicated~\cite{MPV87,CC05,BM80b} Feynman's representation, a path integral in imaginary time~\cite{FHS10,Kleinert}, 
\begin{equation}\label{eq:repl_Z}
\begin{split}
 -\b F&=\overline{\ln Z}\underset{n\to0}{=}\partial_n\overline{Z^n}\\
 Z^n&=\left(\Tr\,e^{-\b\hat H}\right)^n=\oint\prod_{a=1}^n\mathrm{D}\bm X^a\, \exp\left(-\frac{1}{\hbar}\sum_{a=1}^n\int_0^{\b\hbar}\dd t\,\left[\frac{\MM}{2}(\dot {\bm X}^a)^2(t)+\sum_{\m=1}^M v(h_\m(\bm X^a))\right]\right)
\end{split}
\end{equation}
$\oint\mathrm{D}\bm X$ means a sum over all trajectories constrained on the sphere with periodic boundary conditions $\bm X(0)=\bm X(\b\hbar)$. From the free energy one can get the specific heat $C_V=\frac{\partial}{\partial T}\overline{\Tr\left(\hat H e^{-\b\hat H}/Z\right)}=-\b^2\partial^2(\b F)/\partial\b^2$.

We introduce the $O(1)$ overlaps with the obstacles $r_\m^a(t)=\bm X^a(t)\cdot\bm\xi_\m/\sqrt N$ through delta functions, then exponentiate them with auxiliary variables 
$\hat r_\m^a(t)$, allowing to perform the average over the Gaussian disorder and rewrite the kinetic energy term with an integration by part:
\begin{equation}\label{eq:avdis}
\begin{split}
  \overline{Z^n}&=\oint\prod_{a=1}^n\mathrm{D}\bm X^a\mathrm{D}[r^a_\m,\hat r^a_\m]\, 
  e^{\,\sum_{a,\m}\frac{1}{\b\hbar}\int_0^{\b\hbar}i\hat r^a_\m r^a_\m}\,
  \overline{e^{-\sum_{a,\m}\frac{1}{\b\hbar}\int_0^{\b\hbar}i\hat r^a_\m\,\frac{ \bm X^a\cdot\bm\xi_\m}{\sqrt N}}}\, 	
  e^{\frac{1}{\hbar}\sum_a\int_0^{\b\hbar}\left[\frac{\MM}{2}\bm X^a\cdot\ddot {\bm X}^a-\sum_\m v(r_\m^a-\s)\right]}\\
  &=\oint\prod_{a=1}^n\mathrm{D}\bm X^a\mathrm{D}[r^a_\m,\hat r^a_\m]\, e^{\,\sum_{a,\m}\frac{1}{\b\hbar}\int_0^{\b\hbar}i\hat r^a_\m r^a_\m-\frac12\sum_{a,b,\m}\iint_0^{\b\hbar}\frac{\dd t}{\b\hbar}\frac{\dd t'}{\b\hbar}\hat r_a^\m(t)\frac{\bm X^a(t)\cdot\bm X^b(t')}{N}
  \hat r_b^\m(t')+\frac{1}{\hbar}\sum_a\int_0^{\b\hbar}\left[\frac{\MM}{2}\bm X^a\cdot\ddot {\bm X}^a-\sum_\m v(r_\m^a-\s)\right]}\\
  &=\oint\mathrm{D}\wt Q\,(\det\wt Q)^{\frac N2}e^{N\frac{\MM}{2\hbar}\sum_a\int_0^{\b\hbar}\dd t\,\restriction{\partial_s^2 Q_{aa}(t,s)}{s=t}}\\
  &\hspace{2cm}\times\left[\oint\prod_a\mathrm{D}[r_a,\hat r_a]e^{\sum_a\frac{1}{\b\hbar}\int_0^{\b\hbar}i\hat r_ar_a-\frac12\sum_{a,b}\iint_0^{\b\hbar}\frac{\dd t}{\b\hbar}\frac{\dd s}{\b\hbar}\, \hat r_a(t)Q_{ab}(t,s)\hat r_b(s)-\frac1\hbar\sum_a\int_0^{\b\hbar}v(r_a-\s)}\right]^M
\end{split}
\end{equation}
In the last line we have changed variables by introducing the overlap $Q_{ab}(t,s)=\bm X^a(t)\cdot\bm X^b(s)/N$, noted $\wt Q$. Throughout these notes the hat on matrices refers only to replica indices; 
\eg $\wt Q=\{\hat Q(t,s)\}_{t,s}$ with $\hat Q(t,s)$ having $n^2$ elements $Q_{ab}(t,s)$. $\wt Q$ can, for practical purposes, be viewed as a $n\NN\times n\NN$ matrix (discretizing in $\NN$ time steps) with indices $\{\a,\b\}=\{(a,t),(b,s)\}$.  
The $\b\hbar$-periodicity on $\bm X_a$ propagates by definition to $Q_{ab}(t,s)$ and $r_a(t)$ (we use the same closed-contour integral symbol 
to remind this). 
Note that the summation on times is always performed in non-dimensional units (with $\b\hbar$ chosen as the time unit) to keep track of the $\b\hbar$ factors.
As in~Ref.~\cite{FPSUZ17}, the large $N$ (exponential) contribution of the Jacobian is $(\det\wt Q)^{N/2}$. 
This contribution can be inferred quickly from the computation, with symmetric matrices:
\begin{equation}\label{eq:XXtoQ}
\begin{split}
 \textrm{Jacobian}&=\int \prod_{\a=(a,t)}\mathrm{d}\bm X_\a \,\prod_{\a,\b}\d(N Q_{\a\b}-\bm X^\a\cdot\bm X^\b)
 =\int \prod_\a\mathrm{d}\bm X_\a\dd\wt P\, e^{iN\Tr(\wt P\wt Q)-\sum_{\a,\b}iP_{\a\b}\bm X^\a\cdot\bm X^\b}\\
 &\propto \int \dd\wt P\,e^{iN\Tr(\wt P\wt Q)-\frac N2\ln\det(2i\wt P)}\underset{N\to\io}{\propto}(\det\wt Q)^{N/2}
\end{split}
\end{equation}
The last result is provided by the saddle-point value of the exponent, given by the equation $\wt P^{-1}=2i\wt Q$.
Here we have ignored the $\b\hbar$-periodicity; actually the periodicity on both times of $Q_{ab}(t,s)$ makes 
the corrections vanish\footnote{One can take it into account writing 
$\oint \prod_a\mathrm{D}\bm X_a(t)=\int \prod_a\mathrm{D}\bm X_a(t)\,\d(\bm X_a(0)-\bm X_a(\b\hbar))$ then exponentiating the delta functions 
by introducing another variable $\bm\eta_a$; the integral on the positions $\bm X_a(t)$ is still Gaussian and the extra term 
with respect to~\eqref{eq:XXtoQ} is 
$$ \int \prod_a\mathrm{d}\bm \eta_a\,e^{-\frac12\sum_{a,b}\left[(2i\wt P)^{-1}_{ab}(0,0)+(2i\wt P)^{-1}_{ab}(\b\hbar,\b\hbar)
 -(2i\wt P)^{-1}_{ab}(0,\b\hbar)-(2i\wt P)^{-1}_{ab}(\b\hbar,0)\right]\bm \eta^a\cdot\bm \eta^b}$$
which vanishes considering $(2i\wt P)^{-1}=\wt Q$ is $\b\hbar$ periodic.}.\\
Then, one performs the Gaussian integration on the auxiliary $\hat r$ variables in the bracketted term in the last line of~\eqref{eq:avdis},
whose result can be written with both equivalent representations of Gaussian averages\footnote{The Gaussian 
averages will not be affected by the periodicity of $r_a(t)$ owing to the same argument as in the last footnote.}:
\begin{equation}\label{eq:RSBfullint}
\begin{split}
\int\prod_{a=1}^n\frac{\mathrm{D} r_a}{\sqrt{\det\wt Q}}\,&e^{-\frac12\sum_{a,b}\iint_0^{\b\hbar}\frac{\dd t}{\b\hbar}\frac{\dd s}{\b\hbar}\, r_a(t)Q^{-1}_{ab}(t,s)r_b(s)}
e^{-\frac1\hbar\sum_a\int_0^{\b\hbar}\dd t\,v(r_a(t)-\s)}\\
&=e^{\frac12\sum_{a,b}\iint_0^{\b\hbar}\frac{\dd t}{\b\hbar}\frac{\dd s}{\b\hbar}\, Q_{ab}(t,s)\frac{\d^2}{\d r_a(t)\d r_b(s)}}
\restriction{e^{-\frac1\hbar\sum_a\int_0^{\b\hbar}\dd t\,v(r_a(t))}}{r_a(t)=-\s}
\end{split}
\end{equation}
This can be proven by expanding both exponentials involving $\wt Q$ and using Wick's theorem. We get in the end:
\begin{equation}\label{eq:SS}
\begin{split}
 \overline{Z^n}&=\oint\mathrm{D}\wt Q\,e^{N\AA(\wt Q)}\\
 \AA(\wt Q)&=\frac12\ln\det\wt Q +\frac{\MM}{2\hbar}\sum_{a=1}^n\int_0^{\b\hbar}\dd t\,\restriction{\partial_s^2 Q_{aa}(t,s)}{s=t}+\a\ln\z\\
 \z&=\exp\left(\frac12\sum_{a,b}\iint_0^{\b\hbar}\frac{\dd t}{\b\hbar}\frac{\dd s}{\b\hbar}\, Q_{ab}(t,s)\frac{\d^2}{\d r_a(t)\d r_b(s)}\right)\restriction{\exp\left(-\frac1\hbar\sum_a\int_0^{\b\hbar}\dd t\,v(r_a(t))\right)}{r_a(t)=-\s}
\end{split}
\end{equation}
where $\hat Q(t,s)$ is periodic on both times\footnote{The spherical constraint, 
which is of the form $\bm X_a(t)\cdot \bm X_a(t)/N=Q_{aa}(t,t)=1$, can be enforced with delta functions 
and does not affect the Jacobian. This will be done later through a Lagrange multiplier.} with period $\b\hbar$ and $\forall (a,t),\,Q_{aa}(t,t)=1$. The integral over $\wt Q$ is evaluated by the saddle-point method in $N\to\io$ and the free energy 
is given by the replica analytic continuation $-\b F/N\underset{n\to0}{=}\partial_n\AA(\wt Q^\mathrm{sp})$~\cite{MPV87,CC05,BM80b}.


\section{Time-dependent replica-symmetry breaking formulation}\label{sec:timeRSB}
To go further one needs to solve the saddle-point equation $\partial\AA/\partial{\wt Q}=0$ which yields the value of $\wt Q^\mathrm{sp}$. 
In full generality this is not possible so we turn to a variational ansatz. 
First, following~\cite{CGSS01} one remarks that  the off-diagonal elements must be time independent, because by definition for $a\neq b$
\begin{equation}
 N Q^\mathrm{sp}_{ab}(t,t')=\overline{\la\bm X^a(t)\cdot\bm X^b(t')\ra}=\overline{\la\bm X^a(t)\ra \cdot\la \bm X^b(t')\ra}=\overline{\la\bm X^a\ra \cdot\la \bm X^b\ra}
\end{equation}
as before performing the disorder average, the two different replicas are uncorrelated. In the last equality we simply take advantage of the time-translational invariance 
of the imaginary-time action in~\eqref{eq:repl_Z}. The system may be in a mixed state where $\la\bm X^a\ra=\sum_\g w_\g \la\bm X^a\ra_\g$ where $\g$ 
denotes a pure state and $w_\g$ its statistical weight, implying a dependence of the overlap upon the different replicas $a$ and $b$~\cite{MPV87,CC05}.\\
Then, the replica method restricts the possible saddle-point matrix to a hierarchical one, where the time dependence appears only on the diagonal (which is time-translational and replica invariant following the same argument)
\begin{equation}\label{eq:hierarchical}
 Q_{ab}^\mathrm{sp}(t,s)=q_d(t-s)\d_{ab}+Q^*_{ab}
\end{equation}
where $Q^*_{ab}$ is a static hierarchical matrix.
A review of their construction can be found in \eg~\cite{MPV87,CKPUZ14}.
In the replica-symmetry breaking (RSB) formalism and $n\to0$ limits, it is parametrized by a 0 diagonal and an increasing function $q(x)$ defined on $[0,1]$, being non-trivial on $[x_m,x_M]\subset[0,1]$ and 
constant away from it with $q(x_m)=q(0)=q_m$, $q(x_M)=q(1)=q_M$. We also work with its reciprocal $x(q)$.

In the following we will consider the most general hierarchical ansatz of continuous RSB (or fullRSB), which relies on a recurrence construction with $k$ blocks within $\hat Q^*$, with $k\to\io$. 
This ansatz is required to described the landscape marginally stable (LMS) phase. Replica symmetry and other cases of RSB can be recovered as special cases.

\subsection{The interaction term}\label{sub:int}
The term containing the interaction potential can be treated in a similar way to the classical case~\cite{FP16,FPSUZ17}.
Here we just give the main modifications with respect to the usual procedure based on recurrence equations over $k$-RSB matrices~\cite{MPV87,CKPUZ14},
with block indices\footnote{We used the same notations as~\cite{FPSUZ17}, beware that $M$ here has nothing to do with the number of obstacles.} $1=m_k\leqslant m_{k-1}=M\leqslant\dots\leqslant m_0=m\leqslant n$. The technical details are explained in Duplantier's comment~\cite{Du81}, 
and we emphasize here only key differences with the classical case. 
The first ``innermost'' step in Duplantier's recurrence is modified with respect to the classical case due to the time dependence on the diagonal and reads
\begin{equation}\label{eq:intterm}
 g(1=m_k,h)=\exp\left(\frac12\iint_0^{\b\hbar}\frac{\dd t}{\b\hbar}\frac{\dd s}{\b\hbar}\,[q_d(t-s)-q_M]\frac{\d^2}{\d r(t)\d r(s)}\right)\restriction{\exp\left(-\frac1\hbar\int_0^{\b\hbar}\dd t\,v(r(t)+h)\right)}{r(t)=0}
 =\la e^{-\frac1\hbar\int_0^{\b\hbar}\dd t\,v(r(t)+h)}\ra_r
\end{equation}
We shifted $r_a(t)\longrightarrow r_a(t)+h$ with a constant $h$ for the needs of the recurrence procedure where at the endpoint of the recurrence $h$ must be set to $-\s$ as at all times $r_a(t)$ is required to be set 
to $-\s$ in~\eqref{eq:SS}. 
The second line is an average over the Gaussian centered process $r(t)$ with variance 
\begin{equation}\label{eq:Gdef}
 G(t-s)=\la r(t)r(s) \ra_r=q_d(t-s)-q_M
\end{equation}
The proof is akin to~\eqref{eq:RSBfullint}. 

The variable $r$ may be interpreted as an effective one-dimensional quantum particle whose evolution describes the typical overlap with an obstacle, a fluctuating quantity even for large $N$. 
The average~\eqref{eq:intterm} is the analog of its partition function.
This particle evolves with the shifted potential $v(r+h)$. Unlike the potential, the Gaussian average, which is the other component determining the evolution, cannot be in general written in the form 
of a Feynman propagator, \ie cannot be described by a standard one-particle Hamiltonian. It is common that mean-field models can be mapped into such a one-dimensional evolution with non-trivial features resulting from tracing out all degrees of freedom.

Then the recurrence on the innermost blocks proceeds as in the classical case through the identity (or its second-derivative version, as in~Ref.~\cite[Eq. (6)]{Du81}), for any functional $f$ of the $n$ scalar fields $r_a(t)$,
\begin{equation}\label{eq:relationf}
\begin{split}
 \restriction{\sum_{a=1}^{n}\int\frac{\dd t}{\b\hbar}\,\frac{\d f[r_1,\dots,r_n]}{\d r_a(t)}}{\forall (j,s),\,r_j(s)=h}&=\frac{\dd }{\dd h}f[h,\dots,h]\\
 \end{split}
\end{equation}
where on the right hand side all fields have a time-independent value $h$. The Duplantier recurrence equation is thus the same for $i\in\llbracket 0,k-1\rrbracket $
\begin{equation}\label{eq:duprec}
 g(m_i,h)=\exp\left(\frac{q_{i+1}-q_i}{2}\frac{\dd^2}{\dd h^2}\right)	g(m_{i+1},h)^{m_i/m_{i+1}}=\g_{q_{i+1}-q_i}\star g(m_{i+1},h)^{m_i/m_{i+1}}
\end{equation}
where\footnote{The last equality is obtained in the same way as~\eqref{eq:intterm} by expanding or simply by noticing the propagator of the diffusion equation.} 
we define the Gaussian weight and the convolution as
\begin{equation}
\begin{split}
 \g_a(z)&=\frac{e^{-z^2/2a}}{\sqrt{2\p a}}\ ,\qquad
  \g(z)=\g_1(z)=\frac{e^{-z^2/2}}{\sqrt{2\p }}\ ,\qquad
 f\star g(h)=\int \dd x\, f(h-x)g(x)
 \end{split}
\end{equation}
Taking the continuous $k\to\io$ 
and $n\to0$ limits one finally arrives at the other end (outermost block) of the recurrence 
\begin{equation}
\a\ln\z\underset{n\to0}{\sim}n\a\g_{q_m}\star f(q_m,-\s)
\end{equation}
where $f(x,h)=(1/x)\ln g(x,h)$ satisfies Parisi's partial differential equation, here written in terms of the $q=q(x)$ variable instead, with 'initial condition' in $x=1$ given by~\eqref{eq:intterm},
\begin{equation}\label{eq:eqf}
  \dot f(q,h)=-\frac12\left[f''(q,h)+x(q)f'(q,h)^2\right]\qquad \textrm{with}\qquad
  f(q_M,h)=\ln\la e^{-\frac1\hbar\int_0^{\b\hbar}\dd t\,v(r(t)+h)}\ra_r
\end{equation}
where the usual notation has a dot standing for partial derivatives over the static overlap $q$ and respectively a prime over $h$.

\subsection{Fourier modes}

We define Fourier components in the following way: the Matsubara frequency modes are for $n\in\mathbbm Z$
\begin{equation}\label{eq:Matsubaradef}
 \om_n=\frac{2\p}{\b\hbar}n
\end{equation}
The real and even function $q_d$ is expressed as
\begin{equation}
  q_d(t)=\sum_{n\in\mathbbm Z}\wt{q_d}(\om_n) e^{i\om_n t} \qquad \textrm{and}\qquad
  \wt{q_d}(\om_n)=\int_0^{\b\hbar}\frac{\dd t}{\b\hbar}\, q_d(t)e^{-i\om_n t}
\end{equation}
with $\wt{q_d}(-\om_n)=\wt{q_d}(\om_n)$ and $\wt{q_d}(\om_n)\in\RRR$.
For simpler notations we will refer to the mode zero as:
\begin{equation}\label{eq:qbar2}
 \bar q=\wt{q_d}(0)
\end{equation}

\subsection{The Jacobian}

We may compute the Jacobian term $\ln\det\wt Q$ as follows. Since we are interested in the logarithm of the determinant we may discard irrelevant constants. 
We write an integral over one-dimensional periodic paths $r_a(t)$ and use their Fourier representation:
\begin{equation}
\begin{split}
 \frac{1}{\sqrt{\det\wt Q}}&\propto \oint \prod_{a=1}^n\mathrm{D}r_a\,
 e^{-\frac12\sum_{a,b}\iint_0^{\b\hbar}\frac{\dd t}{\b\hbar}\frac{\dd s}{\b\hbar}\, r_a(t)Q_{ab}(t,s)r_b(s)}\\
 &=\oint \prod_{a=1}^n\mathrm{D}r_a\,
 e^{-\frac12\sum_{a,b}\iint_0^{\b\hbar}\frac{\dd t}{\b\hbar}\frac{\dd s}{\b\hbar}\, r_a(t)\left(q_d(t-s)\d_{ab}+Q^*_{ab}\right)r_b(s)}\\
& \propto\int\prod_{a=1}^n\prod_{p\neq0}\dd \wt r_a(\om_p)\,e^{-\frac12\sum_a\sum_{p\neq0}\wt q_d(\om_p)\wt r_a(\om_p)^2}
 \int\prod_{a=1}^n\dd \wt r_a(0)\,e^{-\frac12\sum_{a,b}\wt r_a(0)\left(\bar q\d_{ab}+Q^*_{ab}\right)\widetilde r_b(0)}\\
 &\propto\left[\prod_{p\neq0}\frac{1}{\sqrt{\wt q_d(\om_p)}}\right]^n\frac{1}{\sqrt{\det\left(\bar q\d_{ab}+Q^*_{ab}\right)}}
 \end{split}
\end{equation}
We used that the transformation from the time-dependent paths to their Fourier coefficients is an isometry (due to Parseval's identity~\cite{Appel}), hence the Jacobian of this 
transformation is 1. We conclude:
\begin{equation}\label{eq:lndetJ}
 \ln\det \wt Q=n\sum_{p\in\ZZZ}\ln\wt{q_d}(\om_p)-n\ln\bar q+\ln\det(\bar q\hat\id+\hat Q^*)
\end{equation}
The last term in~\eqref{eq:lndetJ} is similar to the classical case, \ie $\bar q\id+\hat Q^*$ is a hierarchical
matrix parametrized by its diagonal $\bar q$ and the function $q(x)$, 
 the trace of its logarithm is computed in~Ref.~\cite[App. II]{MP91} and has the same form as in the classical Jacobian~\cite{FPSUZ17}.  Thus we conclude
\begin{equation}\label{eq:lambda}
 \frac1n\ln\det \wt Q\underset{n\to0}{\sim}\sum_{p\in\ZZZ}\ln\wt{q_d}(\om_p)-\ln\bar q+\ln\l(q_M)+\frac{q_m}{\l(q_m)}+\int_{q_m}^{q_M}\frac{\dd q}{\l(q)} \qquad
 \textrm{with}\qquad\l(q):=\bar q-q_M+\int_{q}^{q_M}\dd p\, x(p)
\end{equation}

\subsection{The unregularized free energy}\label{sec:freediverg}

The kinetic energy term is simple to express in terms of $q_d$. We conclude that all terms in $\AA(\wt Q^\mathrm{sp})\propto n$ for $n\to0$, which gives the free energy:
\begin{equation}\label{eq:unregularizedF}
 -\frac{\b F}{N}=\frac12\left[\sum_{p\in\ZZZ}\ln\wt{q_d}(\om_p)-\ln\bar q+\ln\l(q_M)+\frac{q_m}{\l(q_m)}+\int_{q_m}^{q_M}\frac{\dd q}{\l(q)}\right] +\frac{\b\MM}{2}q_d''(0)+\a\g_{q_m}\star f(q_m,-\s)
\end{equation}
We have nonetheless overlooked $Q$-independent factors in the partition function, and shall soon see that the latter expression has spurious divergences inherent to the continuous-time limit. 
To solve both issues, one should regularize this free energy \eg with respect to the one of the free particle ($v=0$) to retrieve the correct free energy; this is the subject of the next section.

\subsection{The free particle case}\label{sub:free}

The terms left out from the leading-order saddle-point approach are independent of both $\wt Q$ and the potential $v$ but may be temperature dependent. 
This needs to be fixed for the computation of the thermodynamic energy or the specific heat. We may recover them by writing the total free energy $\FF$ as
\begin{equation}\label{eq:freeregul}
 \FF=\FF_{\rm sph}-F_0+F
\end{equation}
where $\FF_{\rm sph}$ is the free energy of a free particle on the $N$-dimensional sphere of radius $\sqrt N$, and $F_0$ is the unregularized free energy~(\ref{eq:unregularizedF}) for $v=0$. 
The other advantage of this formulation is that it regularizes the divergences appearing in~\eqref{eq:unregularizedF}.\\
A comment about the choice of $\FF_{\rm sph}$ is in order, since this term depends on the actual way one quantizes a free particle on the sphere. We calculated $F$ by quantizing the model with constrained position and momentum operators, enforcing the spherical constraint through delta functions. It is called $F_0$ in the free-particle case, and $\FF_{\rm sph}$ when one discretizes properly the path integral, \ie removing divergences from the expression $F_0$. Alternatively, there is an arguably more intuitive quantization scheme, which consists in defining the model through the \textit{correct} quantum Hamiltonian, expressed in terms of the angular momentum.
This is discussed in the appendices: the latter quantization scheme yields a free-particle free energy noted $\FF_{\rm free}$, computed in App.~\ref{app:freesphere}, 
while $\FF_{\rm sph}$ is calculated in App.~\ref{app:discrete}. The derivation of $\FF_{\rm free}$ is done through the knowledge of the energy spectrum, and may be performed also in a path integral formalism using spherical harmonics~\cite[Sec. 8.9.]{Kleinert}. Unfortunately these derivations do not generalize easily in the interacting case, and as a consequence we sticked to the quantization on the sphere with constrained position and momentum. The two quantizations induce different energy spectra, only through a different zero-point energy; thus the specific heat is the same, which is what we are interested in here.

\subsubsection{Zero potential}\label{sub:zeroV}

Let us now inspect the zero-potential free energy $F_0$, \ie the continuous-time one provided by~\eqref{eq:unregularizedF} for $v=0$.
For a free particle, replica symmetry is preserved which translates in the fullRSB formulation in a constant off-diagonal $q(x)=q$~\cite{MPV87}. Then the free energy depends only on $q$ and $q_d$ and reads
\begin{equation}\label{eq:freev0}
 -\frac{\b F_0}{N}= \frac12\left[\sum_{p\in\ZZZ}\ln\wt{q_d}(\om_p)-\ln\bar q+\ln(\bar q -q)+\frac{q}{\bar q-q}\right]+\frac{\b\MM}{2}q_d''(0)-\frac{\b\m_0}{2}\left[q_d(0)-1\right]
\end{equation}
$\m_0$ is a Lagrange multiplier enforcing the spherical constraint. Optimization with respect to $q$ gives directly $q=0$. This corresponds to $\la\bm X^a\cdot \bm X^b\ra=0$ for $a\neq b$ which must be the result at $v=0$ for two particles on the sphere that are completely uncorrelated. 
Next, the saddle-point equation for $q_d$ can be studied by extremizing~\eqref{eq:freev0} with respect to $\wt{q_d}(\om_n)$:
\begin{equation}\label{eq:SP_F0}
\wt{q_d}(\om_n)=\frac{1}{\b}\frac{1}{\MM\om_n^2+\m_0}
\end{equation}
The value of the Lagrange multiplier ensures $q_d(0)=1$. With the latter expression for the Fourier components and the identity
\begin{equation}
 \sum_{n\in\ZZZ}\frac{1}{n^2+A^2}=\frac{\p}{A}\coth(\p A)
\end{equation}
one has
\begin{equation}\label{eq:Lag0}
\begin{split}
\sum_{n\in\ZZZ}\wt{q_d}(\om_n)&=1 \hskip15pt \Leftrightarrow\hskip15pt
 \frac{\coth a}{a}=\frac{8\p}{\Lambda_{\rm dB}^2}\\
 \Lambda_{\rm dB}&=\sqrt{\frac{2\p\b\hbar^2}{\MM}}\ ,\hskip15pt a=\frac{\b\hbar}{2}\sqrt{\frac{\m_0}{\MM}}
\end{split}
\end{equation}
which is solvable graphically. $\Lambda_{\rm dB}$ is the de Broglie thermal wavelength. Note that for a large $\Lambda_{\rm dB}$ (\eg at low temperature), the unknown $a$ must also be large, hence the left-hand side 
of the latter equation is approximated by $1/a$, which gives an analytical expression for $\m_0$:
\begin{equation}\label{eq:m0lowT}
 \m_0\underset{\Lambda_{\rm dB}\to\io}{\sim}\frac{\hbar^2}{4\MM}
\end{equation}
In the opposite limit $\Lambda_{\rm dB}\to0$, the right-hand side diverges which means $a\to0$, therefore
$\coth a\sim1/a$, implying 
\begin{equation}\label{eq:m0highT}
 \m_0\underset{\Lambda_{\rm dB}\to0}{\sim} T
\end{equation}
This is the classical value.\\

Putting back the solution of the saddle-point equations into~\eqref{eq:freev0} we obtain the free energy:
\begin{equation}\label{eq:F0_final}
  -\frac{\b F_0}{N}=-\frac12\sum_{n\in\ZZZ}\ln(\b\MM\om_n^2+\b\m_0)-\frac{1}{2} \sum_{n\in\ZZZ}\frac{\MM\om_n^2}{\MM\om_n^2+\m_0}
\end{equation}

Both series are divergent; this must also happen for the free energy $F$ at $v\neq0$.
Note that the divergences not only affect the thermodynamic energy but the specific heat as well. 
The subtraction $F-F_0$ gets rid of the divergences and gives a well-defined finite free energy $\FF$. 
This is due to the continous-time formalism~\cite{FHS10,Kleinert}. 
A time discretization of Feynman's path integral 
can be performed to show that in a proper continuous-time limit no such divergences appear\footnote{Equivalently one could apply analytic regularization aiming at recovering the discrete-time results~\cite[Sec. 2.15]{Kleinert}.}. This correct result for the free energy $\FF_{\rm sph}$ is derived in App.~\ref{app:discrete}. The related energy is shown there to be $\UU_{\rm sph}/N=\m_0/2$ (\eqref{eq:sphenergy}).

We stress that~\eqref{eq:SP_F0} is equal to $\wt G(\om_n)$ for $n\neq0$. 
Therefore, in this particularly simple case, the partition function of the effective quantum particle $\exp[f(q_M=q=0,h)]$ in~\eqref{eq:eqf} (with $v=0$) is the Feynman propagator of a harmonic oscillator with frequency $\sqrt{\m_0/\MM}$, 
\ie here the effective quantum particle $r$ evolves with a standard one-particle Hamiltonian. 
This is expected as the spherical constraint implemented via a Lagrange multiplier plays the role of a restoring force on the system. See also App.~\ref{app:discrete} for a related discussion.

\subsubsection{A free quantum particle on the large-dimensional hypersphere}\label{sub:freesphere}

Alternatively, as mentioned before one may compute directly the free energy 
from the exact energy spectrum of a free particle on the $N$-dimensional sphere of radius $\sqrt N$, $\FF_{\rm free}$, in order to check independently the result. This is detailed in App.~\ref{app:freesphere}.

\subsubsection{Conclusion}\label{sub:conclufree}

We have computed directly the partition function in App.~\ref{app:freesphere}, and by a path integral approach suitable to a generalization to the interacting case. The divergences appearing in the latter can be regularized by proper discretization, and the different quantization schemes 
give the same results (up to a zero-point energy), see App.~\ref{app:discrete}. \\
Namely in the large- and small-temperature limits:
\begin{itemize}
 \item when $T\to\io$, one gets the ideal gas law $C_V=N/2$
 \item for $T\to0$, the specific heat is exponentially small $C_V\propto\exp(-\D_{\rm free}/T)$. 
 This is due to the presence of a gap $\D_{\rm free}=\hbar^2/(2\MM)$ in the energy spectrum\footnote{If there is a non-zero gap in the spectrum of the Hamiltonian 
 between the first excited states and the ground state, then in a general fashion one can write the logarithm of the partition function as 
 \begin{equation}
  -\b F=-\b E_0+\ln\left(1+\sum_{E\neq E_0}e^{-\b(E-E_0)}\right)\underset{T\to0}{=}-\b E_0+e^{-\b(E_1-E_0)}+\dots
 \end{equation}
where in the sum all energy differences $E-E_0$ are thus strictly positive due to the gap. The last equality here is not general but valid for well separated eigenvalues. It leads to a specific heat $C_V\propto\exp(-\D/T)$, in the latter example one has $\D=E_1-E_0$.}. 
\end{itemize}

\subsection{The full free energy}

As mentioned earlier, the full free energy could be written $\FF=\FF_{\rm sph}-F_0+F$ to regularize the divergences. 
We have previously shown that in the low-temperature limit the free particle contribution has a gap and is hence exponentially suppressed. 
Therefore, any algebraic behaviour of the specific heat must come from the interactions. 
We start from the simpler free energy $F$ from~\eqref{eq:unregularizedF}:
\begin{equation}\label{eq:varfree}
\begin{split}
  -\frac{\b F}{N}=&\frac12\sum_{n\in\ZZZ}\ln\wt{G}(\om_n)- \frac{\b\MM}{2} \sum_{n\in\ZZZ}\om_n^2\wt{G}(\om_n) +\frac12\left[\ln\l(q_M)+\frac{q_m}{\l(q_m)}+\int_{q_m}^{q_M}\frac{\dd q}{\l(q)}\right]+ \a\g_{q_m}\star f(q_m,-\s)\\
  &-\frac{\b\m}{2}\left[G(0)-(1-q)\right]+\a\int\dd h\,\int_{q_m}^{q_M}\dd q\, P(q,h)\left[\dot f(q,h)+\frac12\left(f''(q,h)+x(q)f'(q,h)\right)\right]\\
 &-\a\int\dd h\, P(q_M,h)\left[f(q_M,h)-\ln\la e^{-\frac1\hbar\int_0^{\b\hbar}\dd t\,v(r(t)+h)}\ra_r\right]
  \end{split}
\end{equation}
which will be regularized later.
We recall~\eqref{eq:Gdef} giving the Gaussian statistics $\la r(t) \ra_r=0$, $ \la r(t)r(s) \ra_r=G(t-s)$.\\
Anticipating the fact that we must extremize the free energy, which is the subject of~\secref{sec:var}, we have introduced
the Lagrange multipliers $P(q,h)$ and $\m$ to explicitly impose  within the free energy respectively
the constrained evolution of the Parisi function $f(q,h)$ from~\eqref{eq:eqf} and the spherical constraint $G(0)=(1-q)$, as we express the problem in terms of the variance $G$ defined 
in~\eqref{eq:Gdef} instead of $q_d$ for convenience. 
Notice that if the constraints are properly enforced, the free energy is only expressed by the first line of~\eqref{eq:varfree}. 

\subsection{Classical limit}\label{sec:classical}

In this section, as a check, we recover the classical equations in the limit $\hbar\to0$. \\
We consider~\eqref{eq:varfree}. 
The imaginary time interval shrinks to zero in this limit, 
\ie $q_d(t)=\mathrm{constant}=q_d(0)=1$ due to the spherical constraint. Then $\wt q_d(\om_n)=\d_{n0}$ 
and only the zero frequency counts in the frequency sums: this makes the logarithmic term vanish 
(as it should, since $\det q_d$ becomes 1, or in other words, $\wt Q$ becomes the classical time-independent hierarchical matrix $\id +\hat Q^*$) as well as the momentum term (proportional to the mass).
From its definition $\bar q$ becomes 1, thus $\l(q)\to1-q_M+\int_{q}^{q_M}\dd p\, x(p)$ is the classical function. The only remaining $\hbar$ dependency is on the initial condition for the function $f(q,h)$. 
The Gaussian average shrinks to a single imaginary time $t=0$ ($r(t)\to r$):
\begin{equation}
  \la e^{-\frac1\hbar\int_0^{\b\hbar}\dd t\,v(r(t)+h)}\ra_r=\int_\RRR\dd r\, \frac{e^{-r^2/2(1-q_M)}}{\sqrt{2\p(1-q_M)}}e^{-\frac{1}{\hbar}\b\hbar v(h+r)}\underset{r\to-r}{=}\g_{1-q_M}\star e^{-\b v(h)}
\end{equation}
Thus we find the same initial condition for $f(q,h)$ as in the classical case. Reintroducing the Lagrange multipliers $P(q,h)$, we get
\begin{equation}
 \begin{split}
  -\frac{\b F}{N}  \underset{\hbar\to0}{\sim}&
  \,\frac12\left[\ln\l(q_M)+\frac{q_m}{\l(q_m)}+\int_{q_m}^{q_M}\frac{\dd q}{\l(q)}\right]+\a\g_{q_m}\star f(q_m,-\s)-\a\int\dd h\, P(q_M,h)\left[f(q_M,h)-\ln\g_{1-q_M}\star e^{-\b v(h)}\right]\\
  &+\a\int\dd h\,\int_{q_m}^{q_M}\dd q\, P(q,h)\left[\dot f(q,h)+\frac12\left(f''(q,h)+x(q)f'(q,h)\right)\right]\\
 \end{split}
\end{equation}
which is the same as in the classical case~\cite[Eq. (32)]{FPSUZ17}.

\section{Non-dimensional variational equations}\label{sec:var}

The $N\to\io$ saddle-point condition on $\AA(\wt Q)$ translates into an optimization over the parameters of the variational ansatz for the overlap $\wt Q$.
The Lagrange multipliers in~\eqref{eq:varfree} allow to get the variational equations accommodating to the various constraints in a simple way. 
The variational equations are obtained by deriving with respect to the functions $f$, $P$, $x$ as in the classical case~\cite{FP16,FPSUZ17}. In addition for the quantum case this must be done with respect to the 
function $q_d$ as well.\\
Furthermore, for convenience we use dimensionless units from now on, except when explicitly mentioned. 
We choose as the unit of energy the potential's amplitude, $\varepsilon$, and $\b\hbar$ as the unit of time. This choice is guided by the fact that in 
the following these quantities will be mere constants. Consequently, we define the following rescaled quantities:
\begin{equation}
 \begin{split}
  \hat t&=\frac{t}{\b\hbar}\,, \hskip15pt \hat \om_n=\b\hbar\om_n=2\p n\,, \hskip15pt \hat v(h)=\frac{v(h)}{\varepsilon}=\frac{h^2}{2}\th(-h)\\
  \wh G(\hat t)&=\sum_{n\in\ZZZ}\wt{\wh G}(\hat \om_n)e^{i\hat \om_n\hat t} \,,\hskip15pt \wt{\wh G}(\hat \om_n)=\int_0^1 \dd\hat t\, e^{i\hat \om_n\hat t} \wh G(\hat t)\\
  \wh G(\hat t)&=G(\b\hbar\hat t)\,, \hskip15pt \hat r(\hat t)=r(\b\hbar\hat t) \,, \hskip15pt \hat\m=\frac{\m}{\varepsilon}
 \end{split}
\end{equation}
All overlaps are not rescaled in any way since they are already measured in dimensionless units of length (squared). Similarly, we can set $\varepsilon=1$, measuring all energies in units of $\varepsilon$. 
Besides we get rid of all the hats in order to lighten the notation. The dimensionless free energy now reads
\begin{equation}\label{eq:varfreedimless}
\begin{split}
  -\frac{\b F}{N}=&\frac12\sum_{n\in\ZZZ}\ln\wt{G}(\om_n)- \frac{\MM}{2\b\hbar^2} \sum_{n\in\ZZZ}\om_n^2\wt{G}(\om_n) +\frac12\left[\ln\l(q_M)+\frac{q_m}{\l(q_m)}+\int_{q_m}^{q_M}\frac{\dd q}{\l(q)}\right]+ \a\g_{q_m}\star f(q_m,-\s)\\
  &-\frac{\b\m}{2}\left[G(0)-(1-q)\right]+\a\int\dd h\,\int_{q_m}^{q_M}\dd q\, P(q,h)\left[\dot f(q,h)+\frac12\left(f''(q,h)+x(q)f'(q,h)\right)\right]\\
 &-\a\int\dd h\, P(q_M,h)\left[f(q_M,h)-\ln\la e^{-\b\int_0^1\dd t\,v(r(t)+h)}\ra_r\right]
  \end{split}
\end{equation}

\subsection{The ``classical'' variational equations}

Most of the classical-like variational equations are unchanged with respect to Refs.~\cite{FP16,FPSUZ17}:
\begin{itemize}
 \item over $P(q,h)$ ($q_m<q<q_M$) and $P(q_M,h)$ we get back the constraints on $f$~\eqref{eq:eqf}
 \item over  $f(q_M,h)$ the equation is automatically satisfied ($0=0$)
 \item over $f(q,h)$ ($q_m<q<q_M$) and $f(q_m,h)$, using integrations by parts and~\eqref{eq:eqf}, we get the conjugated equations ruling the Lagrange multiplier $P$:
\begin{equation}\label{eq:eqP}
  \dot P(q,h)=\frac12\left[P''(q,h)-2x(q)\left(P(q,h)f'(q,h)\right)'\right]\qquad \textrm{with}\qquad
  P(q_m,h)=\g_{q_m}(h+\s)
\end{equation}
\item over $x(q)$, using $\d\l(q')/\d x(q)=\th(q-q')$:
\begin{equation}\label{eq:dxq}
 \frac{q_m}{\l(q_m)^2}+\int_{q_m}^q\frac{\dd p}{\l(p)^2}=\a \int\dd h\, P(q,h)f'(q,h)^2
\end{equation}
\end{itemize}

\subsection{The time-dependent variational equation}

Effective averages on the dynamics of $r$ may be written either as in~\eqref{eq:intterm} or equivalently
\begin{equation}\label{eq:defrmeasurePI}
  \la e^{-\b\int_0^1\dd t\,v(r(t)+h)}\ra_r=\int \DD_G r\,e^{-\b\int_0^1\dd t\,v(r(t)+h)}\qquad\textrm{with}\qquad
  \DD_G r:=  \mathrm{D}r\,\frac{e^{-\frac12\iint\dd t\dd s\,r(t)G^{-1}(t-s)r(s)}}{\sqrt{\det G}} 
\end{equation}
The variational equation over $\wt q_d(\om_n)$ is more easily derived through the expression~\eqref{eq:intterm}:
\begin{equation}\label{eq:qdvarFourier}
 \begin{split}
 &\frac{1-\d_{n0}}{\wt G(\om_n)}+\d_{n0}\left[\frac{1}{\bar q -q_M}-\frac{q_m}{\l(q_m)^2}-\int_{q_m}^{q_M}\frac{\dd q}{\l(q)^2}\right]=\frac{\MM}{\b\hbar^2}\om_n^2+\b\m+\b\wt\Si(\om_n)\\
\wt\Si(\om_n)= -\frac{\a}{\b}\int&\dd h\,P(q_M,h)\iint_0^1\dd t\dd s\,e^{i\om_n(t-s)}\la \b^2 v'(r(t)+h)v'(r(s)+h)-\beta\d(t-s)v''(r(t)+h)\ra_v
 \end{split}
\end{equation}
where we defined the $h$-dependent normalized averages over the effective quantum particle
\begin{equation}\label{eq:intr}
 \la \bullet \ra_v=\frac{\la \bullet\, e^{-\b\int_0^1\dd t\,v(r(t)+h)}\ra_r}{\la e^{-\b\int_0^1\dd t\,v(r(t)+h)}\ra_r}
\end{equation}
To lighten the notation we do not mention that \textit{the effective averages} $\la \bullet \ra_v$ \textit{depend on} $h$, but one should keep in mind that they do.
\eqref{eq:qdvarFourier} is cast into a Dyson equation with self-energy $\Si$~\cite{AS06}. $\wt G$ is the analog of a Green function~\cite{Mahan,Bruus}. 
Finally, the value of the Lagrange multiplier $\m$ is retrieved through the spherical constraint $G(0)=1-q$.

\subsection{Principle of the solution of the full replica-symmetry breaking equations}\label{sub:solveiter}

Here we recap the global structure of the self-consistent equations to solve, and a procedure to achieve it in principle (\eg numerically).\\
To get $q_m$ and $x(q)$, one derives respectively once and twice~\eqref{eq:dxq} with respect to $q\in]q_m,q_M[$\footnote{Consequently these equations are valid only if $\dot x(q)\neq0$, \ie if there exists a continuum of $q$'s.}, using Eqs.~[\ref{eq:eqf}],[\ref{eq:eqP}]:
\begin{equation}\label{eq:ddxq}
  \frac{1}{\l(q)^2}=\a\int\dd h\,P(q,h)f''(q,h)^2
\end{equation}
\begin{equation}\label{eq:xq}
 x(q)=\frac{\l(q)}{2}\frac{\int\dd h\,P(q,h)f'''(q,h)^2}{\int\dd h\,P(q,h)f''(q,h)^2+\l(q)\int\dd h\,P(q,h)f''(q,h)^3}
\end{equation}
Finally doing the ratio of~\eqref{eq:dxq} and~\eqref{eq:ddxq} for $q\to q_m$ and using the second line of~\eqref{eq:eqP},
\begin{equation}\label{eq:qm}
 q_m=\frac{\int\dd h\,\g_{q_m}(h+\s)f'(q_m,h)^2}{\int\dd h\,\g_{q_m}(h+\s)f''(q_m,h)^2}
\end{equation}
Now, the main problem is to get $G(t)$ from the implicit equation~\eqref{eq:qdvarFourier}. The iterative solution of the variational equations may be achieved through the following procedure:
\begin{enumerate}
 \item Start with a guess for $x(q)$, $q_m$, $q_M$ and $q_d(t)$ or $G(t)$;
 \item Get $f$ and $P$ solving Eqs.~[\ref{eq:eqf}],[\ref{eq:eqP}];
 \item Get a new estimate of $q_m$ from~\eqref{eq:qm};
 \item Get a new estimate of $q_M=\bar q-\l(q_M)=\int_0^1 q_d-\l(q_M)$ from~\eqref{eq:ddxq} in the limit $q\to q_M$;
 \item Get a new estimate of $x(q)$ from~\eqref{eq:xq};
 \item Get a new estimate of $q_d(t)$ from~\eqref{eq:qdvarFourier}, \ie from the first term $\propto 1/\wt q_d(\om_n)$ in the left-hand side for $n\neq0$, that is updated from the rest of the equation (evaluated through the non-updated values), 
 while the $n=0$ equation provides $\bar q$ in a similar procedure;
 \item Iterate from 2;
 \item After convergence of  $x(q)$, $q_m$, $q_M$ and $q_d(\t)$, use~\eqref{eq:eqf} to get $f(q_m,h)$, and finally  the free energy.
\end{enumerate}

\section{Replica-symmetric regime}

From the classical $T=0$ analysis~\cite{FP16,FPSUZ17}, we expect the assumption of replica symmetry (RS) to be valid throughout the whole convex phase ($\s\geqslant0$). This will be further discussed in~\secref{sec:margUNSAT}.
A preserved RS means that in this region the free energy displays a single well. 
In the replica framework this translates into the ansatz $Q_{aa}(t,t')=q_d(t-t')$, $Q_{a\neq b}=q$~\cite{MPV87,KMRSZ07,BM80b,CGSS01}, 
\ie  $q(x)=q~(=q_m=q_M)$, and $x$ is either 0 or 1 (the only block parameters are $m_0=1$ and $n\to0$).  

In this section we derive the main equations of the model in the RS hypothesis, as a starting point for the analysis of the next section.

 \subsection{The replica-symmetric free energy}
 
The interaction term is readily given by~\eqref{eq:eqf} or can be computed directly from~\eqref{eq:SS} writing $Q_{ab}(t,t')=\d_{ab}G(t-t')+q$ and using the second derivative version of~\eqref{eq:relationf}.
Both lead to
 \begin{equation}\label{eq:fRSwG}
  -\frac{\b F_{\mathrm{RS}}}{N}=\frac12\sum_{n\in\ZZZ}\ln\wt{G}(\om_n)+\frac{q}{2\wt G(0)}
 -\frac{\MM}{2\b\hbar^2} \sum_{n\in\ZZZ}\om_n^2\wt{G}(\om_n) + \a\g_{q}\star \ln\la e^{-\b\int_0^1\dd u\,v(r(u)-\s)}\ra_r-\frac{\b\m}{2}\left[G(0)-(1-q)\right]
 \end{equation}
 where the effective quantum process for the typical overlap with an obstacle is Gaussian distributed with $\la r(t) \ra_r=0$, ${ \la r(t)r(s) \ra_r=G(t-s)}$ (see~\eqref{eq:Gdef}).

\subsection{The variational equations}

The variational equations read:
\begin{itemize}
 \item With respect to $q$ one can differentiate directly or use \eqref{eq:dxq}:
 \begin{equation}\label{eq:dxqRS}
 \frac{q}{(\bar q-q)^2}=\frac{q}{\wt G(0)^2}=\a \int\dd h\, \g_q(h+\s)\left[\frac{\dd}{\dd h}\ln\la e^{-\b\int_0^1\dd t\,v(r(t)+h)}\ra_r\right]^2
\end{equation}
\item For the imaginary-time-dependent overlap, likewise~\eqref{eq:qdvarFourier} we get the Dyson equation: 
\begin{equation}\label{eq:Schehrlike}
 \begin{split}
  \forall n\neq0\,,\hskip15pt\wt G(\om_n)&=\frac{1}{\b}\frac{1}{\frac{\MM}{(\b\hbar)^2}\om_n^2+\m+\wt\Si(\om_n)}\\
  n=0:\hskip15pt \frac{\bar q-2q}{\b(\bar q-q)^2}&=\frac{{\wt G(0)}-q}{\b\wt G(0)^2}=\m+\wt\Si(0)
 \end{split}
\end{equation}
The self-energy is
\begin{equation}\label{eq:alternateSi}
 \wt\Si(\om_n)=-\frac{\a}{\b}\int\dd h\, \g_q(h+\s)\iint_0^1\dd t\dd s\,e^{i\om_n(t-s)}\la \b^2 v'(r(t)+h)v'(r(s)+h)-\beta\d(t-s)v''(r(t)+h)\ra_v
\end{equation}
where the effective average $\la \bullet\ra_v$ has been defined in~\eqref{eq:intr}.
\item Finally, the Lagrange multiplier must be determined from the equation $G(0)=1-q$:
\begin{equation}\label{eq:lagrange}
 \bar q+\frac{2}{\b}\sum_{n\geqslant1}\frac{1}{\frac{\MM}{(\b\hbar)^2}\om_n^2+\m+\wt\Si(\om_n)}=1
 \hskip15pt \Leftrightarrow \hskip15pt {\wt G(0)}+\frac{2}{\b}\sum_{n\geqslant1}\frac{1}{\frac{\MM}{(\b\hbar)^2}\om_n^2+\m+\wt\Si(\om_n)}=1-q
\end{equation}
\end{itemize}

\subsection{The thermodynamic energy}\label{sub:thermoU}

To investigate the behaviour of the specific heat, we need to study the energy of the system $\UU_{\rm RS}=\overline{\la  H\ra}=\partial(\b \FF_{\rm RS})/\partial\b$. For this we notice that 
$\b F_{\rm RS}/N$ has an explicit dependence on $\b$ and an implicit one through saddle-point quantities such as $q(\b)$, $G(\b)$ and $\m(\b)$ which optimize it. 
Using the chain rule, we see that we only need to derive with respect to the explicit 
$\b$ dependence (terms like $\frac{\partial(\b F_{\rm RS})}{\partial q}\frac{\partial q}{\partial\b}$ and alike vanish due to the saddle-point condition), 
\begin{equation}\label{eq:energyRSdiv}
 \frac{U}{N}=-\frac{\MM}{2}\sum_{n\in\ZZZ}\left(\frac{\om_n}{\b\hbar}\right)^2\wt{G}(\om_n) +\a\int\dd h\, \g_q(h+\s)\la v(r(t)+h) \ra_v
\end{equation}
where we used time-translation invariance of the action in the second term. We thus have the energy written as a sum of a kinetic energy term and a potential energy one.
The sum is manifestly divergent so we regularize it as mentioned in~\secref{sub:free}:
\begin{equation}\label{eq:enregul}
\UU_{\rm RS}=\UU_{\rm sph}-U_0+U
\end{equation}
$\UU_{\rm sph}$ is the energy of the free-particle model quantized with position and momentum operators constrained on the sphere, analyzed in App.~\ref{app:discrete}.
The divergences are taken care of by the subtraction $U-U_0$ with respect to the free particle case. The Green function $\wt G$ is given by~\eqref{eq:Schehrlike} in the interacting case, and by~\eqref{eq:SP_F0} in the 
free-particle case. This amounts to replace $\wt G\to \wt G-\wt G[v=0]$, and exploiting the spherical constraints, 
one gets the regularized total energy
\begin{equation}\label{eq:energyRS}
\begin{split}
  \frac{\UU_{\rm RS}}{N}=&\frac{\UU_{\rm sph}}{N}-\frac{\m_0}{2}+
 \frac{1}{2\b}\sum_{n\in\ZZZ}\frac{\m+\wt\Si(\om_n)}{\MM\left(\frac{\om_n}{\b\hbar}\right)^2+\m+\wt\Si(\om_n)}
 +\a\int\dd h\, \g_q(h+\s)\la v(r(t)+h) \ra_v\\
 =&\frac{1}{2\b}\sum_{n\in\ZZZ}\frac{\m+\wt\Si(\om_n)}{\MM\left(\frac{\om_n}{\b\hbar}\right)^2+\m+\wt\Si(\om_n)}
 +\a\int\dd h\, \g_q(h+\s)\la v(r(t)+h) \ra_v
 \end{split}
\end{equation}
where we used that $\UU_{\rm sph}/N=\m_0/{2}$, see~\eqref{eq:sphenergy} in App.~\ref{app:discrete}. 
For $\a=0$ or $v=0$, since $\Si=0$ it reduces to $\UU_{\rm RS}/N=\m_0/2=\UU_{\rm sph}/N$ in virtue of the spherical constraint~\eqref{eq:Lag0}.
%

\subsection{Landscape marginal stability}

The continuous breaking of replica symmetry characterizes a \textit{landscape} marginal stability (LMS): the Hessian of the free energy close to the saddle-point solution develops a zero mode (the so-called replicon eigenvalue). 
The vanishing of the replicon thus delimits the fullRSB phase, defining the so-called de Almeida-Thouless line (dAT)~\cite{MPV87}. A simpler equivalent procedure that provides the dAT line, without computing 
explicitly the spectrum of the Hessian~\cite{FPSUZ17}, is to note that $q(x)$ becomes non constant when replica symmetry is broken. Usually, the deviation from a constant is localized around a particular point $x_{\rm bp}$, 
called the breaking point, where $\dot q(x)$ becomes continuously non zero. In the unstable phase, \eqref{eq:ddxq} holds in the unstable phase close to the breaking point $x_{\rm bp}(q_{\rm bp})$. Approaching the instability from the unstable phase, 
$q(x)$ goes to a constant $q$ and the expression reduces to its RS form:
\begin{equation}\label{eq:margstab}
 \frac{1}{(\bar q-q)^2}=\frac{1}{\wt G(0)^2}=\a \int\dd h\, \g_q(h+\s)\left[\frac{\dd^2}{\dd h^2}\ln\la e^{-\b\int_0^1\dd t\,v(r(t)+h)}\ra_r  \right]^2
\end{equation}
This equation, computed on the solution of the RS variational equations, gives an implicit equation for the dAT line $\a_{\rm dAT}(\s)$.

\section{Semi-classical expansion at fixed Matsubara period in the (RS) convex phase}\label{sec:hbar}

We now put to work the Schehr-Giamarchi-Le Doussal (SGLD) strategy~\cite{SGLD04,SGLD05,S05} and expand all quantities in the RS equations in a small $\hbar$ expansion with $\b\hbar$ fixed: $\OO=\sum_{n=0}^{\io}\hbar^n \OO_{(n)}(\b\hbar)$, for example 
we set the value of the static overlap as
\begin{equation}
 q=q_{(0)}+\hbar\, q_{(1)}+O(\hbar^2)
\end{equation}

Classically, at finite $T$ the particle on the sphere is allowed to violate the constraints towards the obstacles, such non-zero energy configurations have a weight in the partition function and the system is in a single-well phase. 
At $T\to0$ the system must be in a ground state where constraints must be enforced as much as possible. Increasing the density $\a$ we thus find a transition from a SAT phase where the constraints are satisfied 
to an UNSAT phase where the constraints cannot be satisfied (SAT-UNSAT transition). The order parameter for this classical transition is the ground-state energy:  
in the SAT phase the constraints are satisfied and therefore do not give any contribution to the energy (the ground-state energy is $e_{\rm GS}=0$), while in the UNSAT phase the constraints cannot be satisfied and $e_{\rm GS}>0$.\\
The lowest order in the expansion coincides with the classical value at $T=0$: indeed in the limit we perform one has $T=O(\hbar)\to0$. 
We then need to specify the starting point of the expansion, \ie whether we are in a SAT or UNSAT phase. 
In the SAT phase there is a finite entropy of configurations satisfying the constraints, thus two different replicas typically occupy distinct states and the classical value is $q_{(0)}<1$~\cite{KMRSZ07,MPV87}. 
In the UNSAT phase, all the replicas fall onto the same state when $T\to0$ if there is a unique minimum and as a result one has $q_{(0)}\to1$. 

Incidentally, in this section we will recover classical $T=0$ results of Refs.~\cite{FP16,FPSUZ17}, and derive the classical low-$T$ behaviour of the specific heat.

Let us make two remarks:
\begin{itemize}
 \item At finite temperature in the classical case, the SAT-UNSAT transition becomes a crossover as the constraints need not be enforced drastically, one can sample configurations violating the 
 constraints with a finite probability weight.
 Analogously, this is also the case at $T=0$ but finite $\hbar$, where $q<1$ due to the quantum fluctuations. 
 Indeed the position $\bm X$ of the particle cannot be a conserved quantity of the ground state since it does not commute with the Hamiltonian~\eqref{eq:originalH}, 
 and therefore cannot take a definite value. When $\hbar$ is finite many quantum paths  have non-zero probability weight, inducing fluctuations of such non-conserved quantities.
 \item If at \textit{fixed} $T>0$ we take the limit $\hbar\to0$, $\wt G(0)=O(\hbar^0)$ (its classical value) while $\wt G(\om_{n\neq0})\underset{\hbar\to0}{\to}0$ is a quantum fluctuation (see~\eqref{eq:Schehrlike}). 
We will see below that this is still true in the SAT phase 
when performing the different limit $\hbar\to0$, $\beta\hbar$ fixed, whereas in the UNSAT phase, $\wt G(\om_n)$ is a quantum fluctuation for every $n$. 
\end{itemize}


\subsection{Starting from the SAT phase}\label{eq:SATschehr}

In the following we will be more interested in the UNSAT phase, but here as an example we get the lowest-order values in the SAT phase, which are compared to the classical results~\cite[Sec. IV.A.]{FPSUZ17} 
and are convenient to derive the location of the RS SAT-UNSAT transition line.\\
The classical limit for $G(t)$ is its time-independent value $G(t)\to q_d(0)=1$ (since as remarked in the last section, 
$\wt G(\om_{n\neq0})\underset{\hbar\to0}{\to}0$), hence we can write at fixed $\b\hbar$
\begin{equation}\label{eq:SATGini}
 \wt G(\om_n)=(1-q_{(0)})\d_{n0}+\hbar\,\wt G_{(1)}(\om_n)+O(\hbar^2)
\end{equation}
Thus, at the lowest order, the following terms in the free energy~\eqref{eq:fRSwG} reduce to their classical counterpart at $T=0$~\cite[Eq.(40)]{FPSUZ17}:
\begin{equation}
\begin{split}
  \frac12\sum_{n\in\ZZZ}\ln\wt{G}(\om_n)+\frac{q}{2\wt G(0)} 
 \sim& \frac12\left[\ln(1-q_{(0)})+\frac{q_{(0)}}{1-q_{(0)}}\right]\\
 -\frac{\MM}{2\b\hbar^2} \sum_{n\in\ZZZ}\om_n^2\wt{G}(\om_n)\longrightarrow&
 -\frac{\b\FF_{\rm sph}(\hbar=0)}{N}
\end{split}
\end{equation}
where in the second line the only $O(1)$ term in the sum is the $n=0$ one which does not contribute to the sum, and we regularized the kinetic term as we did for the energy in~\eqref{eq:energyRS}, providing 
the classical contribution of the free particle on the sphere (see App.~\ref{app:discrete}), \ie for $\hbar\to0$, the ideal gas free energy.
To compute the average over the single-particle process $r(t)$, one can introduce one Gaussian centered process for each order in this way:
\begin{equation}\label{eq:rt}
 r(t)=r_{(0)}+\sqrt\hbar\, r_{(1)}(t)+O(\hbar)
\end{equation}
where each $r_{(n)}(t)$ is a Gaussian random variable defined by the moments:
\begin{equation}
\begin{split}
  \la r_{(n)}(t)\ra_n=&0 \hskip30pt \la r_{(n)}(t)r_{(n)}(t')\ra_n=G_{(n)}(t-t')\\
  \textrm{in~particular}\hskip15pt  \la r_{(0)}\ra_0=&0\,,\hskip30pt \la r_{(0)}^2\ra_0=1-q_{(0)}
\end{split}
\end{equation}
and any two different variables are independent, so that one recovers $\la r(t)r(t')\ra=G(t-t')$. The lowest order is once again time independent, and one gets
\begin{equation}
 \la e^{-\b\int_0^1\dd t\,v(r(t)+h)}\ra_r\sim\la e^{-\b v(r_{(0)}+h)}\ra_0\sim \int \dd r_{(0)}\, \g_{1-q_{(0)}}(r_{(0)})\,\th(r_{(0)}+h)=\Th\left(\frac{h}{\sqrt{2(1-q_{(0)})}}\right)
\end{equation}
which is indeed the classical result~\cite[Eq.(41)]{FPSUZ17} once we determine $q_{(0)}$ through~\eqref{eq:dxqRS}, which reads similarly at the lowest order:
\begin{equation}
 \frac{q_{(0)}}{(1-q_{(0)})^2}=\a \int\dd h\, \g_{q_{(0)}}(h+\s)\left[\frac{\dd}{\dd h}\ln\Th\left(\frac{h}{\sqrt{2(1-q_{(0)})}}\right)\right]^2
\end{equation}
This is the classical variational equation for the inter-replica overlap~\cite[Eq.(43)]{FPSUZ17}.
One approaches the UNSAT phase from below through the limit $q_{(0)}\to1$. This defines the classical jamming line (or SAT-UNSAT transition line) 
\begin{equation}\label{eq:classjammline}
 \a_J^{\rm cl}(\s)=\left[\int_{-\io}^0\dd h\,\g(h+\s)h^2\right]^{-1}
\end{equation}

\subsection{Starting from the UNSAT phase}\label{sub:UNSATexp}

We now turn to the RS UNSAT phase as a starting point.\\
The main aim of this section is to get the lowest order of $\m+\wt \Si(\om_n)$ appearing when studying the quantum thermodynamic energy at low temperature, 
by solving the self-consistent equations through the $\hbar\to0$, fixed $\b\hbar$ expansion. This quantity is the main technical issue of this expansion owing to the effective averages. 
We detail how to compute these here so that one can proceed directly to the expansion of the thermodynamic energy in the next section.

We begin with a few definitions. Because $q_{(0)}=1$ in this phase, we have from the reasoning leading to~\eqref{eq:SATGini} that $\forall n$, $\wt G(\om_n)=O(\hbar)$, \ie
\begin{equation}\label{eq:expGhbar}
 \wt G(\om_n)=\hbar\,\wt G_{(1)}(\om_n)+\hbar^2\wt G_{(2)}(\om_n)+O(\hbar^3)
\end{equation} 
Since in the classical limit $\wt G(\om_n)\to0$, then in this limit $G(t)\to$ constant $=G(0)=1-q^{\rm cl}$, where we note $q^{\rm cl}$ the classical value of the static overlap. 
In the low-temperature limit, in the region where the UNSAT phase is defined (strictly speaking only at $T=0$), one has $q^{\rm cl}=1-\chi T+O(T^2)$~\cite{FPSUZ17}. So, if we \textit{first} take 
$\hbar\to0$ \textit{then} $T\to0$, we get $\b\wt G(\om_n)\sim\chi\d_{n0}$. 
One thus expects that in the SGLD expansion, $\c$ is the lowest order of the quantity $\b\wt G(0)$. We define consequently 
\begin{equation}\label{eq:defchi}
 \b\wt G(0)\sim\b\hbar\wt G_{(1)}(0):=\chi=O(1)
\end{equation}
and we shall show in~\secref{sub:classUNSAT} that indeed $\c$ is the classical deviation rate of $q^{\rm cl}$ from $1$, ruled by the classical $T=0$ equation obtained in~Ref.~\cite{FPSUZ17}.

In the following we shall equivalently take the temperature as the small parameter, writing $\b$ instead of $\b\hbar/\hbar$ for $\hbar\to0$. 

\subsubsection{Asymptotic saddle-point expansion}\label{sub:spexpdiag}

We start from the expression of the self-energy~\eqref{eq:alternateSi} in the time domain
\begin{equation}\label{eq:alternateSi2}
 \Si(t-s)=-\a\int\dd h\, \g_q(h+\s)\la \b v'(r(t)+h)v'(r(s)+h)-\d(t-s)v''(r(t)+h)\ra_v
\end{equation}
The self-energy is a functional of $G$, which in turn is related to the self-energy through the Dyson 
equation~\eqref{eq:qdvarFourier}. The non-Gaussian action of the above effective average is, from Eqs.~[\ref{eq:defrmeasurePI}],[\ref{eq:intr}]:
\begin{equation}\label{eq:actiondyn}
 -\b E[r]:=-\frac12\iint_0^1\dd t\dd t'\,r(t)G^{-1}(t-t')r(t')-\b\int_0^1\dd t\, v(r(t)+h)
\end{equation}
In the UNSAT phase, both terms in the action~\eqref{eq:actiondyn} are 
$O(1/\hbar)$; consequently asymptotic saddle-point expansion of the averages may be achieved and yields the exact result in perturbation. In the remaining we shall need only the first two orders in $\hbar$. 
Indeed one can see from~\eqref{eq:alternateSi2} that the lowest order is $O(1/\hbar)$, which is confined to the $\om_n=0$ mode as we shall see, whereas the next order $O(1)$ has a 
non-trivial $\wt G(\om_n)$ dependence for $n\neq0$ that will close the set of equations with the help of Dyson's. 
The saddle point is determined through the equation of motion 
\begin{equation}\label{eq:exponentr}
 \int_0^1\dd t'\,\hbar\, G^{-1}(t-t')r^*(t')+\b\hbar v'(r^*(t)+h)=0
\end{equation}
for which one must find the periodic solutions $r^*(t)$; as a consequence, if $r^*(t)$ is solution then for any $t_0\in]0,1[$, $r^*(t+t_0)$ is too. 
Note that all values are $O(1)$ at leading order here. \\
We shall show later that there exists a constant solution. Moreover \eqref{eq:exponentr} gives an equation for ${(r^*(t)+h)\th(-r^*(t)-h)}$ 
that we square to plug in the potential term of~\eqref{eq:actiondyn}. This yields the saddle-point value of the exponent in Fourier components
\begin{equation}
  -\b E[r^*]= -\frac12\sum_{n\in\ZZZ}\frac{\abs{\wt r^*(\om_n)}^2}{\wt G(\om_n)}-\frac\b2\left(\frac{\hbar}{\b\hbar}\right)^2\sum_{n\in\ZZZ}\frac{\abs{\wt r^*(\om_n)}^2}{\wt G(\om_n)^2}
\end{equation}
so that, at saddle-point level, any time-dependent solution has an exponentially smaller weight than the constant solution.
We conclude that only the constant solution has to be considered. 

We look for a constant solution of~\eqref{eq:exponentr}. For convenience we define $z^*=r^*+h$, the saddle-point equation reads:
\begin{equation}
 z^*\left[\frac{1}{\b\wt G(0)}+\th(-z^*)\right]=\frac{h}{\b\wt G(0)}\hskip15pt \Rightarrow\hskip15pt
 \textrm{sign}(z^*)=\textrm{sign}(h)
\end{equation}
Hence the value of $z^*$ can be determined separating the cases $h<0$ and $h>0$. One gets
\begin{equation}\label{eq:rstardef}
 r^*(h)=-\th(-h)\frac{h}{1+\frac{1}{\b\wt G(0)}}
\end{equation}
In order to compute time-averages $\la\bullet\ra_r$ appearing in~\eqref{eq:alternateSi2} where not only the leading order (simply given by Laplace's method) but the second-to-leading order 
is needed, it is required to perform an asymptotic expansion around the saddle point. We set the fluctuation around the saddle point ${ \r(t)=r(t)-r^*}$
and expand the exponent $E$ in~\eqref{eq:exponentr}, noticing that here we do not need to go further than the quartic order in $\r$:
\begin{equation}\label{eq:expE}
 \begin{split}
  \restriction{\frac{\d^2 E}{\d r(t_1)\d r(t_2)}}{r^*}&=\frac{1}{\b}G^{-1}(t_1-t_2)+\d(t_1-t_2)v''(r^*+h)\\
  \restriction{\frac{\d^3 E}{\d r(t_1)\d r(t_2)\d r(t_3)}}{r^*}&=\d(t_1-t_2)\d(t_1-t_3)v^{(3)}(r^*+h)\\
  \restriction{\frac{\d^4 E}{\d r(t_1)\d r(t_2)\d r(t_3)\d r(t_4)}}{r^*}&=\d(t_1-t_2)\d(t_1-t_3)\d(t_1-t_4)v^{(4)}(r^*+h)
 \end{split}
\end{equation}
At the saddle point, in the exponent $E$, orders in $\r$ greater than the quadratic one are subdominant, 
hence we may use perturbation theory around the Gaussian action for $\r$, $E_{\rm Gaussian}[\r]$ . Diagrammatically we thus write
\begin{equation}\label{eq:perturbth}
\begin{split}
 \r(t)&:=\begin{tikzpicture}[baseline=-1mm]  \draw (0:0) node[cross=4pt,line width=1pt,rotate=0] {}; \draw(0,-0.3) node {$t$}; \draw[line width=1pt] (0:0) -- (0:0.7); \end{tikzpicture}\ ,\qquad
\la \r(t)\r(t')\ra_\r=\frac{1}{\b}\left(\restriction{\frac{\d^2 E}{\d r(t)\d r(t')}}{r^*}\right)^{-1}:=
\begin{tikzpicture}[baseline=-1mm]  \draw (0:0) node[cross=4pt,line width=1pt,rotate=0] {}; \draw(0,-0.3) node {$t$}; 
\draw[line width=1pt] (0:0) -- (0:1);
 \draw (0:1) node[cross=4pt,line width=1pt,rotate=0] {}; \draw(1,-0.3) node {$t'$};
\end{tikzpicture}=O(\hbar)\\
-\b E[r]&=-\b E_{\rm Gaussian}-\b\frac{v^{(3)}}{3!}
\begin{tikzpicture}[baseline=-1mm]  \draw[fill=black] (0:0) circle (0.1) {}; 
\draw[line width=1pt] (0:0) -- (60:0.5);
\draw[line width=1pt] (0:0) -- (120:0.5);
\draw[line width=1pt] (0:0) -- (0,-0.5);
\end{tikzpicture}
-\b\frac{v^{(4)}}{4!}
\begin{tikzpicture}[baseline=-1mm]  \draw[fill=black] (0:0) circle (0.1) {}; 
\draw[line width=1pt] (0:0) -- (60:0.5);
\draw[line width=1pt] (0:0) -- (120:0.5);
\draw[line width=1pt] (0:0) -- (-60:0.5);
\draw[line width=1pt] (0:0) -- (-120:0.5);
\end{tikzpicture}
+\dots\\
E_{\rm Gaussian}[\r]&=E[r^*]+\frac12\iint_0^1\dd t\dd t'\,\r(t)\restriction{\frac{\d^2 E}{\d r(t)\d r(t')}}{r^*}\r(t')
\end{split}
\end{equation}
where we used~\eqref{eq:expE}. As done in the latter equations, in the following we shall write explicitly all the factors, including symmetry factors, in front of the diagrams, 
and we will omit the argument $r^*+h$ in the potential $v$ and its derivatives, 
\ie $v^{(n)}\leftrightarrow v^{(n)}(r^*+h)$. 
Expanding the observables we are averaging and the exponential of the perturbations around the Gaussian action, at the required order, then applying Wick's theorem, we finally get
\begin{equation}\label{eq:sviluppoSi}
 \begin{split}
  \la  v'(r(t)+h)v'(r(s)+h)\ra_v=&\frac{\int \mathrm{D}r\, v'(r(t)+h)v'(r(s)+h) e^{-\b E[r]}}{\int \mathrm{D}r\, e^{-\b E[r]}}\\
 \underset{\substack{\hbar\to0\\\b\hbar~\textrm{fixed}}}{=}&
 (v')^2+(v'')^2  
 \begin{tikzpicture}[baseline=-1mm]  \draw (0:0) node[cross=4pt,line width=1pt,rotate=0] {}; \draw(0,-0.3) node {$t$}; 
\draw[line width=1pt] (0:0) -- (0:1);
 \draw (0:1) node[cross=4pt,line width=1pt,rotate=0] {}; \draw(1,-0.3) node {$s$};
\end{tikzpicture}
+v'v^{(3)}
\begin{tikzpicture}[baseline=-1mm]
 \draw[line width=1pt] (0:0) circle (0.3) node[cross=4pt,line width=1pt, xshift=-0.3cm]{} ;
\end{tikzpicture}
-\b v'v''v^{(3)}
\begin{tikzpicture}[baseline=-1mm]
\draw (-0.7,0) node[cross=4pt,line width=1pt,rotate=0] {};
\draw[line width=1pt] (-0.7,0) -- (0:0);
 \draw[line width=1pt] (0.3,0) circle (0.3) node[circle,fill=black,inner sep=0pt,minimum size=6pt,xshift=-0.3cm]{} ;
\end{tikzpicture}
 +O(\hbar^2)\\
  \la v''(r(t)+h)\ra_v \underset{\substack{\hbar\to0\\\b\hbar~\textrm{fixed}}}{=}&
  v''
+O(\hbar)
 \end{split}
\end{equation}
The diagrams have the following expressions, from Eqs.~[\ref{eq:expE}],[\ref{eq:perturbth}]:
\begin{equation}\label{eq:diagramexpr}
 \begin{split}
   \begin{tikzpicture}[baseline=-1mm]  \draw (0:0) node[cross=4pt,line width=1pt,rotate=0] {}; \draw(0.5,0.3) node {$n$}; 
\draw[line width=1pt] (0:0) -- (0:1);
 \draw (0:1) node[cross=4pt,line width=1pt,rotate=0] {}; 
\end{tikzpicture}
=&\int_0^1\dd (t-s)\,e^{i\om_n(t-s)}\frac{1}{\b}\left(\restriction{\frac{\d^2 E}{\d r(t)\d r(s)}}{r^*}\right)^{-1}
=\frac1\b\frac{1}{\frac{1}{\b\wt G(\om_n)}+ v''}\\
\begin{tikzpicture}[baseline=-1mm]
 \draw[line width=1pt] (0:0) circle (0.3) node[cross=4pt,line width=1pt, xshift=-0.3cm]{} ;
\end{tikzpicture}
=&\frac{1}{\b}\sum_{p\in\ZZZ}\frac{1}{\frac{1}{\b\wt G(\om_p)}+ v''}\qquad , \qquad
\begin{tikzpicture}[baseline=-1mm]
\draw (-0.7,0) node[cross=4pt,line width=1pt,rotate=0] {};
\draw[line width=1pt] (-0.7,0) -- (0:0);
 \draw[line width=1pt] (0.3,0) circle (0.3) node[circle,fill=black,inner sep=0pt,minimum size=6pt,xshift=-0.3cm]{} ;
\end{tikzpicture}
=\frac{1}{\b^2}\frac{1}{\frac{1}{\b\wt G(0)}+ v''}\sum_{p\in\ZZZ}\frac{1}{\frac{1}{\b\wt G(\om_p)}+ v''}
 \end{split}
\end{equation}
The first equation is the $\om=\om_n$ Fourier mode of the propagator, while the two last are constants (thus only assigned to the $n=0$ mode).\\
A comment is here in order: notice that, for convenience, we chose to stick with a general value of $r^*$ in~\eqref{eq:rstardef} where $\b\wt G(0)$ must also be expanded in power of $\hbar$. This means that lowest-order expressions like $v'(r^*+h)^2$ in the middle line of~\eqref{eq:sviluppoSi} are $O(1)$ but provide also higher powers of $\hbar$ when expanded. These must be taken into account for higher orders.

\subsubsection{Asymptotic expansion of the variational equations}\label{sub:asymptexp}

We now expand systematically the variational equations in order to solve the self-consistency problem at the lowest order. 
Plugging the definition~\eqref{eq:expGhbar} into the second line of~\eqref{eq:Schehrlike} we have
\begin{equation}\label{eq:muplusSi0}
 \m+\wt\Si(0)\underset{\substack{\hbar\to0\\\b\hbar~\textrm{fixed}}}{=}-\frac{1}{\b\hbar^2\wt G_{(1)}(0)^2}
 +\frac{1}{\b\hbar}\left[\frac{\wt G_{(1)}(0)-q_{(1)}}{\wt G_{(1)}(0)^2}+2\frac{\wt G_{(2)}(0)}{\wt G_{(1)}(0)^3}\right]+O(\hbar)
\end{equation}
The variational equation on the static overlap $q$,~\eqref{eq:dxqRS}, necessitates the expansion of
\begin{equation}\label{eq:expavvprime}
\frac{\dd}{\dd h}\ln\la e^{-\b\int_0^1\dd t\,v(r(t)+h)}\ra_r=-\b\la v'(r(t)+h)\ra_v\underset{\substack{\hbar\to0\\\b\hbar~\textrm{fixed}}}{=}
-\b\left(v'+\frac{v^{(3)}}{2}
\begin{tikzpicture}[baseline=-1mm]
 \draw[line width=1pt] (0:0) circle (0.3) node[cross=4pt,line width=1pt, xshift=-0.3cm]{} ;
\end{tikzpicture}
-\frac{\b}{2}v''v^{(3)}
\begin{tikzpicture}[baseline=-1mm]
\draw (-0.7,0) node[cross=4pt,line width=1pt,rotate=0] {};
\draw[line width=1pt] (-0.7,0) -- (0:0);
 \draw[line width=1pt] (0.3,0) circle (0.3) node[circle,fill=black,inner sep=0pt,minimum size=6pt,xshift=-0.3cm]{} ;
\end{tikzpicture}
+O(\hbar^2)
\right)
\end{equation}
so that the expansion of~\eqref{eq:dxqRS} gives the following identities by comparing the first two orders on both sides:
\begin{equation}\label{eq:varqexpanded}
 \begin{split}
  \frac{1}{(\b\hbar\wt G_{(1)}(0))^2}=&\a\int\dd h\,\g(h+\s)\frac{h^2\th(-h)}{[1+\b\hbar\wt G_{(1)}(0)]^2}\\
  \frac{q_{(1)}}{(\b\hbar\wt G_{(1)}(0))^2}-\frac{2}{(\b\hbar)^2}\frac{\wt G_{(2)}(0)}{\wt G_{(1)}(0)^3}=&
  \a\int\dd h\,\g(h+\s)\left[-2h^2\th(-h)\frac{\b\hbar\wt G_{(2)}(0)}{[1+\b\hbar\wt G_{(1)}(0)]^3}+(v')^2q_{(1)}\frac{(h+\s)^2-1}{2}\right.\\
&\hskip150pt\left.+\frac{v'v^{(3)}}{\hbar}\begin{tikzpicture}[baseline=-1mm]
 \draw[line width=1pt] (0:0) circle (0.3) node[cross=4pt,line width=1pt, xshift=-0.3cm]{} ;
\end{tikzpicture}-v'v''v^{(3)}\frac{\b}{\hbar}\begin{tikzpicture}[baseline=-1mm]
\draw (-0.7,0) node[cross=4pt,line width=1pt,rotate=0] {};
\draw[line width=1pt] (-0.7,0) -- (0:0);
 \draw[line width=1pt] (0.3,0) circle (0.3) node[circle,fill=black,inner sep=0pt,minimum size=6pt,xshift=-0.3cm]{} ;
\end{tikzpicture}\right]
 \end{split}
\end{equation}
The first term in the right-hand side of each line comes from the expansion of $v'\left(r^*(h)+h\right)^2$. Indeed, as mentioned in~\secref{sub:spexpdiag}, one has to take into account the expansion of $r^*$.\\
The spherical constraint~\eqref{eq:lagrange} gives us the remaining value of $q_{(1)}$ in terms of $G_{(1)}$:
\begin{equation}\label{eq:sphconsq1}
 q_{(1)}+\sum_{n\in\ZZZ}\wt G_{(1)}(\om_n)=0
\end{equation}
Combining these equations, we can write the lowest-order values of $\m$ and $\Si$:
\begin{equation}\label{eq:muetSi}
 \begin{split}
  \wt\Si(\om_n)=&-\a\d_{n0}\int\dd h\,\g(h+\s)\left[\underbrace{\b(v')^2}+\b v'v^{(3)}
  \begin{tikzpicture}[baseline=-1mm]
 \draw[line width=1pt] (0:0) circle (0.3) node[cross=4pt,line width=1pt, xshift=-0.3cm]{} ;
\end{tikzpicture}
-\b^2v'v''v^{(3)}
\begin{tikzpicture}[baseline=-1mm]
\draw (-0.7,0) node[cross=4pt,line width=1pt,rotate=0] {};
\draw[line width=1pt] (-0.7,0) -- (0:0);
 \draw[line width=1pt] (0.3,0) circle (0.3) node[circle,fill=black,inner sep=0pt,minimum size=6pt,xshift=-0.3cm]{} ;
\end{tikzpicture}
+\b\hbar(v')^2q_{(1)}\frac{(h+\s)^2-1}{2}\right]\\
&-\a\int\dd h\,\g(h+\s)\left(\b(v'')^2
\begin{tikzpicture}[baseline=-1mm]  \draw (0:0) node[cross=4pt,line width=1pt,rotate=0] {}; \draw(0.5,0.3) node {$n$}; 
\draw[line width=1pt] (0:0) -- (0:1);
 \draw (0:1) node[cross=4pt,line width=1pt,rotate=0] {}; 
\end{tikzpicture}
-v''\right)+O(\hbar)\\
\m=&\m_{(0)}+O(\hbar)=
\a\int\dd h\,\g(h+\s)\left(\b(v'')^2
\begin{tikzpicture}[baseline=-1mm]  \draw (0:0) node[cross=4pt,line width=1pt,rotate=0] {}; \draw(0.5,0.3) node {$0$}; 
\draw[line width=1pt] (0:0) -- (0:1);
 \draw (0:1) node[cross=4pt,line width=1pt,rotate=0] {}; 
\end{tikzpicture}
-v''\right)+\frac{1}{\b\hbar\wt G_{(1)}(0)}+O(\hbar)
 \end{split}
\end{equation}
$\wt\Si(\om_n)$ actually has an $O(1/\hbar)$ term (underbraced, which also contributes at higher orders when $r^*$ gets expanded).  This contribution is compensated in~\eqref{eq:muplusSi0} by the first term in the right-hand side of the first line owing to the first line of~\eqref{eq:varqexpanded}. 
As a result, $\m\sim\m_{(0)}$ is $O(1)$ at the lowest order. 
The lowest-order self-consistent Dyson equation reads: 
\begin{equation}\label{eq:Dysexp}
 \frac{1}{\b\hbar\wt G_{(1)}(\om_n)}=\MM\left(\frac{\om_n}{\b\hbar}\right)^2+\m_{(0)}+\wt\Si_{(0)}(\om_n)
 \hskip15pt\textrm{for~}n\neq0
\end{equation}
while the $n=0$ mode is given by~\eqref{eq:varqexpanded}. The lowest-order indices $(0)$ in the latter equation are self-consistently given by~\eqref{eq:muetSi}, hence
\begin{equation}
 \m_{(0)}+\wt\Si_{(0)}(\om_n)\underset{n\neq0}{=}
 \frac{1}{\b\hbar\wt G_{(1)}(0)}+
 \a\int\dd h\,\g(h+\s)(v'')^2\left[\frac{1}{\frac{1}{\b\hbar \wt G_{(1)}(0)}+v''}-\frac{1}{\MM(\om_n/\b\hbar)^2+\m_{(0)}+\wt\Si_{(0)}(\om_n)+v''}\right] 
\end{equation}
which is indeed $O(1)$.

We shall apply, from~\secref{sub:expenerg} on, the above results to the specific heat. But first we shall show that they allow to recover and study classical $T=0$ observables.

\subsubsection{The classical limit}\label{sub:classUNSAT}

Let us now compare the previous results with the classical $T=0$ results in Ref.~\cite{FPSUZ17} as a check, since at lowest order in the previous expansion, observables must agree 
with the $T=0$ classical ones computed in this way, \ie in the UNSAT phase (see~\secref{sec:classical}). 

The saddle-point value in~\eqref{eq:rstardef}, useful for the computation of averages, becomes at lowest order from the definition~\eqref{eq:defchi}
\begin{equation}\label{eq:SPzh}
 z^*(h)=r^*(h)+h\sim h\th(h)+\frac{h}{1+\chi}\th(-h)
\end{equation}
which corresponds to the classical low-temperature saddle point~\cite[Eq.(C3)]{FPSUZ17}. The lowest-order equation for $q$ given 
by the first equation of~\eqref{eq:varqexpanded} reads now
\begin{equation}\label{eq:varchi}
 \frac{1}{\chi^2}=\a\int\dd h\,\g(h+\s)v'(z^*(h))^2=\frac{\a}{(1+\chi)^2}\int_{-\io}^0\dd h\,\g(h+\s)h^2
 \hskip15pt \Leftrightarrow\hskip15pt \left(1+\frac1\chi\right)^2=\frac{\a}{\a_J^{\rm cl}(\s)}
\end{equation}
which is exactly the variational equation for $\chi$ in the classical model~\cite[Eq.(47)]{FPSUZ17}, defined as ${q^{\rm cl}=1-\chi T+O(T^2)}$.

\subsection{Specific heat in the classical regime}

We are now able to derive the low-temperature behaviour of the specific heat in the classical regime.
We shall see that, as in solids~\cite{De12,kittel,AM}, it fails to describe correctly the system at low temperature, demanding a quantum treatment.
Moreover it provides additional checks. For convenience, we will separate again the two phases at $T=0$.

\subsubsection{SAT phase}\label{sub:classCVSAT}

Applying the arguments from subsection~\secref{eq:SATschehr} to~\eqref{eq:energyRS}, we get the classical energy
\begin{equation}\label{eq:UtotSATcl}
\begin{split}
 \frac{\UU_{\rm RS}(\hbar=0)}{N}=&\frac{T}{2}+\a\int\dd h\,\g_{q_{(0)}}(h+\s)\frac{\int\dd u\,\g_{1-q_{(0)}}(h-u)e^{-\b v(u)}v(u)}{\Th(h/\sqrt{2(1-q_{(0)})})}+\dots
 =\frac T2+O(T^{3/2})\\
 \Rightarrow\hskip15pt C_V(\hbar=0)\underset{T\to0}{=}&~\frac{N}{2}+O(\sqrt T)
 \end{split}
\end{equation}
The $n\neq0$ terms of the Matsubara sum vanish in the classical limit, so only the $n=0$ term survives and reads $T/2$. The low-temperature limit of the $\a$-dependent term is easy to expand using Gaussian integrals. We conclude that the classical specific heat goes to a constant $N/2$ in the whole $T=0$ RS SAT phase (non convex and convex), actually given by the free-particle result: each momentum degree of freedom gives a contribution $1/2$ by energy equipartition. 

\subsubsection{UNSAT phase}

\begin{figure}[h!]
\begin{center}
 \begin{tabular}{ccc}
  \includegraphics[width=7cm]{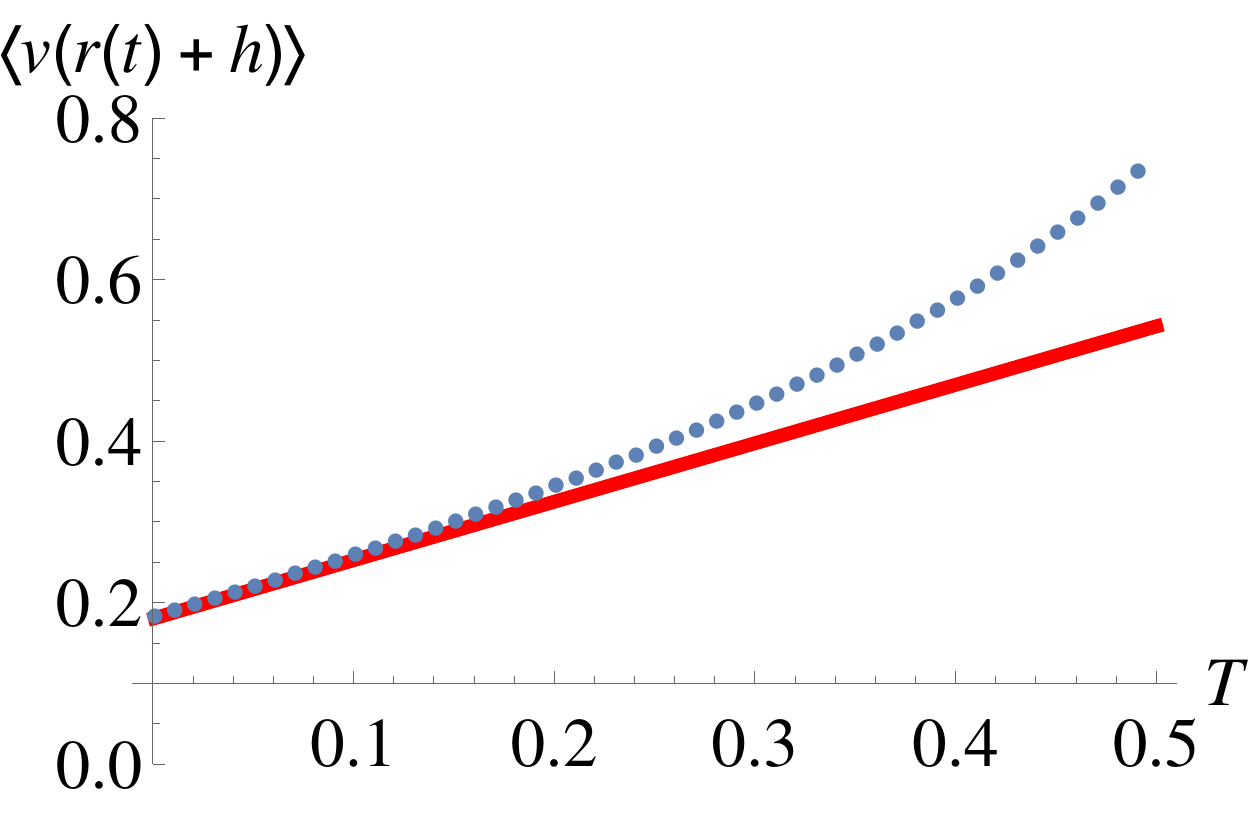}
  & \hspace{1cm} &\includegraphics[width=7cm]{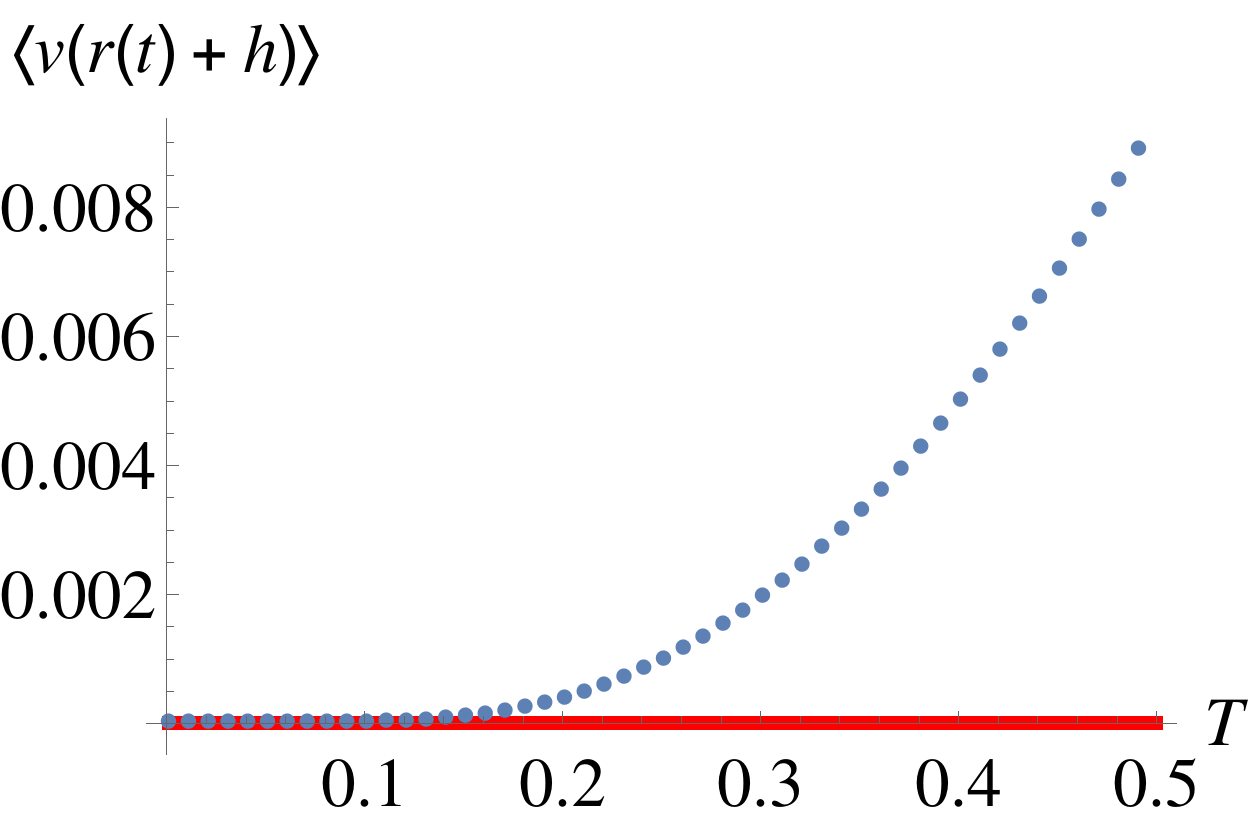}
 \end{tabular}
\caption{Comparison between the numerical computation of the integral (blue dots) and the analytic approximation (red solid line) in~\eqref{eq:approx1D}.\\ 
(Left) $h=-3$, $\chi=4$, $\c'=-4.5$. (Right)  $h=1.5$, $\chi=2$, $\c'=2$.}
\label{fig:approx1D}
\end{center}
\end{figure}

In this phase as well the sum from the kinetic term gives the free-particle result $T/2$; 
however the potential energy must now play a role. We turn to 
the classical analysis of the potential-energy term of~\eqref{eq:energyRS} containing the average 
$\la v(r(t)+h) \ra_v$. In the classical limit, at small temperature, since we know from~\secref{sub:classUNSAT} that $\la \wt r(-\om_n)\wt r(\om_n) \ra_r\sim\chi T\d_{n0}$, we get that only the zero mode contributes to the average. Yet in order to get the correct result, one actually must go beyond this linear order in temperature. In the classical regime it corresponds to expanding the overlap $q=1-\c T-\c'T^2+O(T^3)$. The effective average becomes classical and
\begin{equation}\label{eq:approx1D}
 \la v(r(t)+h) \ra_v^{\hbar=0}\sim\frac{\int \dd r_0\, e^{- r_0^2/[2(\chi T+\c'T^2)]-v( r_0+h)/T}v( r_0+h)}{\int \dd  r_0\, e^{- r_0^2/[2(\chi T+\c'T^2)]-v(r_0+h)/T}}
 \underset{T\to0}{=}v(r^*(h)+h)+\frac{T}{2}\th(-h)\left[\frac{\c}{1+\c}-2\frac{h^2\c'}{(1+\c)^3}\right]+O(T^2)
\end{equation}
where the $T=0$ classical saddle point is given by~\eqref{eq:SPzh}. Here we have performed the same kind of asymptotic saddle-point expansion as in~\secref{sub:spexpdiag}, applied to this simpler one-dimensional integral.
This expansion has been successfully checked numerically in Fig.~\ref{fig:approx1D}.
The saddle-point equation for $q$ (obtained by \eg the limit $\hbar\to0$ of~\eqref{eq:dxqRS} or through~\cite[Eq.(44)]{FPSUZ17}) provides $\c$ (\eqref{eq:varchi}) and $\c'$ by expanding to the next order in temperature:
\begin{equation}\label{eq:chiprime}
 \c'=\frac{\c^4}{2(1+\c)}\a\Th\left(\frac{\s}{\sqrt2}\right)-\frac{\c^2}{2}(1+\c)
\end{equation}
Combining Eqs.~[\ref{eq:varchi}],[\ref{eq:approx1D}],[\ref{eq:chiprime}] we get the low-temperature expansion of the classical energy
\begin{equation}\label{eq:uclassUNSAT}
  \frac{\UU_{\rm RS}(\hbar=0)}{N}=\frac T2+\a\int\dd h\, \g_q(h+\s)\la v(r(t)+h) \ra_v^{\hbar=0}
=\frac T2 +\frac{1}{2\chi^2}+\frac{T}{2}+O(T^2)
=e_{\rm RS}^{\rm cl}+T+O(T^2)
\end{equation}
where we defined the energy
\begin{equation}\label{eq:GSUNSAT}
 e_{\rm RS}^{\rm cl}=\frac{1}{2\chi^2}=\frac12\left(\sqrt{\frac{\a}{\a_J^{\rm cl}(\s)}}-1\right)^2
\end{equation}
which is the ground-state energy of the classical model in the RS UNSAT phase~\cite[Eq.(48)]{FPSUZ17}.
We can note in~Eqs.~[\ref{eq:UtotSATcl}],[\ref{eq:GSUNSAT}] that, by definition of these phases (see~\secref{sec:hbar}), the ground-state energy in the UNSAT phase is strictly positive, whereas in the SAT phase it vanishes.

Although the above analysis is essentially classical, we shall see that this result can be recovered within the SGLD expansion of the quantum thermodynamic energy in an appropriate classical limit in~\secref{sub:expenerg}.

We conclude that the $T=0$ value of the classical specific heat is 
 $C_V(\hbar=0,T=0)=N$, which is Dulong and Petit's law~\cite{DP1819,kittel,AM}.
Indeed, in the UNSAT phase the potential-energy contribution is finite and from replica symmetry the potential-energy landscape is expected at low temperature to be well approximated by a harmonic potential around a single minimum. Therefore the momentum and position degrees of freedom contribute each $N/2$ to the specific heat by equipartition. 
Conversely, in the SAT phase at low $T$ the particle has some finite volume where it can move freely in this volume and the kinetic term dominates. 
We note that in the classical zero-temperature convex phase diagram ($\s\geqslant0$), the specific heat is discontinuous at the jamming transition, while the ground-state energy has no jump. From the thermodynamic point of view, this SAT-UNSAT transition is second order in Ehrenfest's sense.

We know, at least in the case $\a=0$ from the exact free-particle spectrum~\secref{sub:conclufree}, that the above classical results are incorrect and quantum corrections must be accounted for. This is the aim of the next sections.

\subsection{Landscape marginal stability in the UNSAT phase}\label{sec:margUNSAT}

As the SGLD mechanism relies on landscape marginal stability, we shall look for the location of the dAT line in the UNSAT phase.
From~\eqref{eq:margstab}, this line is given by
\begin{equation}\label{eq:marglow}
 \frac{1}{(\b\wt G(0))^2}=\a \int\dd h\, \g_q(h+\s)\left[\frac{\dd}{\dd h}\la v'(r(t)+h)\ra_v  \right]^2
\end{equation}
The lowest order of the average term in the UNSAT phase is, from~\eqref{eq:expavvprime},
\begin{equation}
 \frac{\dd}{\dd h}\la v'(r(t)+h)\ra_v=\frac{\dd}{\dd h} v'(r^*+h)+O(\hbar) \sim\frac{\th(-h)}{1+\b\wt G(0)}\sim\frac{\th(-h)}{1+\chi}
\end{equation}
Therefore~\eqref{eq:marglow} reads at lowest order
\begin{equation}
 \frac{1}{\chi^2}=\frac{\a}{(1+\chi)^2}\int_{-\io}^0\dd h\,\g(h+\s)=\frac{\a\Th(\s/\sqrt2)}{(1+\chi)^2}
\end{equation}
which is, as expected, the classical equation for the dAT line in the convex phase~\cite[Eq.(53)]{FPSUZ17}. We can retrieve the classical location of the dAT line by rewriting the last equation with the help
of~\eqref{eq:varchi} and using $\a\geqslant\a_J^{\rm cl}(\s)>0$,
\begin{equation}\label{eq:RSmargcond}
\frac{\a}{\a_J^{\rm cl}(\s)}=\a\int_{-\io}^0\dd h\,\g(h+\s)\hskip15pt\Leftrightarrow\hskip15pt \int_{-\io}^0\dd h\,\g(h+\s)h^2=\int_{-\io}^0\dd h\,\g(h+\s) 
\hskip15pt\Rightarrow\hskip15pt \s=0
\end{equation}
This unique solution is obtained graphically. This line ${(\s=0,\a\geqslant\a_J^{\rm cl}(0)=2)}$ is the classical dAT line in the UNSAT phase~\cite{FP16,FPSUZ17}. 


In general one may expect that the quantum phase diagram, in particular the marginal phase boundary, changes with respect to the classical one. 
This is the case in the spherical quantum  $p$-spin where the Hamiltonian is the same as the quantum perceptron with a different potential energy (disordered $p$-spin couplings)~\cite{CGSS01};
analogously, the spin-glass transitions in mean-field spherical or Ising quantum spin glasses in a transverse field depend upon the field strength~\cite{NR98,AM12}, 
which plays a similar role to the inverse of the particle mass. 
The LMS property of the model at a given point in parameter space must be checked independently of the SGLD expansion, 
which is able to derive the behaviour of the specific heat \textit{assuming} LMS holds or does not. 
As a matter of fact, we checked that the next order of the expansion of~\eqref{eq:marglow} does not impose further constraints on the location of the LMS line. It is likely that, if a perturbative strategy can be employed to obtain the location of the LMS line in the fully quantum regime, one needs to perturb as well the parameters of the model $\{\s,\a,\MM\}$.

\subsection{The thermodynamic energy in the UNSAT phase}\label{sub:expenerg}

Following~\cite{SGLD04,SGLD05,S05}, we write the expansion for the thermodynamic energy given by~\eqref{eq:energyRS}
\begin{equation}\label{eq:u0u1}
 \frac{\UU_{\rm RS}}{N}=
 \frac{1}{2\b}\sum_{n\in\ZZZ}\frac{\m+\wt\Si(\om_n)}{\MM\left(\frac{\om_n}{\b\hbar}\right)^2+\m+\wt\Si(\om_n)}
 +\a\int\dd h\, \g_q(h+\s)\la v(r(t)+h) \ra_v\\
 =u_{(0)}(\b\hbar)+\hbar\, u_{(1)}(\b\hbar)+O(\hbar^2)
\end{equation}
The idea is to get the temperature behaviour order by order in $\hbar$, by analyzing the scaling of these quantities with the parameter $\b\hbar$, the only parameter containing the temperature.\\
The convolution (``potential-energy'' term) may be expanded as in~\secref{sub:UNSATexp}. One has\footnote{Remember that the potential is evaluated at $r^*(h)+h$.}
\begin{equation}\label{eq:expandv}
 \la v(r(t)+h) \ra_v=v
 +\frac{v''}{2}
\begin{tikzpicture}[baseline=-1mm]
 \draw[line width=1pt] (0:0) circle (0.3) node[cross=4pt,line width=1pt, xshift=-0.3cm]{} ;
\end{tikzpicture}
 -\frac{\b}{2}v'v^{(3)}
 \begin{tikzpicture}[baseline=-1mm]
\draw (-0.7,0) node[cross=4pt,line width=1pt,rotate=0] {};
\draw[line width=1pt] (-0.7,0) -- (0:0);
 \draw[line width=1pt] (0.3,0) circle (0.3) node[circle,fill=black,inner sep=0pt,minimum size=6pt,xshift=-0.3cm]{} ;
\end{tikzpicture}
+O(\hbar^2)
\end{equation}
The lowest-order term $u_{(0)}$ emanates only from the potential-energy term in~\eqref{eq:u0u1}, for which only the first term in~\eqref{eq:expandv} contributes. 
We get
\begin{equation}\label{eq:expu0}
  u_{(0)}=\a\int\dd h\, \g(h+\s)\frac{h^2}{2(1+\c)^2}\th(-h)=\frac{1}{2\c^2}=e_{\rm RS}^{\rm cl}\\
\end{equation}
where we used~\eqref{eq:varchi}. 
The lowest order $u_{(0)}=e_{\rm RS}^{\rm cl}$ coincides with the classical ground state energy in the UNSAT phase~\eqref{eq:GSUNSAT}. This energy does not have any temperature dependence, and thus the first 
low-temperature correction comes from $u_{(1)}(\b\hbar)$. 

For the next order $u_{(1)}$, we note that ${v'v^{(3)}=0}$ which allows to simplify the expansion. 
Furthermore, let us define for convenience (we put an asterisk to avoid any confusion with~\eqref{eq:muetSi}):
\begin{equation}\label{eq:SistarRS}
 \wt\Si_{(0)}^*(\om_n)=-\a\int\dd h\,\g(h+\s)\left(\b(v'')^2
\begin{tikzpicture}[baseline=-1mm]  \draw (0:0) node[cross=4pt,line width=1pt,rotate=0] {}; \draw(0.5,0.3) node {$n$}; 
\draw[line width=1pt] (0:0) -- (0:1);
 \draw (0:1) node[cross=4pt,line width=1pt,rotate=0] {}; 
\end{tikzpicture}
-v''\right)
\end{equation}
$\wt\Si_{(0)}(\om_n)$ and $\wt\Si_{(0)}^*(\om_n)$ coincide for $n\neq0$, and $\m_{(0)}+\wt\Si_{(0)}^*(0)=1/\chi$, see~\eqref{eq:muetSi}. 
In this way we can write a generic relation for $\wt{G}_{(1)}$:
\begin{equation}\label{eq:G1alln}
\frac{1}{\b\hbar\wt{G}_{(1)}(\om_n)}=\MM\left(\frac{\om_n}{\b\hbar}\right)^2+\m_{(0)}+\wt\Si_{(0)}^*(\om_n)
=\MM\left(\frac{\om_n}{\b\hbar}\right)^2+\frac{1}{\chi}+\wt I_{(0)}(\om_n)
\end{equation}
which is valid for any $n\in\ZZZ$, as seen from~\eqref{eq:muetSi} for $n=0$ and~\eqref{eq:Dysexp} for $n\neq0$, and the ``renormalized'' self-energy
\begin{equation}\label{eq:defI0}
 \wt I_{(0)}(\om_n)=\wt\Si_{(0)}^*(\om_n)-\wt\Si_{(0)}^*(0)\qquad \textrm{with} \qquad\wt I_{(0)}(0)=0
\end{equation}
where we separated a \textit{mass term}~\cite{peskin} $1/\chi$ from $\wt I_{(0)}$ that gives the small-frequency dependence. 
Then the first-order correction to the energy reads:
\begin{equation}\label{eq:expu1}
 \begin{split}
  u_{(1)}=&
  \frac{1}{2\b\hbar}\sum_{n\in\ZZZ}\frac{\frac{1}{\chi}+\wt I_{(0)}(\om_n)}{\MM\left(\frac{\om_n}{\b\hbar}\right)^2+\frac{1}{\chi}+\wt I_{(0)}(\om_n)}
   -\frac{\a}{\a_J^{\rm cl}(\s)}\frac{\b\hbar\wt G_{(2)}(0)}{(1+\c)^3}\\
  &+\frac{\a}{2\b\hbar}\Th\left(\frac{\s}{\sqrt 2}\right)\left(\sum_{n\in\ZZZ}\frac{1}{1+\frac{1}{\b\hbar\wt G_{(1)}(\om_n)}}-\frac{1}{(1+\chi)^2}\sum_{n\in\ZZZ}\b\hbar\wt G_{(1)}(\om_n)\right)\\
  =&\frac{1}{2\b\hbar}\sum_{n\in\ZZZ}\frac{\frac{1}{\chi}+\wt I_{(0)}(\om_n)}{\MM\left(\frac{\om_n}{\b\hbar}\right)^2+\frac{1}{\chi}+\wt I_{(0)}(\om_n)}
  +\a\Th\left(\frac{\s}{\sqrt 2}\right)\frac{1}{2\b\hbar}\sum_{n\in\ZZZ}\frac{1}{1+\frac{1}{\chi}+\MM\left(\frac{\om_n}{\b\hbar}\right)^2+\wt I_{(0)}(\om_n)}\\
  &-\left\{\frac{\a\Th(\s/\sqrt 2)}{(1+\chi)^2}+\frac1\c\left[\a_J^{\rm cl}(\s)\Th\left(\frac{\s}{\sqrt 2}\right)-1\right]\right\}\frac{1}{2\b\hbar}\sum_{n\in\ZZZ}\frac{1}{\MM\left(\frac{\om_n}{\b\hbar}\right)^2+\frac{1}{\chi}+\wt I_{(0)}(\om_n)}
 \end{split}
\end{equation}
We performed the integrations, used the variational equation for $\c$~\eqref{eq:varchi} and replaced $ q_{(1)}$ through the spherical constraint~\eqref{eq:sphconsq1}. In the last line we inserted~\eqref{eq:G1alln} and expressed $\wt G_{(2)}(0)$ via the second line of~\eqref{eq:varqexpanded}.
Incidentally, we note that the latter equation is, using $v'v^{(3)}=0$ and~\eqref{eq:varchi},
\begin{equation}\label{eq:G2}
 (\b\hbar)^2\wt G_{(2)}(0)=\left[\frac{\c^3}{2(1+\c)}\a\Th\left(\frac{\s}{\sqrt2}\right)-\frac{\c}{2}(1+\c)\right]\sum_{n\in\ZZZ}\b\hbar\wt G_{(1)}(\om_n)
\end{equation}
As explained in the beginning of~\secref{sub:UNSATexp}, if we now consider the usual classical limit $\hbar\to0$ \textit{without scaling} $T$, then $\wt G(0)=G(0)=1-q^{\rm cl}=\c T+\c'T^2+O(T^3)$, and therefore $\b\hbar\wt G_{(1)}(0)\to\c$ and $(\b\hbar)^2\wt G_{(2)}(0)\to\c'$. Taking first the limit $\hbar\to0$ amounts to retain only the $n=0$ term in the sum of~\eqref{eq:G2} (owing to the Dyson equation~[\ref{eq:G1alln}]), which gives back the classical variational equation for $\c'$ obtained in~\eqref{eq:chiprime}.

Similarly we can retrieve the classical specific heat for $T\to0$ considering the first correction to the classical ground-state energy $\hbar u_{(1)}$ in~\eqref{eq:expu1}.
The $n=0$ term of the kinetic energy sum reads directly $T/2$, while in the potential energy the $n=0$ terms give a contribution, with the expression of $\c$ in~\eqref{eq:varchi},
\begin{equation}
 \frac{1}{2\b}\left\{\a\Th\left(\frac{\s}{\sqrt 2}\right)\frac{1}{1+1/\c}-\frac{\a\Th(\s/\sqrt 2)}{(1+\chi)^2}+\frac1\c\left[\a_J^{\rm cl}(\s)\Th\left(\frac{\s}{\sqrt 2}\right)-1\right]\right\}
 =\frac T2
\end{equation}
\ie we get exactly the first low-$T$ correction to the ground-state energy in the 
classical limit as in~\eqref{eq:uclassUNSAT}, which implies Dulong and Petit's law $C_V=N$.

To go further, we must analyze the behaviour of the Matsubara sums in~\eqref{eq:expu1}, \ie we now need to compute the renormalized self-energy.

\subsection{The renormalized self-energy}

\begin{figure}[h!]
\begin{center}
  \includegraphics[width=8.5cm]{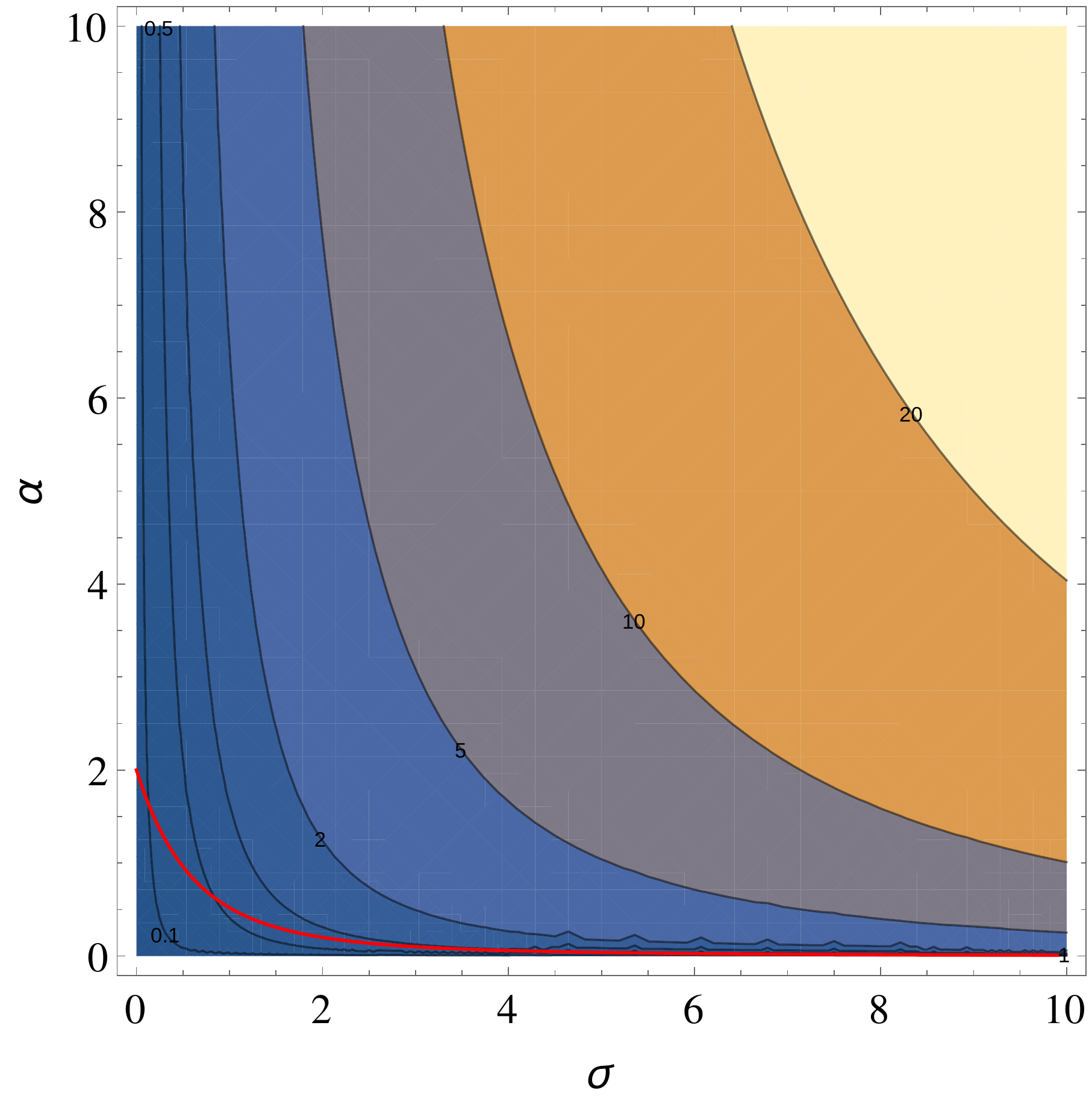}
 \caption{
 Contour plot of $K(\s,\a)$, always strictly positive but $K=0$ for $\s=0$, $\a\geqslant2$ (LMS line). 
 The red solid line represents the classical SAT-UNSAT transition line.}
 \label{fig:mu0}
 \end{center}
\end{figure}

The expression of the thermodynamic energy depends crucially on the renormalized self-energy defined in Eqs.~[\ref{eq:SistarRS}],[\ref{eq:defI0}]. Let us calculate it explicitly:
\begin{equation}\label{eq:computeSi0}
\begin{split}
\wt I_{(0)}(\om_n)&=-\a\int\dd h\,\g(h+\s)\left(\b(v'')^2
\begin{tikzpicture}[baseline=-1mm]  \draw (0:0) node[cross=4pt,line width=1pt,rotate=0] {}; \draw(0.5,0.3) node {$n$}; 
\draw[line width=1pt] (0:0) -- (0:1);
 \draw (0:1) node[cross=4pt,line width=1pt,rotate=0] {}; 
\end{tikzpicture}
-v''\right)
+\a\int\dd h\,\g(h+\s)\left(\b(v'')^2
\begin{tikzpicture}[baseline=-1mm]  \draw (0:0) node[cross=4pt,line width=1pt,rotate=0] {}; \draw(0.5,0.3) node {$0$}; 
\draw[line width=1pt] (0:0) -- (0:1);
 \draw (0:1) node[cross=4pt,line width=1pt,rotate=0] {}; 
\end{tikzpicture}
-v''\right)\\
&=\a\,\Th\left(\frac{\s}{\sqrt 2}\right)\left(1-\frac{1}{1+\frac{1}{\b\hbar\wt G_{(1)}(\om_n)}}\right)-\frac{\a}{1+\chi}\Th\left(\frac{\s}{\sqrt 2}\right)\\
&\underset{\eqref{eq:G1alln}}{=}\a\,\Th\left(\frac{\s}{\sqrt 2}\right)\left(1-\frac{1}{1+\MM\left(\frac{\om_n}{\b\hbar}\right)^2+\frac{1}{\chi}+\wt I_{(0)}(\om_n)}\right)-\frac{\a}{1+\chi}\Th\left(\frac{\s}{\sqrt 2}\right)\\
\Leftrightarrow\hskip15pt 
0&= \wt I_{(0)}(\om_n)^2+\wt I_{(0)}(\om_n)\left[K(\s,\a)+\MM\left(\frac{\om_n}{\b\hbar}\right)^2\right]
-C(\s,\a)\MM\left(\frac{\om_n}{\b\hbar}\right)^2\\
\textrm{where}\qquad K(\s,\a)&=\sqrt{\frac{\a}{\a_J^{\rm cl}(\s)}}\left(1-\a_J^{\rm cl}(\s)\Th\left(\frac{\s}{\sqrt 2}\right)\right)\qquad\textrm{and}\qquad
C(\s,\a)=\sqrt{\a_J^{\rm cl}(\s)\a}\,\Th\left(\frac{\s}{\sqrt 2}\right)
\end{split}
\end{equation}
where we have expressed $\chi(\s,\a)$ through~\eqref{eq:varchi}. Therefore the self-energy at the lowest order is solution of a quadratic equation, and may be expressed in terms of radicals containing the Matsubara frequencies:
\begin{equation}\label{eq:I0solution}
 \wt I_{(0)}(\om_n)=\frac12\left\{-K(\s,\a)-\MM\left(\frac{\om_n}{\b\hbar}\right)^2+\sqrt{\left[K(\s,\a)+\MM\left(\frac{\om_n}{\b\hbar}\right)^2\right]^2+4C(\s,\a)\MM\left(\frac{\om_n}{\b\hbar}\right)^2}\right\}
\end{equation}
We have chosen the only solution consistent with the requirement $\wt G(\om_n)>0$ for all Matsubara frequencies. \\
In the UNSAT phase, $C(\s,\a)>0$ and $K(\s,\a)>0$, except for $\s=0$ (LMS line) where $K(0,\a)=0$. The latter is plotted in Fig.~\ref{fig:mu0}.

\subsection{The low-temperature specific heat from the thermodynamic energy}\label{sec:lowTCVRS}

In this section, for convenience we reinstate the usual definition of the Matsubara frequencies (\ie the frequencies without the hat in~\secref{sec:var}), defined by~\eqref{eq:Matsubaradef}, \ie $\om_n=2\p n/(\b\hbar)$.

Knowing the self-energy, we are now ready to compute the sums appearing in the energy~\eqref{eq:expu1}.
We note that the first sum in $u_{(1)}$ is coming from the regularized kinetic term, and is indeed convergent since $\forall\om$, 
 ${0\leqslant\wt I_{(0)}(\om)\leqslant C(\s,\a)=\underset{\om\to\pm\io}{\lim}\wt I_{(0)}(\om)}$.
Such sums depend strongly on the analytic properties of the self-energy~\cite{Mahan,Bruus}, usually studied through the function $\phi(z)$ where $\phi(i\om_n)=\wt I_{(0)}(\om_n)$, \ie
\begin{equation}\label{eq:analyticphi}
 \phi(z)=\frac{-K(\s,\a)+\MM z^2+\sqrt{\left[K(\s,\a)-\MM z^2\right]^2-4C(\s,\a)\MM z^2}}{2}
\end{equation}
This function is displayed in Fig.~\ref{fig:branch}, and has interesting branch cuts due to the square root of a quartic polynomial. In the low-temperature limit the analytic properties around $z=0$ of 
the self-energy are crucial. It is analytic in 0 out of the dAT line since $K>0$; however, on this line $K$ vanishes and the gap of the two branches on the real axis closes, inducing 
a singularity around zero. This translates into the small-frequency behaviour:
\begin{itemize}
 \item \underline{Out of the LMS line:} $K>0$ from Fig.~\ref{fig:mu0}(b), hence
 \begin{equation}\label{eq:Ksup0}
  \wt I_{(0)}(\om)\underset{\om\to0}{\sim}\frac{C(\s,\a)}{K(\s,\a)}\MM\om^2
 \end{equation}
\item \underline{On the LMS line $\s=0$:} $K=0$ thus the low-frequency behaviour changes as
\begin{equation}\label{eq:gregorylikeRS}
 \wt I_{(0)}(\om)\underset{\om\to0}{\sim}\sqrt{C(0,\a)\MM}\left|\om\right|
\end{equation}
\end{itemize}

The details of the computation of the Matsubara sums are left to App.~\ref{app:Matsubara}.

\subsubsection{Out of the landscape marginal stability line: gapped phase}

The computation of the sums is performed in Apps.~\ref{sub:appmargout} and~\ref{sub:kinmargout}. 
The conclusion is that at this order $u_{(1)}\propto e^{-\b\hbar\om_-} $, where the gap is $\hbar \om_-$. 
This quantity is defined in~\secref{sub:vDOS} and corresponds to the gap obtained in Debye's approximation in~\secref{sec:Debye}. 
The gap's scaling in $\hbar$ changes when adding more orders in the perturbative expansion (see~\secref{sec:finitehbar}).

\subsubsection{At the landscape marginal stability line: gapless phase}

On the dAT line $(\s=0,\a\geqslant2)$, the non-analyticity of the self-energy in zero gives different asymptotic results for the energy:
\begin{equation}\label{eq:gapless}
 \begin{split}
  u_{(1)}\sim&\,\textrm{constant}+u_{(1)}^{\rm kin}+u_{(1)}^{\rm pot}\\
  u_{(1)}^{\rm kin}\sim&\,\MM\int_0^\io\frac{\dd \om}{\p}\frac{\sqrt{C\MM}\om^3\sqrt{1-\MM \om^2/4C}}{(1/\chi-\MM\om^2/2)^2+C\MM\om^2(1-\MM \om^2/4C)}f_{\rm B}(\om)\\
  u_{(1)}^{\rm pot}\sim&\,\a \int_{0}^{\io}\frac{\dd \om}{2\p}\, f_{\rm B}(\om)\left[
 \frac{\sqrt{C\MM}\om\sqrt{1-\MM \om^2/4C}}{(1+1/\chi-\MM\om^2/2)^2+C\MM\om^2(1-\MM \om^2/4C)}\right.\\
  &\hskip100pt\left. -\frac{1}{(1+\chi)^2}\frac{\sqrt{C\MM}\om\sqrt{1-\MM \om^2/4C}}{(1/\chi-\MM\om^2/2)^2+C\MM\om^2(1-\MM \om^2/4C)}\right]
 \end{split}
\end{equation}
where $f_{\rm B}(\om)=\left(e^{\b\hbar\om}-1\right)^{-1}$ is the Bose-Einstein factor. It is the only 
temperature-dependent quantity at this stage.\\
In the low-temperature limit, these integrals are dominated by the small-$\om$ behaviour (as can be seen from a rescaling $\b\hbar\om=\om'$), thus for $1/\chi>0$ the two terms coming from the potential energy, which separately would induce linear in $T$ contributions to the specific heat, cancel each other at leading order. 
This cancellation of terms linear in $T$ was first discovered in the three quantum spin-glass models analyzed in~Ref.~\cite{SGLD04,SGLD05,S05}. 
Yet at the next order (expanding in $\om\to0$) they do give a $T^3$ contribution:
\begin{equation}\label{eq:CVpot}
\begin{split}
  u_{(1)}^{\rm pot}\sim&\frac{\a}{2\p}\left(\frac{T}{\hbar}\right)^4\MM\sqrt{C\MM}\frac{\chi^4[C\chi^2+(2C-1)\chi-1]}{(1+\chi)^4}\underbrace{\int_0^\io\dd x\,\frac{x^3}{e^x-1}}_{=\p^4/15}\\
  \Rightarrow\qquad \frac{C_V^{\rm pot}}{N}\sim&\frac{2\p^3}{15}\a\MM\sqrt{C\MM}\frac{\chi^4[C\chi^2+(2C-1)\chi-1]}{(1+\chi)^4}\left(\frac{T}{\hbar}\right)^3
\end{split}
\end{equation}
The kinetic term contribution to the specific heat is:
\begin{equation}\label{eq:CVint}
\begin{split}
 \frac{C_V^{\rm kin}}{N}=&(\b\hbar)^2\MM\int_0^\io\frac{\dd \om}{\p}\frac{\sqrt{C\MM}\om^3\sqrt{1-\MM \om^2/4C}}{(1/\chi-\MM\om^2/2)^2+C\MM\om^2(1-\MM \om^2/4C)}\frac{\om}{4\sinh^2(\b\hbar\om/2)}\\
 \sim&\frac{\MM}{\p}\sqrt{C\MM}\chi^2(\b\hbar)^{-3}\underbrace{\int_0^\io\dd x\,\frac{x^4}{4\sinh(x/2)^2}}_{=4\p^4/15}
 =\frac{4\p^3}{15}\chi^2\MM\sqrt{C\MM}\left(\frac{T}{\hbar}\right)^3
\end{split}
\end{equation}
The scalings in~\eqref{eq:CVpot} and~\eqref{eq:CVint} are valid on the dAT line except at the jamming point $1/\chi=0$, and provide the low-temperature scaling of $C_V=C_V^{\rm kin}+C_V^{\rm pot}$ 
at this perturbative order of the expansion. As in~Ref.~\cite{SGLD04,SGLD05,S05} we recover the scaling  ${C_V\propto T^3}$ in the LMS phase. \\

Let us end by the analysis of the jamming point $(\s=0,\a=2)$. Here $\chi\to\io$ (\textit{jamming criticality}), the behaviour of the integrands is different, hence with similar considerations and $C=1$ at the jamming point we get 
\begin{equation}\label{eq:CVjammRS}
\begin{split}
u_{(1)}^{\rm kin}\sim&\MM\int_0^\io\frac{\dd \om}{\p}\frac{\om}{\sqrt\MM}f_{\rm B}(\om)=\frac{\sqrt\MM}{\p}(\b\hbar)^{-2}\underbrace{\int_0^\io\dd x\,\frac{x}{e^x-1}}_{=\p^2/6}\\
u_{(1)}^{\rm pot}\sim&2\int_0^\io\frac{\dd \om}{2\p}\sqrt\MM\om f_{\rm B}(\om)=\frac{\sqrt\MM}{\p}(\b\hbar)^{-2}\int_0^\io\dd x\,\frac{x}{e^x-1}\\
  \frac{C_V}{N}=&\frac{C_V^{\rm kin}}{N}+\frac{C_V^{\rm pot}}{N}\sim \frac{2\p}{3}\sqrt\MM\frac{T}{\hbar}
\end{split}
\end{equation}
\ie a \textit{linear} scaling of the specific heat.

\section{Semi-classical expansion at fixed Matsubara period in the fullRSB UNSAT phase}\label{sec:fullRSBschehr}

We consider here the UNSAT phase in the non-convex part of the phase diagram, more related to structural glasses~\cite{FP16,FPSUZ17,FPUZ15,nature,CKPUZ17}. 
Having investigated in detail the RS case, we may extend more directly the results of~\secref{sec:hbar} to the fullRSB case. 

The fullRSB variational equations have been written in~\secref{sec:var}. The lowest order in the $\hbar\to0$ with fixed $\b\hbar$ expansion is, as before, 
given by the classical model at $T=0$; we here assume we start from the fullRSB UNSAT phase. The classical low-$T$ properties of this phase have been investigated
in~Ref.~\cite[Sec. 5.C.]{FPSUZ17}. The innermost overlap, corresponding to the overlap between replicas in the same basin at the very bottom of the free-energy landscape, scales like
\begin{equation}
 q_M^{\rm cl} =1-\chi T+O(T^2)
\end{equation}
and the $T=0$ classical Lagrange multiplier $P_{\rm cl}(q,h)$ is a smooth function of order 1. The lowest order for $\bar q$ and $q_M$ in the quantum case is thus 1. As a consequence we will expand them defining, as now usual,
\begin{equation}
\begin{split}
 P(q,h)&=P_{(0)}(q,h)+\hbar P_{(1)}(q,h)+O(\hbar^2)=P_{\rm cl}(q,h)+O(\hbar)\\
 \wt G(\om_n)&=\wt q_d(\om_n)-q_M\d_{n0}=(1-q_M^{\rm cl}(T=0))\d_{n0}+\hbar\,\wt G_{(1)}(\om_n)+O(\hbar^2)=\hbar\,\wt G_{(1)}(\om_n)+O(\hbar^2)\\
  \b\wt G(0)&\sim\b\hbar\wt G_{(1)}(0):=\chi
\end{split}
\end{equation}
As in the RS UNSAT phase, $\wt G(\om_n)=O(\hbar)$ $\forall n$. 
The time averages on the process $r(t)$ are thus calculated through the exact same saddle-point formulas as in~\secref{sub:UNSATexp}. 

Let us start from the thermodynamic energy by deriving the free energy~\eqref{eq:varfreedimless} and regularizing it:
\begin{equation}\label{eq:thenergyRSB}
 \frac{\UU}{N}=
 \frac{1}{2\b}\sum_{n\in\ZZZ}\frac{\m+\wt\Si(\om_n)}{\MM\left(\frac{\om_n}{\b\hbar}\right)^2+\m+\wt\Si(\om_n)}
 +\a\int\dd h\, P(q_M,h)\la v(r(t)+h) \ra_v\\
 =u_{(0)}(\b\hbar)+\hbar\, u_{(1)}(\b\hbar)+O(\hbar^2)
\end{equation}

The self-energy~\eqref{eq:qdvarFourier} is expanded as in the RS case, 
\begin{equation}\label{eq:Sifull}
 \wt\Si(\om_n)=-\a\int\dd h\, P(q_M,h)\left\{\b\d_{n0}\left[(v')^2
 + v'v^{(3)}
  \begin{tikzpicture}[baseline=-1mm]
 \draw[line width=1pt] (0:0) circle (0.3) node[cross=4pt,line width=1pt, xshift=-0.3cm]{} ;
\end{tikzpicture}
-\b v'v''v^{(3)}
\begin{tikzpicture}[baseline=-1mm]
\draw (-0.7,0) node[cross=4pt,line width=1pt,rotate=0] {};
\draw[line width=1pt] (-0.7,0) -- (0:0);
 \draw[line width=1pt] (0.3,0) circle (0.3) node[circle,fill=black,inner sep=0pt,minimum size=6pt,xshift=-0.3cm]{} ;
\end{tikzpicture} \right]
+\b (v'')^2 
\begin{tikzpicture}[baseline=-1mm]  \draw (0:0) node[cross=4pt,line width=1pt,rotate=0] {}; \draw(0.5,0.3) node {$n$}; 
\draw[line width=1pt] (0:0) -- (0:1);
 \draw (0:1) node[cross=4pt,line width=1pt,rotate=0] {}; 
\end{tikzpicture}
- v''+O(\hbar)\right\}
\end{equation}
Note that we have not expanded so far $P(q_M,h)=P_{\rm cl}(1,h)+O(\hbar)$, as it won't be required. As in the RS case, the term $\b v'(r^*+h)^2$ brings an $O(1/\hbar)$ term that do not enter the Dyson equation, and an $O(1)$ contribution expanding $r^*$. 
To find this out, we combine the $n=0$ variational equation from~\eqref{eq:qdvarFourier}
\begin{equation}
 \b\m+\b\wt\Si(0)=\frac{1}{\wt G(0)}-\frac{q_m}{\l(q_m)^2}-\int_{q_m}^{q_M}\frac{\dd q}{\l(q)^2}
\end{equation}
with the variational equation on $x(q)$,~\eqref{eq:dxq}, in $q=q_M$, to obtain:
\begin{equation}\label{eq:mu+SiRSB}
\begin{split}
 \m+\wt\Si(0)=&\frac{1}{\b\wt G(0)}-\frac\a\b\int\dd h\, P(q_M,h)f'(q_M,h)^2=\frac1\c-\a\b\int\dd h\, P(q_M,h)\la v'(r(t)+h) \ra_v^2\\
 =&\frac1\c-\a\b\int\dd h\, P(q_M,h)\left[
 (v')^2+v'v^{(3)}
\begin{tikzpicture}[baseline=-1mm]
 \draw[line width=1pt] (0:0) circle (0.3) node[cross=4pt,line width=1pt, xshift=-0.3cm]{} ;
\end{tikzpicture}
-\b v'v''v^{(3)}
\begin{tikzpicture}[baseline=-1mm]
\draw (-0.7,0) node[cross=4pt,line width=1pt,rotate=0] {};
\draw[line width=1pt] (-0.7,0) -- (0:0);
 \draw[line width=1pt] (0.3,0) circle (0.3) node[circle,fill=black,inner sep=0pt,minimum size=6pt,xshift=-0.3cm]{} ;
\end{tikzpicture}
+O(\hbar^2)
 \right]
 \end{split}
\end{equation}
where in the last line the first two orders were computed with the help of~\eqref{eq:expavvprime}. Combining~\eqref{eq:Sifull} and~\eqref{eq:mu+SiRSB} provides the lowest order of the Lagrange multiplier 
$\m=\m_{(0)}+O(\hbar)$:
\begin{equation}
 \m_{(0)}+\Si^*_{(0)}(0)=\frac1\c
\end{equation}
where we have defined, similarly to the RS case~\eqref{eq:SistarRS}, the component of the self-energy~\eqref{eq:Sifull} without the $\propto\d_{n0}$ term
\begin{equation}\label{eq:Sistarfull}
 \wt\Si_{(0)}^*(\om_n)=-\a\int\dd h\,P_{\rm cl}(1,h)\left(\b(v'')^2
\begin{tikzpicture}[baseline=-1mm]  \draw (0:0) node[cross=4pt,line width=1pt,rotate=0] {}; \draw(0.5,0.3) node {$n$}; 
\draw[line width=1pt] (0:0) -- (0:1);
 \draw (0:1) node[cross=4pt,line width=1pt,rotate=0] {}; 
\end{tikzpicture}
-v''\right)
\end{equation}
Note that in this definition all terms are $O(\hbar^0)$ which is the reason why we just need the lowest-order approximation for 
$P(q_M,h)\sim P_{\rm cl}(1,h)$.

We may now come back to the Dyson equation, and write at lowest order
\begin{equation}
 \forall n\neq0\,,\hskip15pt \m+\wt\Si(\om_n)\sim \m_{(0)}+\Si^*_{(0)}(0)+\Si^*_{(0)}(\om_n)-\Si^*_{(0)}(0)=\frac1\c+\wt I_{(0)}(\om_n)
\end{equation}
with the renormalized self-energy $\wt I_{(0)}(\om_n)=\Si^*_{(0)}(\om_n)-\Si^*_{(0)}(0)$:
\begin{equation}\label{eq:IexplicitfRSB}
 \wt I_{(0)}(\om_n)=\a\int\dd h\,P_{\rm cl}(1,h)\b(v'')^2\left(
\begin{tikzpicture}[baseline=-1mm]  \draw (0:0) node[cross=4pt,line width=1pt,rotate=0] {}; \draw(0.5,0.3) node {$0$}; 
\draw[line width=1pt] (0:0) -- (0:1);
 \draw (0:1) node[cross=4pt,line width=1pt,rotate=0] {}; 
\end{tikzpicture}
-
\begin{tikzpicture}[baseline=-1mm]  \draw (0:0) node[cross=4pt,line width=1pt,rotate=0] {}; \draw(0.5,0.3) node {$n$}; 
\draw[line width=1pt] (0:0) -- (0:1);
 \draw (0:1) node[cross=4pt,line width=1pt,rotate=0] {}; 
\end{tikzpicture}
\right)
\end{equation}
so that we can write generally the following Dyson equation:
\begin{equation}\label{eq:DysonfullRSB}
 \forall n\in\ZZZ\,,\hskip15pt\frac{1}{\b\hbar\wt G_{(1)}(\om_n)}=\MM\left(\frac{\om_n}{\b\hbar}\right)^2+\frac1\c+\wt I_{(0)}(\om_n)
\end{equation}
Following the RS case, a crucial quantity is the low-frequency scaling of the self-energy. Here as well we get a quadratic equation
using~Eqs.[\ref{eq:IexplicitfRSB}] and~[\ref{eq:DysonfullRSB}]:
\begin{equation}\label{eq:renormselffull}
 \begin{split}
  \wt I_{(0)}(\om_n)=&\a\left(\int_{-\io}^0\dd h\,P_{\rm cl}(1,h)\right)
  \left[\frac{\c}{1+\c}-\frac{1}{1+\MM\left(\frac{\om_n}{\b\hbar}\right)^2+\frac 1\c+\wt I_{(0)}(\om_n)}\right]\\
  \Leftrightarrow\hskip15pt 0=&\wt I_{(0)}(\om_n)^2
  -\a\left(\int_{-\io}^0\dd h\,P_{\rm cl}(1,h)\right)\MM\left(\frac{\om_n}{\b\hbar}\right)^2\\
  &+\wt I_{(0)}(\om_n)\left[1+\frac 1\c-\frac{\c}{1+\c}\a\left(\int_{-\io}^0\dd h\,P_{\rm cl}(1,h)\right)+\MM\left(\frac{\om_n}{\b\hbar}\right)^2\right]
 \end{split}
\end{equation}
We expect, from~\cite{SGLD04,SGLD05,S05}, that $\wt I_{(0)}(\om_n)$ is singular close to $0$ because we are in a phase where the 
landscape is marginal. The marginality condition is given by~\eqref{eq:ddxq}, which, in $q=q_M$ and at the first two lowest order, reads here\footnote{Note that $\l(q_M)=\wt G(0)$ from~\eqref{eq:lambda}.}
\begin{equation}\label{eq:ddxqchi}
\begin{split}
  \frac{1}{\chi^2}=&\frac{\a}{(1+\c)^2}\int_{-\io}^0\dd h\,P_{\rm cl}(1,h)\\
  \b\hbar\wt G_{(2)}(0)=&-\frac{\a\c^3}{2(\c+1)}\int^0_{-\io}\dd h\,P_{(1)}(q_M,h)
  -\frac{\a\c^3}{4(\c+1)}P_{\rm cl}'(1,0)\frac{1}{\b\hbar}\sum_{n\in\ZZZ}\frac{1}{1+\frac1\c+\MM\left(\frac{\om_n}{\b\hbar}\right)^2+\wt I_{(0)}(\om_n)}
\end{split}
\end{equation}
The last term in the second line comes from the diagram contributions of the $h$ derivative of~\eqref{eq:expavvprime} followed by an integration by parts. This contribution is absent in the RS regime since there $P_{\rm cl}(1,h)\to\g(h)$ (at $\s=0$). \\
Note that at the classical jamming line the behaviour is critical ($\c\to\io$) and therefore for $\s\leqslant0$ the extension of~\eqref{eq:classjammline} which defines the jamming line is now replaced by the implicit equation
\begin{equation}\label{eq:jammfRSB}
 \frac{1}{\a_J^{\rm cl}(\s)}=\restriction{\int_{-\io}^0\dd h\,P_{\rm cl}(1,h)}{\a=\a_J^{\rm cl}(\s)}
\end{equation}
which coincides with~\eqref{eq:classjammline} for $\s=0$.\\
\eqref{eq:ddxqchi} provides the same cancellation as on the $\s=0$ UNSAT line. Indeed, the renormalized self-energy in~\eqref{eq:renormselffull} becomes:
\begin{equation}\label{eq:eqforI0full}
 \wt I_{(0)}(\om_n)^2+\wt I_{(0)}(\om_n)\MM\left(\frac{\om_n}{\b\hbar}\right)^2
  -\MM\left(\frac{\om_n}{\b\hbar}\right)^2\sqrt{\a\int_{-\io}^0\dd h\,P_{\rm cl}(1,h)}=0
\end{equation}
Once again this coincides with the RS result~\eqref{eq:computeSi0} on the line $\s=0$.
As in~\secref{sec:lowTCVRS}, with $\om\leftrightarrow \om_n/(\b\hbar)$ the frequency argument of the function $\wt I_{(0)}$, we have the non-analytic 
scaling at low frequency
\begin{equation}\label{eq:gregorylikeRSB}
 \wt I_{(0)}(\om)\underset{\om\to0}{\sim}\left|\om\right|\sqrt{\MM}\left(\a\int_{-\io}^0\dd h\,P_{\rm cl}(1,h)\right)^{\frac14}
\end{equation}

Coming back to the energy~\eqref{eq:thenergyRSB}, and expanding similarly to the RS case~\eqref{eq:expu1} we get:
\begin{equation}
 \begin{split}
  u_{(0)}(\b\hbar)=&\frac\a2\int_{-\io}^0\dd h\,P_{\rm cl}(1,h)\frac{h^2}{(1+\c)^2}=e_{\rm fRSB}^{\rm cl}\\
  u_{(1)}(\b\hbar)=&
  \frac{1}{2\b\hbar}\sum_{n\in\ZZZ}\frac{\frac{1}{\chi}+\wt I_{(0)}(\om_n)}{\MM\left(\frac{\om_n}{\b\hbar}\right)^2+\frac{1}{\chi}+\wt I_{(0)}(\om_n)}\\
  &+\left(1+\frac1\c\right)^2\frac{1}{2\b\hbar}\sum_{n\in\ZZZ}\frac{1}{1+\frac{1}{\chi}+\MM\left(\frac{\om_n}{\b\hbar}\right)^2+\wt I_{(0)}(\om_n)}
  +\frac{\a}{2(1+\c)^2}\int_{-\io}^0\dd h\,h^2P_{(1)}(q_M,h)\\
  &+\frac{\a^2\c^3}{(\c+1)^4}\left(\int_{-\io}^0\dd h\,h^2P_{\rm cl}(1,h)\right)\left[\int_{-\io}^0\dd h\,P_{(1)}(q_M,h)+\frac{P_{\rm cl}'(1,0)}{2\b\hbar}\sum_{n\in\ZZZ}\frac{1}{1+\frac{1}{\chi}+\MM\left(\frac{\om_n}{\b\hbar}\right)^2+\wt I_{(0)}(\om_n)}\right]
 \end{split}
\end{equation}
$u_{(0)}$ is the ground-state energy, independent of temperature. 
Here a complication arises from the next order of the Lagrange multiplier ${P_{(1)}(q_M,h)}$, which we could not compute analytically from~\eqref{eq:eqP}. 
Nonetheless, from~\cite{S05} and since the same general mechanism has been identified here on the marginal line $\s=0$, 
we assume that the scaling of the kinetic and potential terms are identical. Next, the kinetic term is computed in a similar manner to Apps.~\ref{sub:appdAT} and~\ref{sub:kinmargin}, with the same 
low-$\om$ behaviour of the self-energy expressed in~\eqref{eq:gregorylikeRSB}, apart from irrelevant prefactors.

We conclude, as in the RS UNSAT phase, that $C_V\propto T^3$ in the whole fullRSB UNSAT phase except at jamming where the \textit{mass} $1/\chi$ vanishes, yielding a linear $C_V\propto T$, at lowest order of the present expansion.

\section{Comparison to Debye's approximation in the UNSAT phase}\label{sec:Debye}
\subsection{Vibrational density of states}\label{sub:vDOS}

The aim of this section is to compare the 
expansion of the previous sections to Debye's approximation. The latter amounts to approximate the potential energy by a quadratic form around a minimum $\bm X_{\rm min}$ of the potential-energy landscape
\begin{equation}
 H_{\rm cl}=\sum_{\m=1}^M v(h_\m)={\rm constant} +\left.\frac12\sum_{i,j}^{1,N}\frac{\partial^2H_{\rm cl}}{\partial X^i\partial X^j}\right|_{\rm min} (X^i-X^i_{\rm min})(X^j-X^j_{\rm min})+\dots
 \simeq{\rm constant}+\frac12\sum_{i=1}^N \l_i (Y^i)^2
\end{equation}
where $\l_i$ are the eigenvalues of the Hessian of the classical Hamiltonian $H_{\rm cl}$, and $\bm Y$ the coordinates in the diagonal basis.

The Hessian is naturally a random matrix due to the quenched obstacles. When $N\to\io$ one gets a continuous set of eigenvalues described by the density 
$\r(\l)=\frac1N\overline{\sum_{i=1}^N\d(\l-\l_i)}$. This spectrum has been computed in the UNSAT phase in~\cite{FPUZ15}. Defining the usual phononic frequency $\om$ by $\l=\MM\om^2$, one gets from $\r(\l)$ the probability density of vibrational modes,
or density of states (DOS) $D(\om)$, such that the measure $\r(\l)\dd\l=D(\om)\dd\om$. It has its support in a positive interval $[\om_-,\om_+]$ and reads
\begin{equation}\label{eq:DOSpropto}
 D(\om)\propto \om\frac{\sqrt{(\om^2-\om_-^2)(\om_+^2-\om^2)}}{\om^2+\zeta\varepsilon/\MM}
\end{equation}
where $ \MM\om_\pm^2/\varepsilon=\left(\sqrt{[1]}\pm1\right)^2-\zeta$ and $\zeta=[h^2]+\s[h]$ with the averaged moments of the configuration in a potential-energy minimum
$[h^n]=\overline{\la\frac1N\sum_{\m=1}^M h_\m^n\th(-h_\m)\ra}$. We have reinstated the mass $\MM$ and energy of the potential $\varepsilon$ (set to 1 in~\cite{FPUZ15}), but in order to compare with the results 
of~\secref{sec:lowTCVRS}, we will only keep the $\MM$ dependence and set $\varepsilon=1$ as energy unit. 
One can show (see footnote~\ref{footnot} and~\cite{FPUZ15}) that the following formula holds in the RS UNSAT phase
\begin{equation}
 [h^n]=\frac{\a}{(1+\chi)^n}\int_{-\io}^0\dd h\, \g(h+\s)h^n
\end{equation}
which allows to compute $\om_\pm$ and $\z$. Note that for $\s\leqslant0$, $\zeta\geqslant0$~\cite{FPUZ15,FPSUZ17} and one can then define a cutoff frequency $\om_*=\sqrt{\zeta\epsilon/\MM}$. 
It is associated in the literature to a plateau in the DOS, since the latter is roughly flat for $\om_*\ll\om\ll\om_+$. This cutoff frequency goes to zero at jamming~\cite{FPUZ15,OHSLN03,WSNW05,XVLN10,DGLFDLW14}.

The quantization of these harmonic vibrations brings an energy $\hbar\om f_{\rm B}(\om)$ per mode $\om$ and relates the thermodynamic energy $\UU_{\rm Debye}$ to the DOS as
\begin{equation}\label{eq:UDeb}
 \frac{\UU_{\rm Debye}}{N}=\int_0^\io \dd \om\,  D(\om)\,\hbar\om f_{\rm B}(\om)
\end{equation}

\subsection{Gapped phase}

The stability of the single-well free energy (RS) is related to the fact that $\om_->0$ in the corresponding region $\s>0$. 
There, the spectrum has a gap $\hbar\om_-$, and one sees easily that~\eqref{eq:UDeb} yields for $T\to0$ an exponentially small specific heat scaling as $\exp(-\hbar\om_-/T)$.\\
This is also the scaling of the Matsubara sums appearing in the expression of $u_{(1)}(\b\hbar)$ 
in~\eqref{eq:expu1}, computed in App.~\ref{app:Matsubara}. Thus the lowest-order expansion agrees with the Debye result. However, anticipating~\secref{sec:finitehbar}, the higher orders of the expansion naturally perturb $\omega_-$ with corrections in power series of $\hbar$.

\subsection{Gapless phase}
Conversely, the LMS phase implies that the landscape has flat directions along which soft modes can flow, and the gap closes, \ie $\om_-=0$. 
In order to make closer contact with the results of~\secref{sec:lowTCVRS}, we focus on the LMS UNSAT phase\footnote{\label{footnot} The rest of the fullRSB UNSAT region 
may be analyzed similarly by computing the moments $[h^n]$. This is achieved~\cite{FPSUZ17,FP16,FPUZ15} by computing the gap probability distribution 
 \begin{equation*}
  \overline{\la\frac 1M \sum_{\m=1}^M\d(h-h_\m(\bm X))\ra}=\frac1\a\frac{\d f^{\rm cl}}{\d v(h)}
 \end{equation*}
 where $f^{\rm cl}$ is the classical free energy per dimension. Only 
the knowledge of $P_{\rm cl}(1,h)$ is required.}, and more precisely on its RS boundary $\s=0$. 
Since there the fraction of contacts, or of unsatisfied constraints, is $[1]\geqslant1$ (the system is isostatic at jamming, and hyperstatic above it), the normalization is given by $\int\dd\om\,D(\om)=1$~\cite{FPUZ15,FP16,FPSUZ17}, \ie it is the probability of finding a phonon 
mode $\om$ in the spectrum, and from~\eqref{eq:DOSpropto}
\begin{equation}\label{eq:DOSnorm}
 D(\om)=\frac{4/\p}{\MM\om_+^2-2\z \left[\sqrt{\frac{\MM\om_+^2+\z}{\z}}-1\right]}\,\MM^{\frac32}\om^2\frac{\sqrt{\MM\om_+^2-\MM\om^2}}{\MM\om^2+\z}\qquad \textrm{for}~0\leqslant\om\leqslant\om_+
\end{equation}
and $D(\om)=0$ otherwise. 
Away from the jamming line $\zeta\neq0$ (\ie $\om_*\neq0$). Performing the same low-$T$ approximation of the integral as in~\secref{sec:lowTCVRS}, one gets a cubic specific heat
\begin{equation}\label{eq:DebCV}
\frac{C_V^{\rm Debye}}{N}\underset{T\to0}{\sim}\frac{16\p^3}{15\z}\frac{\sqrt{\MM\om_+^2}}{\MM\om_+^2-2\z \left[\sqrt{\frac{\MM\om_+^2+\z}{\z}}-1\right]} \MM^{\frac32} \left(\frac T\hbar\right)^3
\end{equation}
On the jamming line, the constraints are on the verge of satisfiability $h_\m=0$, making $\zeta\propto\om_*^2$ vanish. This implies $C_V\propto T$. 
For a specific comparison of the prefactor on the jamming point for $\s=0$ we have ${\sqrt{\MM}\om_+(\s=0,\a=2)=2}$ and one gets the linear specific heat
\begin{equation}\label{eq:Debyejamm}
 \frac{C_V^{\rm Debye}}{N}\underset{T\to0}{\sim}\frac{2\p}{3}\sqrt\MM\frac T\hbar
\end{equation}
The different scaling is caused by the emergence of a large number of soft modes at jamming related to isostaticity, 
which induce a flat DOS at small frequency instead of $D(\om)\propto\om^2$ in the rest of the LMS phase~\cite{FPUZ15}.\\

The Debye approximation gives back the same results as the lowest order $\hbar\to0$ expansion at fixed $\b\hbar$ in the RS UNSAT phase. This is manifest at jamming from the expressions~[\ref{eq:CVjammRS}] and~[\ref{eq:Debyejamm}]. Out of the jamming point, Eqs.~[\ref{eq:CVpot}],~[\ref{eq:CVint}] add up to retrieve~\eqref{eq:DebCV}. Notice that the scalings $C_V\propto T$ at jamming and $C_V\propto T^3$ out of jamming are predicted for the whole LMS UNSAT phase by Debye's approximation, since the vibrational DOS~\eqref{eq:DOSnorm} is valid throughout this phase.

\section{Effect of finite $\hbar$ and the avoided jamming transition}\label{sec:finitehbar}

Summarizing, we performed an expansion for $\hbar\to0$ \textit{first}, with 
\begin{equation}
\wh T=\frac{T}{\hbar} 
\end{equation}
fixed, \ie $T=O(\hbar)$ (unlike the usual semiclassical expansion).\\

The question now is to discuss what happens for finite $\hbar$. For this we will use Schehr's results about the expansion to all orders, hinting at a more direct low-$T$ expansion, detailed in~Ref.~\cite{S05}. \\
We notice first that we computed the first order in the semiclassical expansion, and all the features of the three quantum glassy models studied in~Ref.~\cite{S05} have been recovered at this order, namely:
\begin{itemize}
 \item a \textit{gapped} scaling of the self-energy $\wt I_{(0)}(\om)\underset{\om\to0}{\propto} \om^2$ and thus of the specific heat out of the LMS phase
 \item in the LMS phase, due to the marginality condition, a \textit{gapless} scaling of the self-energy $\wt I_{(0)}(\om)\underset{\om\to0}{\propto} |\om|$ gives a cubic power-law behaviour of the specific heat. We will discuss 
 later the case of the jamming line, which anyway has no analog in the models analyzed in~Ref.~\cite{S05}
 \item in the LMS phase, both kinetic and potential energy terms scale with the same power law in temperature.
\end{itemize}
Having recovered the basic mechanism (at lowest order\footnote{Note that in principle, following the strategy of~\secref{sec:hbar} and~\ref{sec:fullRSBschehr}, one could check directly if this SGLD mechanism generalizes to higher orders of the $\hbar$ expansion with fixed $\b\hbar$, especially coming from the RS UNSAT phase as in~\ref{sec:hbar}; nonetheless this is a non-trivial task, as the next order already requires a computation of the self-energy to 3-loop order.}) described in~Ref.~\cite{S05}, whose validity is due only to the LMS condition applied on a similar thermodynamic energy, we will \textit{assume} here that this general 
mechanism is at work also in this quantum glassy model, which allows us to extrapolate to the regime $\hbar$ finite, $T\to0$. 

\subsection{Extrapolation to finite $\hbar$}

Both kinetic and potential energy terms being expected to scale in the same way, as discussed above, in the following we focus on the kinetic term in the regularized expression of the energy~\eqref{eq:thenergyRSB}:
\begin{equation}
u_{\rm kin}=\frac{1}{2\b}\sum_{n\in\ZZZ}\frac{\m+\wt\Si(\om_n)}{\MM\left(\frac{\om_n}{\b\hbar}\right)^2+\m+\wt\Si(\om_n)}
\end{equation}
This term is similar to the one studied in the quantum periodic elastic manifold in a random potential or the $p$-spin quantum glass~\cite{SGLD04,SGLD05,S05}. 
We rewrite the self-energy\footnote{In~\cite{S05}, this is the same notation except that $\Xi$ is called $\Si$.} as
\begin{equation}\label{eq:newq}
 \m+\wt\Si(\om_n)\underset{n\neq0}{=}\Xi+\wt I(\om_n) \ , \qquad \wt I(0)=0
\end{equation}
where the definition is similar to the ones in~\secref{sec:fullRSBschehr}: define $\wt\Si^*(\om_n):=\wt\Si(\om_n)$ for any $n\neq0$, which is a function of $n$, but due to $\d_{n0}$ contributions 
in $\wt\Si(\om_n)$, we have that $\wt\Si^*(\om_n\to0)\neq\wt\Si(0)$. Then the new quantities in~\eqref{eq:newq} are defined $\forall n$ by
\begin{equation}
 \Xi:=\m+\wt\Si^*(0) \ ,\qquad \wt I(\om_n):=\wt\Si^*(\om_n)-\wt\Si^*(0)
\end{equation}
The same kind of avoidance of the $\d_{n0}$ contributions is present in~Ref.~\cite{S05}, because the ``correct'' analytic continuation of $\d_{n0}$ for $\om_n\to -i\om+0^+$ is not obvious. 
Here we may just remember that they do not play any role in the thermodynamics of the system.
Now we have
\begin{equation}\label{eq:kineticMatsubara}
u_{\rm kin}=\frac{1}{2\b}\sum_{n\in\ZZZ}\frac{\Xi+\wt I(\om_n)}{\MM\left(\frac{\om_n}{\b\hbar}\right)^2+\Xi+\wt I(\om_n)}
\end{equation}
Out of the LMS phase, following the generalization of~\cite{S05}, the self-energy is analytic close to $\om=0$ (here we have calculated its first order in the semiclassical expansion), consequently 
the phase is gapped and the specific heat is exponentially damped for $T\to0$, by a computation similar to~\secref{sub:kinmargout}. 

Let us now consider the LMS phase. Due to the LMS condition~\cite{S05}, at low $T$ \textit{independently of the semiclassical expansion}, the self-energy becomes non-analytic close 
to zero: $\wt I(\om)\sim B |\om|+B'\om^2$, with $B>0$. 
In other words, all orders $\wt I_{(k)}(\om)$ in the $\hbar\to0$, $\wh T$ fixed expansion follow the same low-$\om$ behaviour as the one we computed for $\wt I_{(0)}(\om)$ 
in~\eqref{eq:gregorylikeRSB} (apart from the prefactor $B_{(k)}$, where $B=\sum_{k\geqslant0}B_{(k)}\hbar^k$, which may be $k$ dependent).
Similarly one gets 
\begin{equation}
 \Xi=\frac1\chi+\sum_{k=1}^\io\Xi_{(k)}\hbar^k
\end{equation}
where again we have only computed here the first order $1/\chi$ in the semiclassical expansion in~\eqref{eq:ddxqchi}.
From~\cite{S05} we know that $\Xi$, $B$ and $B'$ are $T=0$ constants, \ie \textit{they hold no $T$ dependence}. \\
As in App.~\secref{sub:kinmargin}, we may write the Matsubara sum as an integral for $T\to0$
\begin{equation}\label{eq:intapproxukin}
 u_{\rm kin}\underset{T\to0}{\sim}   \int_0^\io\frac{\dd \om}{\p}\, 
 \frac{B\MM\om^2}{\Xi^2+[B^2-2\Xi(\MM+B')]\om^2}\hbar\om f_{\rm B}(\om)
 \underset{ x:=\om/\wh T}{=}\ \hbar B\MM\wh T^4\int_0^\io\frac{\dd x}{\p}\, 
  \frac{ x^3}{\Xi^2+(b\wh T)^2 x^2}\frac{1}{e^x-1}
\end{equation}
where we defined in the last equality $b:=\sqrt{B^2-2\Xi(\MM+B')}$. 
Only in the case $\Xi=0$ does the actual value of $b$ affect the prefactor of the power law in $T$, albeit without modifying the exponent (see App.~\secref{sub:kinmargin}). In any case this is just an unknown non-zero constant at this stage. 

\subsection{Temperature cutoff: avoided jamming criticality}

For $T\to0$,~\eqref{eq:intapproxukin} provides the leading order in the power-law dependence of the (kinetic) energy. Higher orders necessitate the inclusion of the full $\om$-dependence of the renormalized self-energy. 
The low-$T$ scaling of~\eqref{eq:intapproxukin}  is understood by comparing the mass term $\Xi$ to the renormalized self-energy's contribution $b\wh T$, as it  depends on a single parameter formed by their ratio. This defines a cutoff temperature $T_{\rm cut}$:
\begin{equation}
 T_{\rm cut}=\frac{\hbar\,\Xi}{b}
\end{equation}
which has indeed the dimension of an energy. As a consequence, the specific heat at low temperature scales \textit{cubically} $C_V\propto T^3$ for $T\ll T_{\rm cut}$, and \textit{linearly} for $T\gg T_{\rm cut}$ (then at higher temperatures, higher orders start to play a role).

The regime in between is when $T_{\rm cut}$ scales linearly with $T$. Indeed~\eqref{eq:intapproxukin} reads
\begin{equation}\label{eq:scalingGG}
\begin{split}
    u_{\rm kin}\underset{T\to0}{\sim} 
   \hbar \frac{B\MM}{b^2}\left(\frac{T}{\hbar}\right)^2\GG\left(\frac{T}{T_{\rm cut}}\right)\qquad
  \textrm{with}\qquad\GG(\xi)=&\int_0^\io\frac{\dd x}{\p}\,   \frac{ x^3}{\xi^{-2}+ x^2}\frac{1}{e^x-1}\\
  \GG~\textrm{increases monotonically, with asymptotics}\qquad \GG(\xi)\underset{\xi\to 0}{\sim}&\frac{\p^3}{15}\xi^2
  \quad ,\quad \GG(\xi)\underset{\xi\to \io}{\sim}\frac{\p}{6}
 \end{split} 
\end{equation}

Now let us focus on the value of $T_{\rm cut}$. 
Above the jamming line, the leading order in $\Xi$ is a mere constant $1/\c$, which likely dominates its value, $\hbar$ be small enough. We note suggestively that upon approaching jamming from the non-convex UNSAT phase ($\s<0$), 
$1/\c\propto \sqrt{\d\epsilon}$ where $\d\epsilon$ is the distance to the jamming line\footnote{Conversely, the scaling is different in the convex UNSAT phase~\cite{FPUZ15,FP16,FPSUZ17}, and is easily deduced from~\eqref{eq:varchi}: $1/\c\propto\d\epsilon$.}; this has been deduced from the numerical solution of the $T=0$ classical equations in~\cite{FPUZ15,FP16,FPSUZ17}. For instance $\d\epsilon=\a-\a_J^{\rm cl}(\s)$ 
where $\a_J^{\rm cl}(\s)$ is the jamming line fixed by~\eqref{eq:jammfRSB}; any other direction from the UNSAT phase would do. 
We also note similarly that since close to jamming ($\c\gg1$), $[h^n]\propto\chi^{-2n}$~\cite[Sec. 5.6.]{FPSUZ17}, therefore $\z\sim\s[h]$ and $\om_*\propto1/\c$.  
Thus, $\hbar$ be small enough, one has $T_{\rm cut}\propto\Xi\propto\om_*\propto \sqrt{\d\epsilon}$.

Let us temporarily reinstate the dimensionful constants $k_B$, $\varepsilon$ (not to be confused with $\d\epsilon$) and $\DD$ of the problem to estimate an order of magnitude of the cutoff temperature. 
$\DD$ is the unit length (set so far to 1) and has physically the meaning of the typical interaction 
range with the obstacles (the diameter of the particle). In a real liquid in Euclidean space it would be roughly the inter-particle distance / 
particle diameter / typical interaction length. Since $\Xi$ cannot have an $\hbar$ nor $T$ dependence as mentioned above, and its leading order has been shown not to contain any $\MM$ dependence, dimensional analysis\footnote{The dimensions are $[\Xi]={\rm mass}\cdot{\rm time}^{-2}$, $[B]={\rm mass}\cdot {\rm time}^{-1}$ and $[B']={\rm mass}$.} implies that  $\Xi\propto\varepsilon\sqrt{\d\epsilon}/\DD^2$ at leading order. 
Similarly, we know that the $\MM$ dependence of $B$ (respectively $B'$) at lowest order is $B\propto\sqrt\MM$ (respectively $B'\propto\MM$) from~Eqs.~[\ref{eq:eqforI0full}],~[\ref{eq:gregorylikeRSB}]. 
Thus assuming no other dependence in $\hbar$ or $\MM$, \ie assuming that this leading order dominates its value, gives an order of magnitude for $b\propto\sqrt{\MM\varepsilon/\DD^2}$. We then get
\begin{equation}\label{eq:Tc_estim}
  T_{\rm cut}\propto \sqrt{\frac{\varepsilon}{\MM}}\frac{\hbar}{k_B\DD}\sqrt{\d\epsilon}\propto 1\  {\rm K}\cdot\sqrt{\d\epsilon}
\end{equation}
up to a purely numerical factor. 
Here we estimated values of the dimensionful constants from Zeller and Pohl's experimental data~\cite{ZP71} of a typical glassformer (${\rm Si}{\rm O}_2$). 
Namely, $\DD\approx10^{-9}$ m and $\MM\simeq 60$ u $\approx 10^{-25}$ kg. 
Note that the order of magnitude of the mass $\MM$ and the inter-particle distance $\DD$ 
is rather universal and would not change significantly by considering other molecular glasses. 
Besides we took $\varepsilon/k_B\approx 100$ K, a typical value\footnote{In this paper the soft-spheres pair potential $v$ is an inverse power law with exponent 12, but this does not change the typical order of magnitude $\varepsilon$ taken for the pair potential.} given by~\cite{BRH90} for Argon. 
Clearly this is just an extrapolation from the real-liquids parameters in this more abstract spherical perceptron model; nevertheless it gives an order of magnitude which seems reasonable, in view of the fact that the onset of low-temperature anomalies is usually and universally detected below $10$~K~\cite{ZP71,PCRTRVR14,PCJRR14}. 

The crucial point we wish to emphasize here is that the cutoff temperature scales like $\om_*\propto\sqrt{\d\epsilon}$ at leading order, and thus as one gets closer to the unjamming transition (one may think about density as a control parameter in a liquid instead of $\d\epsilon$ here), the cutoff temperature is lowered; this might provide a mechanism for the specific heat anomaly.

\begin{figure}[h!]
\begin{center}
\begin{tabular}{cc}
 \includegraphics[height=6.5cm]{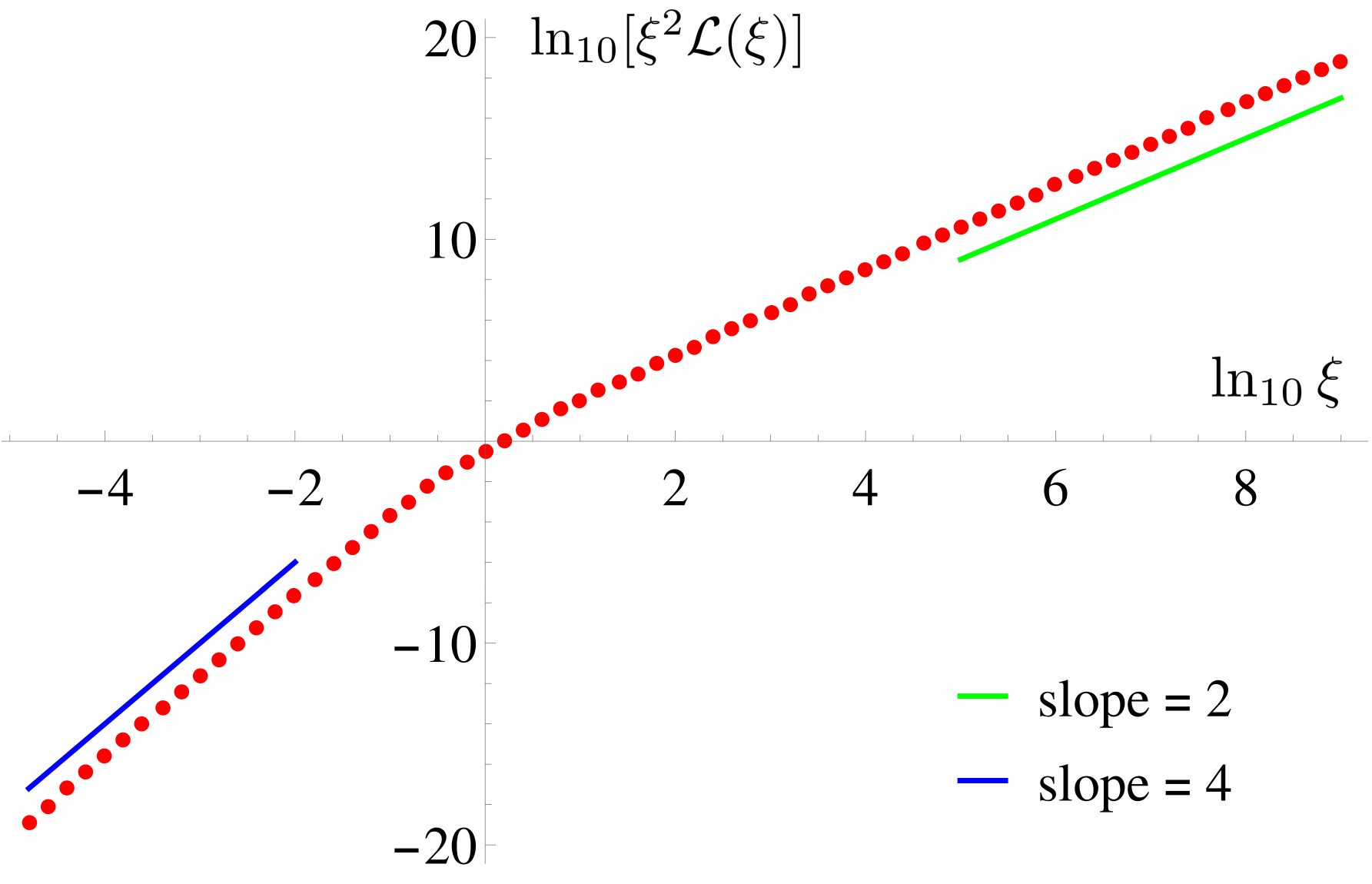}&
 \includegraphics[height=7.5cm]{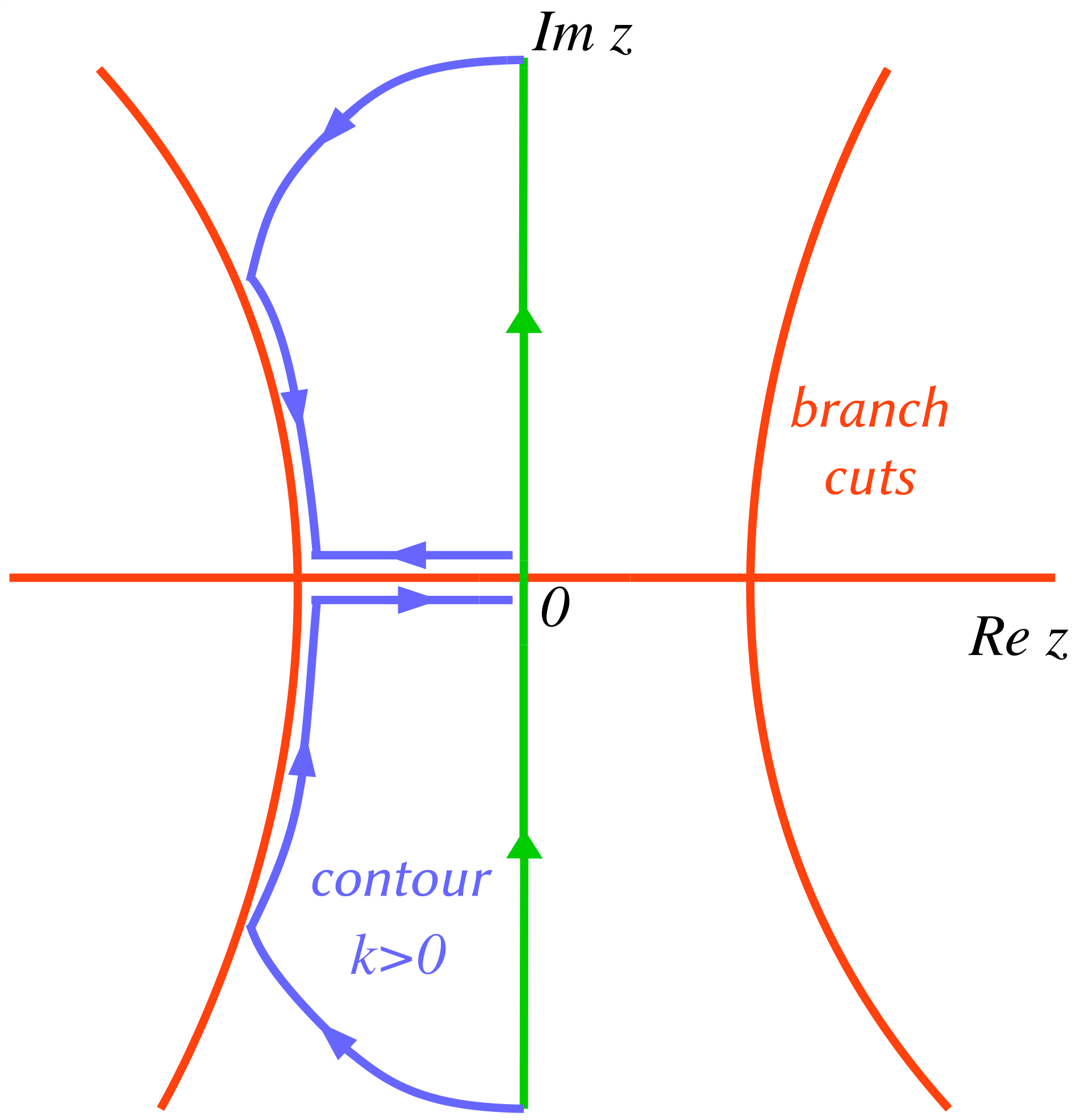}\\
 (a)&(b)
\end{tabular}
 \caption{(a) Log-log plot of the integral $\xi^2\LL(\xi)$ appearing in the scaling form of $C_V^{\rm kin}$ in~\eqref{eq:Lscaling}.\\ (b) Contours used to compute $\SS_k$ (green line). The green line is closed using the blue ones, wherein the integrand is analytic and the closed integral vanishes.}
 \label{fig:scalingcontour}
 \end{center}
\end{figure}

We conclude from~\eqref{eq:intapproxukin} that the contribution of the kinetic term to the low-temperature specific heat is given by the following scaling function:
\begin{equation}\label{eq:Lscaling}
\frac{C_V^{\rm kin}}{N}\propto \frac{B\MM}{b^2} \frac{T}{\hbar}\LL\left(\frac{T}{T_{\rm cut}}\right)
\qquad \textrm{with} \qquad \xi^{2}\LL(\xi)=\int_0^\io\frac{\dd x}{\p}\,   \frac{ x^3}{1+ x^2}\frac{e^{x/\xi}}{(e^{x/\xi}-1)^2}
\end{equation}
where $\LL(\xi)\underset{\xi\to0}\propto\xi^2$ and reaches a constant for $\xi\to\io$, its behaviour is displayed in Fig.~\ref{fig:scalingcontour}(a). 
If $T_{\rm cut}$ turns out to be below the experimental temperature, then the specific heat scales roughly linearly as
\begin{equation}
 \frac{C_V^{\rm kin}}{N}\propto \frac{B\MM }{b^2\hbar}\,T
\end{equation}
Conversely in the temperature range $T\ll T_{\rm cut}$ it obeys the scaling 
\begin{equation}
 \frac{C_V^{\rm kin}}{N}\propto \frac{B\MM}{\Xi^2\hbar^3}\,T^3
\end{equation}

The prefactors are of course expected to be renormalized by the potential-energy contribution, although the $T^3$ prefactor should blow up for $T_{\rm cut}\to0$ in any case.

Contrary to the prediction of the semiclassical Debye analysis of~\secref{sec:Debye}, the linear scaling is \textit{always avoided} at extremely low temperature due to $\hbar$ corrections to the \textit{mass} $\Xi$ (\ie jamming criticality is avoided), making $T_{\rm cut}$ finite. 
This reflects the physical observation that the classical $T=0$ jamming transition itself must be rounded-off in the quantum regime $\hbar\neq0$, becoming a crossover, as mentioned in the main text. 

A crude  estimate of $T_{\rm cut}$ on the jamming line $\d\epsilon=0$ can be done in a similar way to~\eqref{eq:Tc_estim}. 
Taking into account only the next order in the mass term, one has $\Xi\propto\hbar$. Since we have not computed $\Xi_{(1)}$, this time we cannot exclude a $\MM$ dependence. Yet we can exclude again both further dependence on $\hbar$ or $T$. It turns out that dimensional analysis indeed needs the $\MM$ dependence to be conclusive and $\Xi\propto(\hbar/\DD^3)\sqrt{\varepsilon/\MM}$, while we keep the same expressions for $B,B'$ given by their leading order. 
Thus one gets a correction to the leading order of the cutoff temperature ${T_{\rm cut}(\d\epsilon=0)\propto {\hbar^2}/({k_B\MM\DD^2})\approx 0.01}$~K, 
again up to a purely numerical prefactor.  As expected one gets a lower estimate than~\eqref{eq:Tc_estim}, but of course numerical prefactors may turn out to be large. 


\subsection{Status of the Debye approximation}

In~\secref{sec:Debye} we have seen that the Debye approximation in the RS gapless phase yields the same results as the lowest order in the $\hbar\to0$, fixed $\b\hbar$ expansion. This is natural as in a RS regime we expect the Debye approximation to hold, since the system should be well approximated at low temperature by harmonic motions around an energy minimum. Nonetheless we saw that corrections appear when adding higher orders in $\hbar$, which even destroy the linear scaling predicted by the Debye approximation at $T=0$. 
Besides by construction, in a gapped phase, the Debye approximation cannot predict anything else than a gap which scales \textit{linearly} with $\hbar$, whereas perturbative corrections to this linear scaling are expected to arise when adding more orders in the $\hbar\to0$, fixed $\b\hbar$ expansion. This would happen either by a motion of the branch cuts of the self-energy at higher orders, or by the emergence of new poles. We conclude that the Debye approximation can be interpreted as providing correct semiclassical  results (\ie the right scaling and prefactors, or the right gap, in the limit $\hbar\to0$), if the system is replica symmetric. It is not clear if this interpretation extends to the bulk of the LMS phase since we could not compute exactly the prefactors.

In App.~\ref{app:p=2} we show that in a spherical spin-glass model with a quadratic Hamiltonian the imaginary-time formalism yields the same exact result as the Debye approach, the latter holding strictly in such a model by definition. The SGLD expansion truncates exactly to its first order in this model, as expected from the correspondence discussed above.

\clearpage
\appendix
\renewcommand{\thesubsection}{\thesection.\arabic{subsection}}

\section*{\textit{NOTES}}

\vspace{1cm}

\section{A free quantum particle on a large-dimensional hypersphere}

In this section we study the free energy of the non-interacting model ($v=0$), \ie a free particle motion constrained on the {$(N-1)$-dimensional} sphere of radius $N$. 
In~\secref{app:freesphere}, we first study it from the \textit{exact} spectrum of the free-particle Hamiltonian, to get the exact result. Secondly, in~\secref{app:discrete}, we compare it to the Feynman path-integral representation~\cite{FHS10,Kleinert} 
used in this paper. The motivation is twofold: on the one hand to test that the way we treat the problem, allowing to study the effect of disorder and interactions, does not introduce spurious features in the partition function, in particular yields the correct 
energy spectum. This is true \textit{up to a zero-point energy}. 
On the other hand, divergences that do not only affect the zero-point energy appear in the field theory. They must be regularized accordingly; by discretizing the path integral, we show that one retrieves 
the same variational equations as in the continuous-time limit (\secref{sub:free}), and that the free energy is finite as it should when properly discretized.

\subsection{Computation of the partition function from the exact energy spectrum}\label{app:freesphere}

The (quantum) Hamiltonian for the free particle on the {$(N-1)$-dimensional} sphere of radius $N$ must be constructed from the angular momentum, as it is the generator of the only symmetry of the system~\cite{FG70}. It reads
\begin{equation}
 \hat H_{\rm free}=\frac{\hat L^2}{2\MM N}
\end{equation}
whose eigenvalues are given by a discrete quantum number $l\in\mathbbm{N}$~\cite[Sec.8.9.]{Kleinert}\cite{KS97,EF14}, similar to the well-known ${N=3}$ case~\cite{Cohen-Tannoudji}:
\begin{equation}\label{eq:eigenL}
 L^2_l=\hbar^2 l(l+N-2)
\end{equation}
Here, a comment is in order since the formulation on the sphere is a subtle matter. In~\cite{Kleinert}, it is shown that constructing the Feynman path integral on the sphere by defining position and momentum operators spherically constrained and accounting of the constraint through an exponential 
representation of the Dirac function $\d(\bm X^2-N)$, which is the procedure we follow in~\secref{sec:Z} and~\secref{sec:timeRSB} using a Lagrange multiplier, 
does not yield the \textit{exact} energy spectrum on the sphere, but replaces the angular momentum eigenvalues (\eqref{eq:eigenL}) by $\hbar^2[(l+N/2-1)^2-1/4]=L_l^2+\hbar^2N^2/4$ up to subdominant constants for large $N$.
This spectrum anyway coincides with~\eqref{eq:eigenL}, up to an additional extensive constant $N\hbar^2/8\MM$ to the energy for large $N$ (see also~\cite{Sc02} on this point). \\ 
Ultimately, these subtleties depend crucially on the definition of the quantum model; 
an ambiguity arises here since the perceptron is originally defined in a classical setting without any concept of momentum. 
Even classically a proper definition of the perceptron in a dynamical setting must be realized paying attention to this issue,  
as one must constrain the momentum to remain tangent to the sphere~\cite{ABUZ18}. 
Here we will use the arguably \textit{correct} definition of the Hamiltonian on a sphere through the angular momentum and make a direct computation. 
The path integral version of the computation is equivalent to a quantum harmonic oscillator model because of the introduction of a Lagrange multiplier for the spherical constraint; in this setting one may use 
the canonical momentum and position operators to define the Hamiltonian (see App.~\ref{app:discrete} for more details).

In order to compute the free energy we also need the degeneracy of the levels. The number of spherical harmonics for a given quantum number $l$ is given by~\cite[Sec. 4.1.]{EF14}\cite[Eq. (8.113)]{Kleinert}:
\begin{equation}\label{eq:gl}
 g_l=\frac{(2l+N-2)(l+N-3)!}{l!(N-2)!}=\frac{2l+N-2}{l}\begin{pmatrix}l+N-3\\l-1\end{pmatrix}
\end{equation}
Note that $g_0=1$. For $N=3$ it yields the well-known result $g_l=2l+1$. For $N=2$ it yields $g_{l\neq0}=2$ which corresponds to the fact that the angular momentum is only along the $z$-axis in two dimension. The eigenvalues 
of $\hat L_z$ are indeed $m\hbar$ with $m=0,\pm1,\pm2,\dots$ hence the twofold degeneracy~\cite{Cohen-Tannoudji}.\\
Then we may compute the partition function. 
At large $N$, a first attempt could be to write the energies as $E_l={\hbar^2}l/{2\MM}$, which amounts to assume that the $l\ll N$ dominate the series. 
This corresponds also to the limit of large $\b$. One has in this approach
\begin{equation}
  \sum_{l\geqslant0}g_l e^{-\b E_l}\underset{N\to\io}{\approx}\sum_{l\geqslant0}\begin{pmatrix}l+N-1\\l\end{pmatrix}e^{-\frac{\b\hbar^2}{2\MM}l}
\end{equation}
Using derivatives of the Taylor series representation of $1/(1-x)$ for $\abs{x}<1$, one gets the following identity
\begin{equation}\label{eq:binome}
 \frac{1}{(1-x)^N}=\sum_{l\geqslant0}\begin{pmatrix}l+N-1\\l\end{pmatrix}x^l   
\end{equation}
we then have under this hypothesis
\begin{equation}
  \sum_{l\geqslant0}g_l e^{-\b E_l}\underset{N\to\io}{\approx}\frac{1}{\left(1-e^{-\frac{\b\hbar^2}{2\MM}}\right)^N}=\exp\left[-N\ln\left(1-e^{-\frac{\b\hbar^2}{2\MM}}\right)\right]
\end{equation}
This means that the free energy would be:
\begin{equation}\label{eq:freelinear}
 \FF_{\rm free}\sim NT\ln\left(1-e^{-\frac{\b\hbar^2}{2\MM}}\right)
\end{equation}
which has, by construction, the correct low-temperature behaviour (the energy gap is ${\D_{\rm free}=\b\hbar^2/2\MM}$), but gives an energy $\la \hat H_{\rm free}\ra\sim NT$ at large temperature, missing a 1/2 factor since one expects that in the large $N$ limit we should recover a free particle 
in a {$\sim N$-dimensional} space (\ie where the effect of the spherical constraint would be negligible for large $N$). 

In the opposite approach, one may keep only the quadratic term in the energy, assuming the $l\gg N$ dominate the partition function, \ie $E_l=(\hbar l)^2/2\MM N$. 
We already know that this cannot be a good approximation for large $\b$, \eg since this spectrum is gapless for large $N$, but it is suitable for small $\b$. 
In this approximation we have $g_l\sim l^{N-2}$. Hence, the energy density of states is $\rho(E)\sim E^{N/2}$, the partition function may be approximated by the integral $\int_0^\io \dd E\, E^{N/2}e^{-\b E}$, giving the same result as the classical ideal gas, \ie the 
temperature dependence of the free energy is $\FF_{\rm free}\sim \frac N2 \ln T $, which is indeed fine at large temperature.

From the previous discussion one can infer that the partition function is dominated by the quantum orbital numbers $l=O(N)$, which make the energy eigenvalues extensive and allow to retain in the energy both linear and quadratic dependences upon the quantum orbital number $l$. Defining $\l=l/ N$ and using Stirling's approximation one has
\begin{equation}
 g_l\sim e^{N\phi(\l)}\hskip15pt\textrm{with}\hskip15pt \phi(\l)=(1+\l)\ln(1+\l)-\l\ln\l
\end{equation}
Then the partition function takes the form of a Riemann sum and we get:
\begin{equation}
  \sum_{l\geqslant0}g_l e^{-\b E_l}=  \sum_{l\geqslant0}e^{N\left[\phi(\l)- \frac{\b\hbar^2}{2\MM}\l(\l+1)\right]}\underset{N\to\io}{\sim}N\int_0^\io\dd\l\,e^{N\left[\phi(\l)- \frac{\b\hbar^2}{2\MM}\l(\l+1)\right]}
\end{equation}
The latter integral can be evaluated via the saddle-point method:
\begin{equation}
 \ln(1+\l^{\rm sp})-\ln\l^{\rm sp}=\frac{\Lambda_{\rm dB}^2}{4\p}(2\l^{\rm sp}+1)
\end{equation}
whose unique solution for each value of the parameter (the de Broglie thermal length, defined in~\eqref{eq:Lag0}) can be obtained graphically. 
In the limit $\Lambda_{\rm dB}\ll1$, the saddle-point solution comes from the large values of $\l$, and the saddle-point equation becomes
\begin{equation}
 \frac{1}{\l^{\rm sp}}\sim\frac{\Lambda_{\rm dB}^2}{2\p}\l^{\rm sp} \hskip15pt \Longrightarrow \hskip15pt \l^{\rm sp}\sim\sqrt{\frac{\MM}{\b\hbar^2}}
\end{equation}
At large $\l$ one has $\phi(\l)=\ln\l+1+O(1/\l)$ thus the partition function becomes $Z_{\rm free}\sim (\MM/\b\hbar^2)^{N/2}$, which gives the classical ideal gas partition function, as it should at high temperature. \\
In the opposite limit $\Lambda_{\rm dB}\gg1$, $\l^{\rm sp}$ becomes small and $\phi(\l)=\l(1-\ln\l)+O(\l^2)$, implying the saddle-point value
\begin{equation}\label{eq:finalappA}
\begin{split}
 &\l^{\rm sp}\sim \exp\left(-\frac{\b\hbar^2}{2\MM}\right)\hskip15pt\Rightarrow\hskip15pt 
 \FF_{\rm free}\sim -NT\exp\left(-\frac{\b\hbar^2}{2\MM}\right)\\
  \hskip15pt\Rightarrow\hskip15pt
  &\frac{\UU_{\rm free}}{N}=\frac{\hbar^2}{2\MM}\exp\left(-\frac{\b\hbar^2}{2\MM}\right)
 \hskip15pt\Rightarrow\hskip15pt
 \frac{C_V^{\rm free}}{N}=\left(\frac{\b\hbar^2}{2\MM}\right)^2\exp\left(-\frac{\b\hbar^2}{2\MM}\right)
 \end{split}
\end{equation}
This corresponds to the small temperature behaviour and we see that it agrees with the same limit of the previous free energy~\eqref{eq:freelinear} which neglects the quadratic term in the eigenvalues of the Hamiltonian. 
This is the expected result at low temperature since the Hamiltonian has a gap $\D_{\rm free}=\hbar^2/(2\MM)$ (at large $N$).

\subsection{Computation from an alternative quantization scheme and regularization of the field theory}\label{app:discrete}

Here we take a closer approach to the one used in the interacting case: we employ the same quantization scheme with constrained position and momentum on the sphere.  Let us start from the path integral similar to~\eqref{eq:repl_Z}:
\begin{equation}\label{eq:partfsph}
 Z_{\rm sph}=\oint\mathrm{D}\bm X\, \exp\left(-\frac{1}{\hbar}\int_0^{\b\hbar}\dd t\,\frac{\MM}{2}(\dot {\bm X})^2(t)\right)
 =\int_{\bm X(0)=\bm X(\b\hbar)}\mathrm{D}\bm X\, \exp\left(-\frac{1}{\hbar}\int_0^{\b\hbar}\dd t\,\left[\frac{\MM}{2}(\dot {\bm X})^2(t)+\frac{\m_0}{2}(\bm X^2(t)-N)\right]\right)
\end{equation}
This path integral can be computed by adding a Lagrange multiplier $\m_0$ for convenience to enforce the spherical constraint and discretizing time in $M$ time slices, with standard techniques~\cite{FHS10,Kleinert}. Equivalently one can follow the same formalism as in the interacting case~\secref{sec:Z} defining a $M\times M$ time-overlap matrix $Q_{mn}=\bm X_m\cdot\bm X_n/N$ which is calculated via a saddle-point method for large $N$. Note that in this discrete setting one can show that the integration by parts used in~\eqref{eq:avdis} $\int_0^{\b\hbar}\dd t\, \dot{\bm X}^2(t)=-\int_0^{\b\hbar}\dd t\,\bm X(t)\cdot\ddot{\bm  X}(t)$ holds in its discretized version, although the paths are not differentiable. We do not report the full derivation as it is somewhat lengthy but its outcomes are:
\begin{itemize}
 \item In the continuous-time limit $M\to\io$ one retrieves the time-dependent overlap function~\eqref{eq:SP_F0} (describing the translation-invariant saddle-point value of $Q_{mn}$) and the spherical constraint providing the value of the Lagrange multiplier~\eqref{eq:Lag0}. 
 This is an important point: the saddle-point equations are not affected in any way by divergences, even in the direct continuous-time formalism of~\secref{sec:Z}.
 \item The free energy is not anymore divergent owing to this proper continuous-time limit and the well-defined path integral measure. 
 It thus has a different expression from $F_0$ of~\eqref{eq:F0_final}; the latter can be recovered in a \textit{naive} (and incorrect) large-$M$ limit. 
\end{itemize}
Let us now compute equivalently but straightforwardly the free energy $\FF_{\rm sph}$. We remark that the Hamiltonian we have to deal with in~\eqref{eq:partfsph} is the one of a collection of $N$ independent harmonic oscillators:
\begin{equation}\label{eq:Hamiltoniandiscr}
 \frac{\hat H_{\rm sph}}{N}=\frac{\hat P^2}{2\MM}+\frac{\m_0}{2}(\hat X^2-1)=\hat H_{\rm harmonic}-\frac{\m_0}{2}
\end{equation}
where the harmonic oscillator has frequency $\sqrt{\m_0/\MM}$. 
The only difference with a harmonic oscillator here is that the frequency is fixed in terms 
  of the basic physical parameters by the saddle-point equation at large $N$, \ie the spherical constraint (\eqref{eq:Lag0}). 
The free energy thus reads
\begin{equation}\label{eq:finaldiscrete}
 \frac{ \FF_{\rm sph}}{N}=\frac{1}{\b}\ln\left[2\sinh\left(\frac{\b\hbar}{2}\sqrt{\frac{\m_0}{\MM}}\right)\right]-\frac{\m_0}{2}
\end{equation}
One can obtain explicit results in the small and large temperature limits, determining $\m_0$ in these limits which 
has been done in~\secref{sub:zeroV}:
\begin{itemize}
 \item for high temperature, $\b\m_0\sim1$ and one gets the $T$ dependence of the free energy from~\eqref{eq:finaldiscrete} (up to irrelevant constants to $\ln Z_{\rm sph}$)
 \begin{equation}
  \frac{\b \FF_{\rm sph}}{N}\sim\ln\L_{\rm dB}
   \hskip15pt\Leftrightarrow\hskip15pt
   \frac{\UU_{\rm sph}}{N}=\frac{\la \hat H_{\rm sph}\ra}{N}\sim\frac{T}{2}
 \end{equation}
which is the correct result from the classical energy equipartition.
\item at low temperature, with~\eqref{eq:m0lowT}, the energy reads
\begin{equation}\label{eq:UUsph}
 \frac{\UU_{\rm sph}}{N}=\frac{\la\hat H_{\rm sph}\ra}{N}\sim\frac{\hbar^2}{8\MM}+\frac{\hbar^2}{2\MM}\exp\left(-\frac{\b\hbar^2}{2\MM}\right)
\end{equation}
up to higher exponentially small corrections. The term $\hbar^2/(8\MM)$ actually coincides with the extra contribution at order $O(N)$ of the incorrect angular momentum eigenvalues
 with respect to the exact energy spectrum~\eqref{eq:eigenL} (as discussed in~\secref{app:freesphere}). 
Nevertheless the constants in the energy may be discarded and the only temperature dependence is in the exponential term, yielding an exponentially vanishing specific heat given by~\eqref{eq:finalappA}. 
We recover the right gap $\D_{\rm free}=\hbar^2/(2\MM)$ ruling the low-temperature quantum value.
\end{itemize}


We now prove a final convenient relation $\UU_{\rm sph}/N=\m_0/2$. From the harmonic oscillator spectrum or the free energy in~\eqref{eq:finaldiscrete}, one has
\begin{equation}\label{eq:sphenergy}
  \frac{\UU_{\rm sph}}{N}=\frac{\la\hat H_{\rm sph}\ra}{N}=
  \left(\frac{\hbar}{2}\sqrt{\frac{\m_0}{\MM}}+\frac{\b\hbar}{4\sqrt{\m_0\MM}}\frac{\partial\m_0
  }{\partial\b}\right)\coth\left(\frac{\b\hbar}{2}\sqrt{\frac{\m_0}{\MM}}\right)-\frac{\m_0}{2}-\frac{\b}{2}\frac{\partial\m_0 }{\partial\b}=\frac{\m_0}{2}
\end{equation}
where we eliminated the $\coth$ factor through~\eqref{eq:Lag0} defining $\m_0$.

\section{Computations of the sums over Matsubara frequencies}\label{app:Matsubara}

Here we aim at evaluating sums of the type 
\begin{equation}\label{eq:SS2}
 \SS=\frac{1}{\b\hbar}\sum_{n\in\ZZZ}\frac{1}{\MM\om_n^2+A+\wt I_{(0)}(\om_n)}\qquad\textrm{and}\qquad
 \SS'=\frac{1}{\b\hbar}\sum_{n\in\ZZZ}\frac{A+\wt I_{(0)}(\om_n)}{\MM\om_n^2+A+\wt I_{(0)}(\om_n)}
\end{equation}
with $A>0$, $\om_n=2\p n/(\b\hbar)$ and $\wt I_{(0)}$ defined by~\eqref{eq:I0solution}. We will first compute $\SS$; we deal with $\SS'$ in~\secref{sub:SSprime}.\\
We recall that the structure of the complex function $\phi(z)=\wt I_{(0)}(-iz)$  is given by Fig.~\ref{fig:branch}, 
and that we are focusing on the small temperature limit $\b\to\io$. We proceed by applying the Poisson summation formula~\cite[Chap.11]{Appel}, which identifies~\eqref{eq:SS2} with the sum of the Fourier transform of the summand, 
\ie it amounts to write
\begin{equation}\label{eq:Poisson}
  \frac{2\p}{\b\hbar}\sum_{n\in\ZZZ}\d(\om-\om_n)=\sum_{k\in\ZZZ}e^{i\b\hbar k\om}\qquad
  \Rightarrow \qquad\SS=\sum_{k\in\ZZZ}\SS_k \qquad\textrm{with}\qquad \SS_k=\int_{-\io}^\io\frac{\dd \om}{2\p}\frac{e^{i\b\hbar k\om}}{\MM \om^2+A+\wt I_{(0)}(\om)}
\end{equation}
For $\b\to\io$, the behaviour is controlled by the small-$\om$ limit of the integrand, which changes qualitatively in or out of the dAT line. 
We already note that the term $k=0$ can be discarded, since it is purely a constant to the energy $\hbar u_{(1)}(\b\hbar)$, independent on $\b\hbar$.

\subsection{At the landscape marginal stability line}\label{sub:appdAT}

\begin{figure}[h!]
\begin{center}
 \begin{tabular}{ccc}
  \includegraphics[width=8cm]{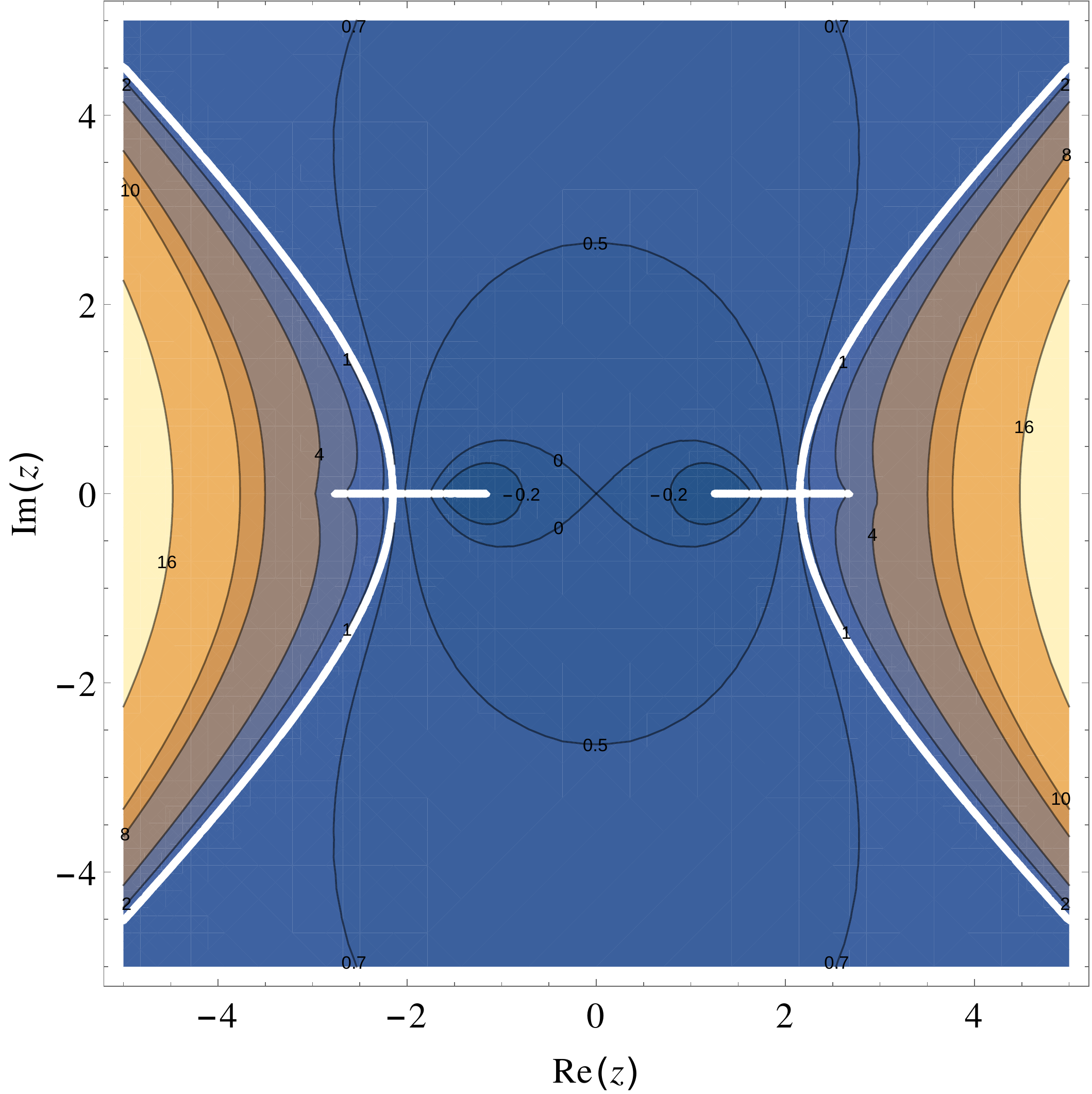}
  & \hspace{1cm} &\includegraphics[width=8cm]{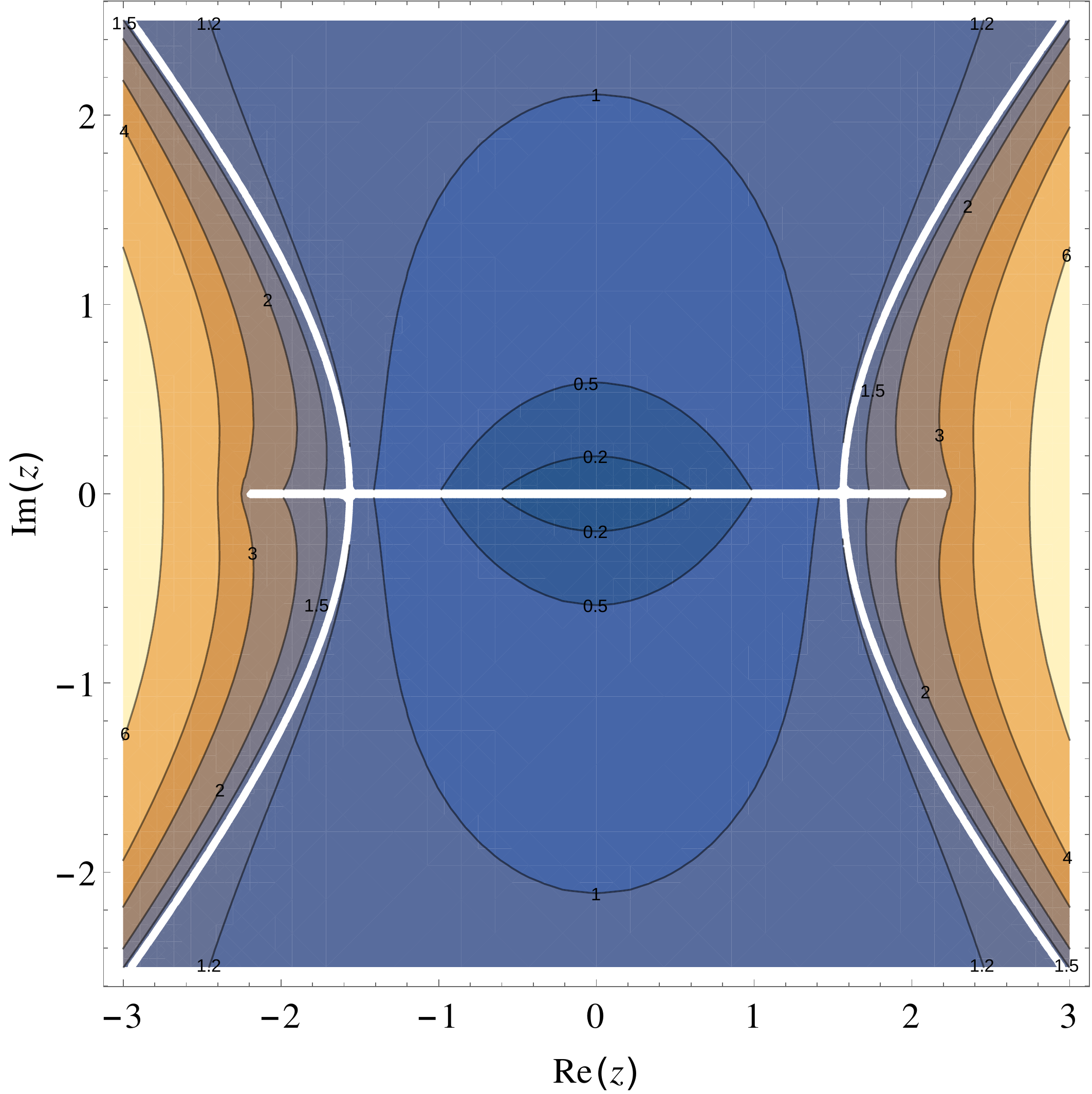}\\
  (a) & \hspace{1cm} & (b)
 \end{tabular}
\caption{Representative plots of $\Re[\phi(z)]$ for different values of $(\s,\a)$ with $\MM=1$, exhibiting the branch cuts of $\phi$ (white solid lines). \\
(a) $(\s,\a)=(2,3)$, in the replica-symmetric UNSAT phase. There is a gap between the two branch cuts on the real axis, so that $\phi$ is analytic around $z=0$. \\
(b) $(\s,\a)=(0,3)$, on the dAT line. The latter gap closes and $\phi$ is not analytic around $z=0$.}
\label{fig:branch}
\end{center}
\end{figure}

Here $K(\s=0,\a)=0$ and the two branch cuts on the real axis in Fig.~\ref{fig:branch}(a) join, resulting in a single branch cut in a symmetric interval on the real axis including $z=0$ 
(Fig.~\ref{fig:branch}(b)). For $A>0$ the integral $\SS_k$ is still convergent but now there is a non-analyticity in zero. For $k\neq0$ again only the analytic properties of 
$\phi(z)$~\eqref{eq:analyticphi} around $z=0$ matter, the parabolic branches being sent to infinity for $\b\to\io$. Let us rewrite the integral with $z=i\om$ in order to refer to Fig.~\ref{fig:branch}
\begin{equation}
 \SS_{k\neq0}=\int_{-i\io}^{i\io}\frac{\dd z}{2i\p}\frac{e^{\b\hbar kz}}{A-\MM z^2+\phi(z)}
\end{equation}
First we note that, since $\phi(-z)=\phi(z)$, the change of variables $z\to -z$ provides $\SS_k=\SS_{-k}$, so we can assume $k>0$. In order to compute this integral we close the contour both above and below the branch cut 
on the real axis, see Fig.~\ref{fig:scalingcontour}(b). We have checked numerically that there are no poles inside the contour.

For large $\b$ we may approximate with the $z\to0$ behaviour, nevertheless to go to higher orders in $T$ we will retain the whole dependence, \ie
$A-\MM z^2+\phi(z)=A-\MM z^2/2+\sqrt{-C\MM z^2}\sqrt{1-\MM z^2/4C}$ and we have\footnote{Notice the slight notation abuse here: we extended the interval of integration to $-\io$ whereas it should be extended to only $\om=-2\sqrt{C/\MM}$ owing to the square root in $\phi(z)$. But we will be interested in the limit $T\to0$ in which $\om$ is rescaled by $\b\hbar$, defining a new variable $\om'=\b\hbar\om$ which extends to $-2\b\hbar\sqrt{C/\MM}\underset{T\to0}{\to}-\io$, see~\secref{sec:lowTCVRS}.}, taking a small $\epsilon>0$
\begin{equation}\label{eq:SSkposapp}
\begin{split}
 \SS_{k>0}=&-\int_{-\io}^0\frac{\dd \om}{2i\p}\frac{e^{\b\hbar k(\om+i\epsilon)}}{A-\frac\MM2(\om+i\epsilon)^2+\sqrt{-C\MM(\om+i\epsilon)^2}\sqrt{1-\MM (\om+i\epsilon)^2/4C}}\\
 &-\int_{0}^{-\io}\frac{\dd \om}{2i\p}\frac{e^{\b\hbar k(\om-i\epsilon)}}{A-\frac\MM2(\om-i\epsilon)^2+\sqrt{-C\MM(\om-i\epsilon)^2}\sqrt{1-\MM (\om-i\epsilon)^2/4C}}
 \end{split}
\end{equation} 
The branch cut introduces a discontinuity in the imaginary part of $\phi$, since if we choose the square root to have a branch cut on the negative real axis we get  
$\sqrt{-C\MM(\om\pm i\epsilon)^2}\underset{\epsilon\to0^+}{=}{\pm i\sqrt{C\MM}|\om|}$. Hence for $\epsilon\to0^+$
\begin{equation}
  \SS_{k>0}\underset{\om\to-\om}{=}\int_{0}^{\io}\frac{\dd \om}{\p}e^{-\b\hbar k \om}\,\Im\left(\frac{1}{A-\MM\om^2+\wt I_{(0)}(\om_n\to-i\om+0^+)}\right)
\end{equation}
Once again performing the total sum for $\om>0$, the Bose-Einstein factor shows up:
\begin{equation}
 \sum_{k=1}^\io e^{-\b\hbar k \om}=\frac{1}{e^{\b\hbar  \om}-1}=:f_{\rm B}(\om)
\end{equation}
from which we conclude
\begin{equation}
\begin{split}
  \SS=&\SS_0+\frac2\p\int_{0}^{\io}\dd \om\, f_{\rm B}(\om)\Im\left(\frac{1}{A-\MM\om^2+\wt I_{(0)}(\om_n\to-i\om+0^+)}\right)\\
  =&\SS_0+\frac2\p\int_{0}^{\io}\dd \om\, f_{\rm B}(\om)\frac{\sqrt{C\MM}\om\sqrt{1-\MM \om^2/4C}}{(A-\MM\om^2/2)^2+C\MM\om^2(1-\MM \om^2/4C)}
\end{split}
\end{equation}
where $\SS_0$ is a constant.

\subsection{Out of the landscape marginal stability line}\label{sub:appmargout}

The computation is similar to the the above case. The $k>0$ contour is nearly the same as in Fig.~\ref{fig:scalingcontour}(b), except that here the horizontal branch cut does not extend up to $z=0$ (see Fig.~\ref{fig:branch}(a)) but extends up to the branch point where it is closed by a small half circle around it. This branch point is easily calculated by imposing that the square root in $\phi(z)$ be zero (marginally negative), 
which gives 
\begin{equation}
 \abs{z_{\rm bp}}=\frac{\sqrt{C+K}-\sqrt C}{\sqrt\MM}=\omega_-
\end{equation}
with $\omega_-$ defined in~\secref{sub:vDOS}, the lower edge of the vibrational DOS (remember that here $\varepsilon=1$). As we have checked numerically that there are no poles inside the contour, the integral can be computed neglecting the closing parts of the contour which are sent to infinity, and also the part along the non-horizontal branch cut, as it is damped for $T\to0$ by an exponential factor much smaller than the one provided by the contour around the horizontal branch cut. The small half circle is parametrized by $z=-\om_-+re^{i\th}$ with $\th\in[-\p/2,\p/2]$, which gives a vanishing contribution when the half circle shrinks to zero ($r\to0^+$). Then the only remaining integrals to compute are given by~\eqref{eq:SSkposapp} with $-\om_-$ instead of $0$ in the integral boundaries. Performing the same manipulations one arrives to the conclusion that $\SS_{k>0}\propto e^{-\b\hbar k\om_-}$, and then summing over $k$ one gets that $\SS\propto e^{-\b\hbar\omega_-}$ for $T\to0$ up to an additive constant $\SS_{k=0}$ (not depending upon temperature).


\subsection{Sum from the kinetic term}\label{sub:SSprime}

We can follow very similar steps for the sum $\SS'$ in~\eqref{eq:SS2}.

\subsubsection{On the dAT line}\label{sub:kinmargin}

As in App.~\ref{sub:appdAT}, one can write $ \SS'=\sum_{k\in\ZZZ}\SS'_k $ with:
\begin{equation}
\begin{split}
 \SS'_{k>0}=\SS'_{-k}&=\int_{-i\io}^{i\io}\frac{\dd z}{2i\p}\frac{A+\phi(z)}{A-\MM z^2+\phi(z)}e^{\b\hbar kz}\\
 &\sim \int_0^\io\frac{\dd \om}{2i\p}e^{-\b\hbar k\om}\left[\frac{A+\MM \om^2/2-i\sqrt{C\MM}\om\sqrt{1-\MM \om^2/4C}}{A-\MM\om^2/2-i\sqrt{C\MM}\om\sqrt{1-\MM \om^2/4C}}
-\frac{A+\MM \om^2/2+i\sqrt{C\MM}\om\sqrt{1-\MM \om^2/4C}}{A-\MM\om^2/2+i\sqrt{C\MM}\om\sqrt{1-\MM \om^2/4C}}\right]\\
  &=\int_0^\io\frac{\dd \om}{\p}e^{-\b\hbar k\om}\MM\om^2\,\Im\left(\frac{1}{A-\MM\om^2+\wt I_{(0)}(\om_n\to-i\om+0^+)}\right)\\
 &=\int_0^\io\frac{\dd \om}{\p}e^{-\b\hbar k\om}\frac{\MM\sqrt{C\MM}\om^3\sqrt{1-\MM \om^2/4C}}{(A-\MM\om^2/2)^2+C\MM\om^2(1-\MM \om^2/4C)}
\end{split}
\end{equation}
where we follow the same contours to evaluate the integrals. The difference here is that the cut also affects the numerator. We conclude
\begin{equation}
 \SS'\sim\SS_0'+2\int_0^\io\frac{\dd \om}{\p}\MM\om^2\,\Im\left(\frac{1}{A-\MM\om^2+\wt I_{(0)}(\om_n\to-i\om+0^+)}\right)f_{\rm B}(\om)
\end{equation}

\subsubsection{Out of the dAT line}\label{sub:kinmargout}

The calculation is identical to the one in App.~\ref{sub:appmargout}, with extra terms on the numerator of the integrand, which can be treated as in App.~\ref{sub:kinmargin}.
We obtain similarly for $T\to0$, up to an additive constant, $\SS'\propto e^{-\b\hbar\omega_-}$.

\section{Benchmarking the approach with the spherical Sherrington-Kirkpatrick model}\label{app:p=2}

In this section we consider a spherical version of the Sherrington-Kirkpatrick (SK) model~\cite{SK75} described by the Hamiltonian\footnote{We may define the Hamiltonian with a sum over spins that is either on the pairs $\sum_{i<j}$ or with self-interactions $\frac12\sum_{i,j}$. In the large $N$ limit the spectrum of the Hessian in the Debye approach, and respectively the partition function of the system, are not affected by this convention.}
\begin{equation}
  H_2=\frac{\bm P^2}{2\MM}+\frac12\sum_{i,j}^{1,N}J_{ij}S_i S_j+\frac{\m}{2}\left(\sum_{i=1}^N S_i^2-N\right)
\end{equation}
where $J_{ij}$ are i.i.d. Gaussian couplings with zero mean and variance $\overline{J_{ij}^2}=\tilde J^2/N$. As usual $\m$ enforces the spherical constraint $\sum_{i=1}^N S_i^2=N$. We added momenta $P_i$, $i\in\llbracket1,N\rrbracket$, in order to define a quantum version of the model following Refs.~\cite{CGSS01,SS81}: the spins are commuting position operators with commuting canonical momenta associated to them, obeying the commutation relations $[\hat S_i,\hat P_j]=i\hbar\d_{ij}$.

In the following we compute the specific heat of the model by the two methods used in this paper with the spherical perceptron: the Debye approximation and a direct calculation in the imaginary-time formalism. 
Since the model is quadratic the Debye approximation must be exact, thus providing a benchmark of the approach. 

This model is solved by a replica-symmetric ansatz and is marginal at low temperature~\cite{CGSS01,SS81}; subsequently we expect from Ref.~\cite{S05} 
that $C_V\propto T^3$ for $T\to0$. We find that both approaches gives the latter law with the same prefactor, contrary to the situation with the spherical perceptron, where it happens \textit{only at lowest order in} $\hbar$ in the RS UNSAT phase.
This is because, as we shall see, for the spherical SK model the SGLD expansion truncates to the first order, which is not the case for the spherical perceptron. 

\subsection{Debye's approximation}

The Hessian is $ \partial^2 H_2/\partial S_i\partial S_j=J_{ij}+\mu\d_{ij}$, \ie a random matrix with GOE statistics for large $N$ whose eigenvalues get shifted by $\mu$. 
The value of this Lagrange multiplier must be computed for an absolute minimum of the Hamiltonian $\bm S^*$. The condition $\partial H_2/\partial S_i=0$ together with the spherical constraint give the value of the Lagrange multiplier $\m N=-\sum_{i,j}J_{ij}S_i^*S_j^*$. This is difficult to compute for a single minimum or a single realization of the disorder, yet here one can calculate its value for a typical minimum, important for the thermodynamics. This static value can be obtained through the saddle-point evaluation of the classical free energy, which can be expressed by diagonalizing the interaction Hamiltonian or through the replica trick as a function of the replica overlap $Q_{ab}=\overline{\la\bm S^a\cdot\bm S^b\ra}/N$~\cite{KTJ76}. The RS ansatz $Q_{aa}=1$, $Q_{a\neq b}=q$ is always stable in this model: for high $T$ the model is paramagnetic ($q=0$) and at $T=\tilde J$ there is a critical transition to a spin-glass value $q=1-(T/\tilde J)^2$ (see \eg Ref.~\cite[Sec. 2.3.]{these} for the derivation). The spherical constraint reads 
\begin{equation}
 \b\m=(\b\tilde J)^2+\frac{1-2q}{(1-q)^2}
\end{equation}
yielding $\m=2\tilde J$ in the spin-glass phase. The density of eigenvalues follows therefore a shifted semicircle law~\cite{vivo,mehta}
\begin{equation}
\r(\l)=\frac{\sqrt{4\tilde J\l-\l^2}}{2\p\tilde J^2}\ ,\qquad \l\in[0,4\tilde J]
\end{equation}
This low-$T$ spectrum turns out to be marginally stable. As in~\secref{sec:Debye} through $\l=\MM\om^2$ we get the vibrational DOS 
\begin{equation}
 D(\om)=\frac2\p\left(\frac{\MM}{\tilde J}\right)^{\frac32}\om^2\sqrt{1-\frac{\MM\om^2}{4\tilde J}}\ ,
 \qquad 0\leqslant\om\leqslant2\sqrt{\tilde J/\MM}
\end{equation}
We apply the Debye approximation relating the DOS to the thermodynamic energy through~\eqref{eq:UDeb} and consequently get via the same analysis as in~\secref{sec:Debye} the following scaling
\begin{equation}\label{eq:UDebp=2}
 \frac{U_{\rm Debye}}{N}=\textrm{constant}+\frac{2\p^3}{15}\hbar\left(\frac{\MM}{\tilde J}\right)^{\frac32}\left(\frac{T}{\hbar}\right)^{4}+O(T^6)
\end{equation}

\subsection{Solution of the quantum thermodynamics}

The quantum thermodynamics can be studied in a similar way to the classical model, with the overlap order parameter $Q_{ab}(t,t')=\overline{\la\bm S^a(t)\cdot\bm S^b(t')\ra}/N$ which is RS stable: $Q_{aa}(t,t')=q_d(t-t')$ and $Q_{a\neq b}=q$. We refer to Refs.~\cite{SS81,CGSS01} for the derivation.
It is convenient to express the result with unit of energy $\tilde J$ and unit of time $\hbar/\tilde J$ \ie $\tilde J=\hbar=1$. We simply quote the result~\cite{S05}, \ie the energy and the saddle-point equations determining the overlaps and self-energy in the large-$N$ limit:
\begin{equation}\label{eq:p=2SP}
\begin{split}
 \frac{\overline{\la H_2\ra}}{N}=&\frac{z'}{2}+\frac12\int_0^\b\dd t\, [q_d(t)-q]-\int_0^\b\dd t\, [q_d(t)^2-q^2]\\
 q=&1-\frac1\b\sum_{n\in\ZZZ}\frac{1}{\MM\om_n^2+z'+\wt\Si(\om_n)}\\
 q_d(t)-q=&\frac1\b\sum_{n\in\ZZZ}\frac{\cos(\om_n t)}{\MM\om_n^2+z'+\wt\Si(\om_n)}\\
 \wt\Si(\om_n)=&\int_0^\b\dd t\, [1-\cos(\om_n t)][q_d(t)-q]
 \end{split}
\end{equation}
$\om_n=2\p n/\b$ are the Matsubara frequencies. $z'$ is related to the Lagrange multiplier; for a generic spherical $p$-spin model the spherical constraint reads ${(z')^{-2}=2q^{2-p}/[p(1+x_p)]}$. 
The marginality condition here reads $x_p=p-2$~\cite{CGSS01,S05}, hence here for $p=2$ in the spin-glass phase $q\neq0$ one has $z'=1$ ($z'>0$).
With~\eqref{eq:p=2SP} one arrives at the energy
\begin{equation}\label{eq:energyp=2}
  \frac{\overline{\la H_2\ra}}{N}=\frac2\b\sum_{n\in\ZZZ}\frac{1}{\MM\om_n^2+1+\wt\Si(\om_n)}-\frac1\b\sum_{n\in\ZZZ}\frac{1}{\left[\MM\om_n^2+1+\wt\Si(\om_n)\right]^2}
\end{equation}
and the self-energy can be straightforwardly computed
\begin{equation}\label{eq:SEp=2}
 \wt\Si(\om_n)^2+\MM\om_n^2\wt\Si(\om_n)=\MM\om_n^2 \qquad\Rightarrow\qquad
 \wt\Si(\om_n)=\frac12\left(-\MM\om_n^2+\sqrt{\MM^2\om_n^4+4\MM\om_n^2}\right)
\end{equation}
We notice it is exactly the (renormalized) self-energy of the perceptron in the RS UNSAT phase at lowest-order in the SGLD expansion given in~\eqref{eq:I0solution} with $K=0$ (marginality) and $C=1$. A difference here is that we don't need to perform this expansion as the model is exactly solvable. Moreover the latter expansion is \textit{exactly} given by its first order \ie $\wt\Si=\wt\Si_{(0)}$ as~\eqref{eq:SEp=2} is of the form $f(\wt\Si(\om_n),\MM\om_n)=0$ implying no extra dependence in $\hbar$ than the one contained in the Matsubara frequencies, in other words vanishing higher orders $\wt\Si_{(i)}(\om_n)=0$ $\forall i>0$. It thus has the same analytic properties in the complex plane, in particular it is singular around $\om=0$ ($\wt\Si(\om)\sim\sqrt\MM\abs{\om}$), which implies a gapless phase and power-law dependence of the specific heat.

The Matsubara sums appearing in the energy~\eqref{eq:energyp=2} are evaluated at low temperature as in App.~\ref{app:Matsubara}. 
One has
\begin{equation}
\begin{split}
   \frac{\overline{\la H_2\ra}}{N}\underset{T\to0}{\sim}&\textrm{constant}+\frac4\p\int_0^\io\dd\om\,\frac{\sqrt\MM\om\sqrt{1-\MM\om^2/4}}{(1-\MM\om^2/2)^2+\MM\om^2(1-\MM\om^2/4)}f_{\rm B}(\om) \\
  &- \frac4\p\int_0^\io\dd\om\,\frac{\sqrt\MM\om(1-\MM\om^2/2)\sqrt{1-\MM\om^2/4}}{\left[(1-\MM\om^2/2)^2+\MM\om^2(1-\MM\om^2/4)\right]^2}f_{\rm B}(\om) 
\end{split}
\end{equation}
The two integrals come respectively from the two sums in~\eqref{eq:energyp=2}. Each term is $O(T^2)$ but this first order cancels from the sum, as in the general mechanism identified in~\cite{S05}. The low-temperature behaviour is therefore, reinstating the dimensionful constants $\tilde J$ and $\hbar$, 
\begin{equation}
 \frac{\overline{\la H_2\ra}}{N}=\textrm{constant}+\frac{2\p^3}{15}\hbar\left(\frac{\MM}{\tilde J}\right)^{\frac32}\left(\frac{T}{\hbar}\right)^{4}+O(T^6)
\end{equation}
which yields the exact same specific heat as in the Debye approximation~\eqref{eq:UDebp=2}, like it should for this quadratic Hamiltonian.

\newpage

\section*{\textit{FIGURES}}
\vspace{1cm}

%

%

\vspace{1cm}
\newpage

\clearpage
\bibliography{HS}